\documentclass{aa}  
\usepackage{graphicx}
\usepackage{lscape}
\usepackage{natbib}
\usepackage{txfonts}
\begin{document}
   \title{Internal kinematics of spiral galaxies in distant clusters
   \thanks{Based on observations collected at the European Southern 
   Observatory (ESO), Cerro Paranal, Chile (ESO Nos. 66.A--0546 , 66.A-0547 \& 74.B--0592)}
   }

   \subtitle{III. Velocity fields from FORS2/MXU spectroscopy}

   \author{E. Kutdemir\inst{1,2}
          \and
          B. L. Ziegler\inst{3,1}
          \and
          R. F. Peletier\inst{2}
          \and
          C. Da Rocha\inst{3,1}	  
          \and
          T. Kronberger\inst{4}
          \and
          W. Kapferer\inst{4}
          \and
          S. Schindler\inst{4}
          \and
          A. B\"ohm\inst{5}
          \and
          K. J\"ager\inst{6}
          \and
	  Harald Kuntschner\inst{7}
          \and
          M. Verdugo\inst{1}
          }

   \offprints{E. Kutdemir}

   \institute{Institut f\"{u}r Astrophysik, Georg-August-Universit\"{a}t,
              Friedrich--Hund--Platz 1, 37077 G\"{o}ttingen, Germany\\
              \email{kutdemir@kapteyn.astro.rug.nl}
         \and
         Kapteyn Astronomical Institute, P.O. BOX 800, 9700AV, Groningen, 
         The Netherlands
	 \and 
	 European-Southern Observatory, Karl-Schwarzschild Str. 2, 85748 Garching, Germany 
         \and
         Institut f\"{u}r Astro- und Teilchenphysik, Universit\"{a}t Innsbruck,
	 Technikerstrasse 25,6020 Innsbruck, Austria
         \and
	 Astrophysikalisches Institut Potsdam, An der Sternwarte 16, 14482 Potsdam, Germany
	 \and
	 Max-Planck-Institut f\"{u}r Astronomie, D-69117 Heidelberg, Germany
         \and
	 ST-ECF, Karl-Schwarzschild Str. 2, 85748 Garching, Germany
             }

   \date{Received ; accepted }

% \abstract{}{}{}{}{} 
% 5 {} token are mandatory
 
  \abstract
   {We continue our investigation on how the cluster environment affects the evolution
of galaxies.}
   {By examining both galaxy structure and internal kinematics of cluster galaxies at
lookback times of $\sim5$\,Gyr we study the nature and impact of possible
interactions at the peak epoch of cluster assembly. } 
   {Going beyond our previous measurements of two-dimensional rotation curves, we here
observe the whole velocity field of the galaxies of the sample.  We achieve a complete
coverage and optimal spatial sampling of galaxy sizes by placing three adjacent and parallel
FORS2 MXU (Mask eXchange Unit) slits onto each object yielding simultaneously several
emission and absorption lines.  We reconstruct the gas velocity field and decompose it into
circular rotation and irregular motions using a harmonic decomposition method called
kinemetry. To measure the irregularity in the gas kinematics, we define 3 parameters:
$\sigma_{PA}$ (the standard deviation of the kinematic position angle within a galaxy),
$\Delta \phi$ (the average misalignment between kinematic and photometric position angles)
and $k_{3,5}$ (squared sum of the higher order Fourier terms).}   
 { We present the analysis of the velocity fields and morphology of 22 distant galaxies in the
MS\,0451.6--0305 field with 11 members at $z=0.54$ and a local sample from SINGS.  Using local,
undistorted galaxies the three parameters $\sigma_{PA}$, $\Delta \phi$ and $k_{3,5}$ can be used
to establish the regularity of the gas velocity fields.  Among the galaxies for which we could
measure these parameters, we find both field ones (4 of 8)  and  cluster members (3 of 4), which
have a velocity field that we consider both irregular and asymmetric.  We show that these
fractions are underestimates of the total number of objects with irregular velocity fields.  The
values  of the irregularity parameters for cluster galaxies are not very different from those of
the field galaxies, implying that there are isolated field galaxies that are as distorted as the
cluster members.  None of the deviations in our small sample correlate with photometric/structural
properties like luminosity or disk scale length in a significant way. }
   {We have demonstrated that our 3D-spectroscopic method successfully maps the
velocity field of distant galaxies.  Together with a structural analysis the
importance and efficiency of cluster specific interactions can be assessed
quantitatively.}


   \maketitle
%
%________________________________________________________________

\section{Introduction}

The motion of stars and gas clouds within a galaxy are important measurable
characteristics representative of the whole system.  Since the internal
kinematics are subject to the overall gravitational potential they provide a
proxy for the total mass.  In addition to the baryonic mass of gas and stars,
that may be inferred from photometric observations, velocities trace the dark
matter distribution \citep[e.g.][]{SR01}.  The (ir)regularity of its
3-dimensional velocity field can provide clues about possible distortions of a
galaxy, such as warps.  Peculiar velocity fields may also be indicators of recent
or ongoing interaction processes.  This is particularly important in the
environment of galaxy clusters, where specific interactions occur in addition to
merging and accretion events that are observed in the field population and that
are elements of the hierarchical growth of structure in the Universe
\citep[e.g.][]{Gorko04,P04}.  Cluster specific processes are presumably rather
frequent at redshifts $z\approx0.4-1$, because the assembly of galaxy clusters is
expected to peak at these epochs under the conditions of the concordance
cosmology \citep[e.g.][]{B91,KSABLPST07}.  That is also the reason why fundamental
properties of galaxies at these redshifts are measured and compared to those of
local objects in order to explore galaxy formation and evolution.

There are considerable advantages in using 3-dimensional spectroscopy over
conventional long-slit data.  Apart from a better assessment of the interaction
origin of galaxies it also improves the accuracy of the establishment of
scaling relations.  For example, an important tool to measure the cosmological
evolution of disk galaxies is the Tully--Fisher relation \citep[TFR,][]{TF77}
where the parameter next to the intrinsic luminosity is the maximum velocity
$V_{\rm max}$ of the flat part of the rotation curve.  Only for very regular
Rotation Curves (RC's), where the turn-over and the flat part are clearly
visible, $V_{\rm max}$ can be derived with sufficient accuracy to include the
galaxy under scrutiny into a TF analysis.  This is particularly important in the case
of distant, small and faint galaxies
\citep[e.g.][]{ZBFJN02,BZSBF03,CBRVP05,BAM06,KWFKL07,BZ07}.  The assessment of RC
quality may also be the main reason why recent studies of the cluster TFR
evolution differ from each other with respect to sign and amount of offset
between distant galaxies in the cluster and the field environment
\citep{ZBJHM03,BMAS05,NAMAI05,MKSP06}.

Apparently smooth RCs can nevertheless result in wrong estimates of $V_{\rm
max}$, if for example the position angle of the major axis, as measured from
the photometry, is not the same as the kinematic position angle.  The same
holds if the photometric center does not agree with the center measured from
the kinematics.  A way to resolve the situation is to obtain 3-dimensional
kinematical information.  For example, \citet{MAPB03} use velocity field of
galaxies in compact groups obtained using Fabry--P\'{e}rot spectroscopy  to
show that smooth RCs can be derived for most galaxies that were previously
judged to be distorted on the basis of the limited  information given by
2D-spectra \citep{RHF91}.  Alternatively, in some cases the rotation curve
along the photometric axis may look regular even though the velocity field
is distorted.  Because of this, 3-dimensional information is clearly
preferable to long-slit data.  It remains a requirement that the observed
velocity fields should cover the flat part of the rotation curve.

Several comprehensive studies of galaxies in the Local Universe exist that
explore velocity fields using optical observations.  Fabry--P\'erot
interferometry of the H$\alpha$ emission line is the basis of the GHASP
\citep{GMABG05}, SINGS \citep{DCAHCBK06} and Virgo \citep{CBCCA06} surveys, for
example.  3D--spectroscopy is regularly performed with the SAURON integral
field unit, delivering spectra within a limited wavelength range, with which
stellar absorption lines can be investigated in addition to gaseous emission
lines \citep[e.g.][]{GFPCE06,FBBCD06,SFDBBCZEFKKMP06}.  Integral--field
spectroscopy of HII regions in nearby disk galaxies was established, as another
example, with DensePak on the WIYN 3.5m telescope \citep{ABSGW06}.  While
spatially resolved H\,\textsc{i} measurements with radio telescopes are quite
common locally, \citep[e.g.][]{BOFHS05,NHSSA07} such observations at higher
redshift are just becoming feasible with new instrumentation (e.g. EVLA, APEX).

At $z\gtrsim0.2$, studies of the global velocity field of galaxies in the optical and
near-infrared regimes are also quite challenging.  While in the NIR a high spatial
resolution can be achieved thanks to the combination of 3D--spectroscopy with
adaptive--optics techniques \citep{FGBVESSDL06}, one is hampered by seeing effects in
the optical \citep{KKSZ07}.  One of the more comprehensive optical studies has made
use of the 15 deployable small IFUs (integral field units) of the FLAMES instrument
at the VLT. That way, 35 field galaxies at $0.4<z<0.7$ were examined and
analyzed using the TFR \citep{FHPAB06}.  Eliminating the galaxies that have perturbed
or complex kinematics, they found no evolution in the Tully Fisher Relation since
$z=0.6$.  Other investigations are restricted to a small number of objects.  Using
the GMOS integral field spectrograph and exploiting the light magnification of a
foreground cluster, \citet{SBSSK06} for example, probed the emission--line properties
of six $z=1$ field galaxies in the background.  Investigating where these galaxies
lie on the B and I band TFR compared to local galaxies, they reported that in the
B-band the galaxies are brightened by $0.5\pm0.3$ mag, while in the I band they are
in agreement with the local relation.

To determine the nature and efficiency of interaction processes in the cluster environment
we continue our study of galaxies at $z\approx0.5$ with new spectroscopy using FORS2 at
the VLT and HST/ACS imaging.  Our project involves observations of four different cluster
fields (MS0451.6--0305 at $z=0.54$, MS1008.1--1224 at $z=0.301$, F1557.19TC at $z=0.510$,
MS2137.3--2353 at $z=0.313$) with about 20 galaxies in each of them.  While in our
previous campaigns we derived the internal kinematics along the photometric major axis of
the galaxies \citep[]{ZBJHM03,JZBHM04} this time we measure the three-dimensional velocity
field of the galaxies.  We achieve this by placing three adjacent, parallel FORS2 MXU
(Mask eXchange Unit) slits onto the same galaxy.  In HST Cycle\,14 the clusters were
observed using the ACS camera (PID 10635), covering the full field-of-view of the
spectroscopy by a $2\times2$ mosaic allowing a detailed morphological and structural
analysis of all target galaxies.  In this paper, we present the results of our kinematic
and photometric analysis of the cluster MS0451 at $z=0.54$ \citep{D96}, classify the
galaxies according to the regularity of their gas kinematics and investigate whether this
is related to the environment (cluster/field) and some other properties like galaxy type,
luminosity, disk scale length, etc.  In follow-up papers, similar data for the other three
clusters will be investigated.  In addition, we will compare the kinematic and structural
analysis with hydrodynamic N--body simulations of both isolated \citep{KKSBZ05} and
interacting galaxies \citep{KKSBK06}.  Using these simulations we already examined
systematically what is the effect of observational constraints like a low spatial sampling
in the case of distant small galaxies on the measured velocity fields \citep{KKSZ07}.  In
a final paper we will combine all our results to address the question of the interaction
history of cluster galaxies. 

In Sect.2, we outline our unique approach of  ``matched IFU simulations'' using MXU
masks with the FORS2 spectrograph at the VLT and give an account of our spectral
reduction, explain how the velocity fields are constructed and describe the observed
sample in the field of MS0451.  Our kinematic analysis of the gas velocity fields and
how we measure the stellar kinematics are also explained in this Section.  In Sect.3,
we describe how we derive structural parameters from HST images and luminosities in
different filters from ground-based photometry.  In Sect.4, we explain how we quantify
the deviations from regularity in the gas kinematics, present the results of that
analysis, make a comparison between the field and cluster galaxies and investigate the
existence of a relation between the kinematic regularity and some photometric
properties.  In Sect.5, we discuss our results.  In Sect.6 we summarize the paper and
draw our conclusions.  In Appendix A, we discuss the effects of the spatial
resolution on our kinematical analysis.  Kinematic and photometric information on
individual galaxies in the MS0451 cluster field are given in Appendix B.  Throughout
this paper we use a standard cosmology with: $H_0 = 70$\,km\,s$^{-1}$\,Mpc$^{-1}$,
$\Omega_{\rm m}=0.3$, $\Omega_{\lambda}=0.7$
\citep{TSBCC03}.

\section{3D--spectroscopy with FORS2/MXU}
\subsection{Observing strategy}
Our aim was to obtain spatially resolved emission line rotation curves for
galaxies with typical surface brightnesses of $V\approx22.5$\,mag/$\arcsec^2$. 
This is feasible in a reasonable time only with large ground based telescopes. 
The natural approach to observe 3D-velocity fields would be to use IFUs to achieve
some kind of 3D-spectroscopy.  Inspecting the performance of all such instruments
available at the ESO-VLT, it turned out that none of them are optimally suited and
efficient for our purpose.  One of the main goals of our study, although not of
the current paper, is to unambiguously determine the maximum rotation velocity
$V_{\rm max}$ of galaxies with regular velocity fields.  From our previous
campaigns, we knew that many spirals at $z=0.3-0.6$ have rather large sizes with
the turn-over points of their RCs at radii around $1\farcs5-2\arcsec$,
corresponding to disk scale lengths of $\approx3R_d$ on either side of the center
of the galaxies \citep[see][Fig.1]{ZBJHM03}.  Thus, the FLAMES/ARGUS field-of-view
in high-resolution mode ($4\farcs2\times6\farcs6$) would just have been adequate,
but given the low throughput of the high-resolution spectrograph GIRAFFE, the
necessary exposure times would have been so high that even a full night would not
have been sufficient to observe just one galaxy.  With the VIMOS IFU, still only 2
or 3 objects could have been observed per night due to the small field of view. 
Suitable to some extent are only the 15 deployable IFUs of FLAMES, whose octagonal
lens array has a spatial coverage of $\approx2\arcsec\times3\arcsec$.  This does
not probe the outer regions of rotation curves, which is vital for an accurate
determination of $V_{\rm max}$.  Only extrapolation of the measurements from the
inner parts of the VF gives an estimate of $V_{\rm max}$ \citep[see, e.g., Fig.7
in][]{FHPAB06}.  No real way out is the possibility to use a mosaic of four
pointings per field, because then the necessary observing time would be too large.

\begin{figure}
   \includegraphics[width=\columnwidth]{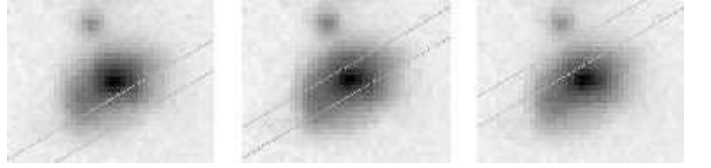}
      \caption{Example of the positioning of three slits onto one galaxy (galaxy C8).}
         \label{slits}
   \end{figure}

For this reason and for other reasons stated below, we conceived a method to
``simulate IFUs'' that are exactly matched to our purpose using the FORS2 focal
reducer spectrograph.  This instrument offers a mode (MXU) using custom-made slit
masks, in which slitlets can be rotated with respect to the y-axis of the CCD
\citep{SNHHH00}.  We designed three masks for each cluster field, so that each
object was subsequently covered by three different, but parallel slits of
$1\arcsec$ width.  One slit was placed along the photometric major axis of a galaxy, the
other two were shifted by $1\arcsec$ along the minor axis on both sides of the
central slit, respectively (Fig.\ref{slits}).  This resulted in an appropriate
aperture of $\sim7\arcsec\times3\arcsec$ for each target (the actual dimension of
the slits along the major axes are much longer to retrieve the velocities at
further distances to the center than $3\farcs5$ in case there is still signal and
to allow accurate sky subtraction).  Although each cluster field must be observed
three times (with three different masks), this method is still very efficient due
to two reasons.  First, the high-throughput VPH grisms of FORS2 provide the
necessary $S/N$ in the spectra with an integration time of 2.5\,h only, i.e. for
all three setups together 7.5\,h would be needed (comparable/lower than
shutter-open times needed with FLAMES).  Second, the multiplex capability of
designing many slits across the full field of FORS2 ($6\arcmin\times6\arcmin$)
allows to observe simultaneously a rather high number of objects (in case of
MS0451 there were 20 slits for 22 galaxies).

%-------------------------------------------------------------
%
%_____________________________________________________________

%-------------------------------------------------------------
   \begin{figure}
   \includegraphics[width=\columnwidth]{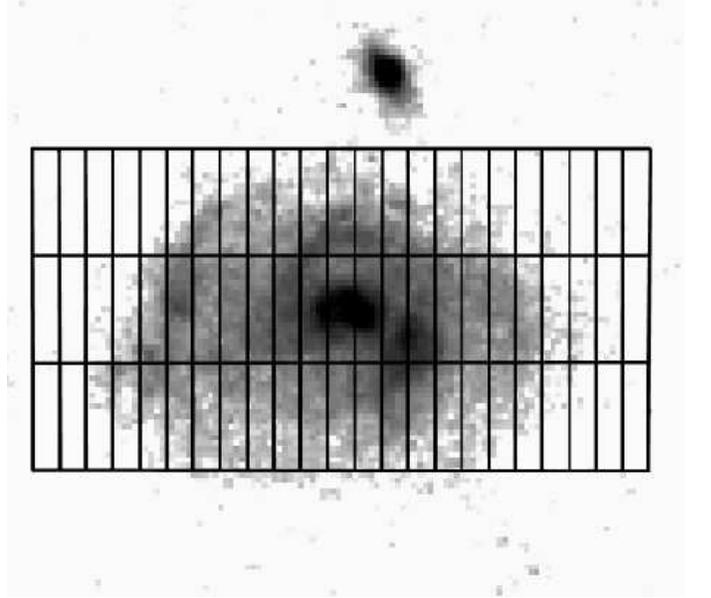}
 \caption{HST/ACS image of a member galaxy (galaxy C8, $z=0.5326$) of cluster MS0451.
  % It is probably undergoing a merger. 
  The overlaid grid indicates the 3 slit positions with each slit
  ($1\arcsec$ width) subdivided into FORS pixels ($0\farcs25$) that
  correspond to spectral rows.
              }
         \label{grid}
   \end{figure}
%
%_____________________________________________________________

Our approach has several additional advantages: Since the pixel scale of the
two FORS2 MIT--CCD chips is $0\farcs25/$pixel, our spatial resolution of the
resulting velocity fields is very adequate for the investigated redshifts
(Fig.\ref{grid}) and is in practise only limited by the seeing.  An adequate
spectral resolution of $R\approx1000$ is achieved by using the holographic
grism grism\_600RI.  The instrumental resolution we measure from the data is
$5.6\AA$ (FWHM) which corresponds to $\sigma_{inst} \sim 120$ km s$^{-1}$ at
$6000\AA$.  Here we are able to measure the velocities down to $\sim \sigma/10
$km s$^{-1}$, i.e. the typical accuracy of our velocities is $\sim 12$ km
s$^{-1}$.  With this grism a rather large wavelength coverage was obtained,
making it possible to simultaneously observe several strong emission lines,
allowing independent determinations of the velocity field (VF), as well as many
absorption lines, that can be used to derive, for example, stellar rotation
curves (see Sect.2.5.).  For the cluster members of MS0451, we probe the
emission lines of [O\,\textsc{ii}]3727, H$\gamma$, H$\beta$,
[O\,\textsc{iii}]4959 and [O\,\textsc{iii}]5007.  Line ratios can then be used,
for example, to derive the gas-phase metallicities of the galaxies or to detect
possible contaminations from AGN.  To summarize, our observational strategy
provides a matched aperture size, large target coverage and a wide range of
wavelength for efficiently low exposure times.  Spatial sampling is sufficient
(0\farcs25) along the spatial axis while it is 1\arcsec along the spectral axis
(Fig.\ref{grid}).

\subsection{Data Reduction}

Each cluster field was observed with three separate masks to be able to
place  three parallel slits onto each galaxy; the central one along the
photometric major axis and the other two 1\arcsec offset along the
minor axis in opposite directions.  For cosmic ray removal, the integration
time for each mask was split into three exposures.  The observations for
MS0451 were spread across 6 nights in January and February 2005.  In the
case of one mask, all exposures were taken during the same night.  For that
mask, all science frames were combined right after the overscan and bias
correction process, since the alignment of the frames was excellent.  For
the rest, the order of the reduction steps was the following:

\begin{itemize}
\item Overscan and bias correction of all science and calibration frames 
\item Extracting 2D spectra of all objects, corresponding arc spectra and flat 
fields from the main frames
\item Normalization of each flat field frame 
\item Flat field correction of each science and arc spectrum
\item Wavelength calibration using the arc lamp spectra first and then using
around 10 strong sky lines to correct for the shift caused by the movement of
the  telescope during the observation 
\item Sky subtraction 
\item Rebinning the spectra to ensure that all the spectra start at the same
wavelength and have the same wavelength interval per pixel 
\item Checking the alignment of the spectra taken with the same mask 
(the difference between the photometric center at a certain wavelength in 3 
different exposures for each galaxy spectrum was on the subpixel level, so no 
shift had to be applied before combining them) 
\item Median combining the three exposures for each mask 
\item Correcting for the curvature of the spectrum caused by the optics of the
instrument.  After summing up each column with around 100 surrounding columns,
a parabola was fitted to the positions of maxima at each column to trace the
curvature. 
\end{itemize}

\subsection{Galaxies in the MS0451 Sample}

In this paper, we present the analysis of the cluster field of MS0451 at
$z=0.54$.  Twenty-six galaxies in this cluster had already  been spectroscopic
targets in our previous campaigns, similar to the MOS observations described by
\citet{ZBJHM03} and \citet{JZBHM04}.  In these previous runs, only one MXU slit
was placed onto each galaxy, mostly along the photometric major axis as determined
from (ground-based) FORS images (ESO PID 66.A--0547).  Due to geometric
constraints (mask setup) or when two objects were forced into the same slit other
slit orientations were chosen.  The results from this campaign will be presented
in another paper of this series.  Examples of four galaxies have already been
given in \citet{KKSBK06}.

Since for these galaxies accurate redshifts and equivalent widths of visible emission
lines were available, they formed the primary candidate list for target selection for
our new study presented here.  First priority was given to cluster members
with significant emission in [OII]3727, second priority to field galaxies
with a clear detection of any emission line that would fall onto our chosen observed
wavelength range.  Note that no morphological information was used for the sample
selection.  Further objects were drawn from a catalog provided by the CNOC survey
\citep{EYAMC98} with either redshift information or measured $(g-r)$ color that
matches expectations for spiral templates at $z\approx0.5$.  If there was space in the
MXU setup, and no suitable candidate at all was available, a galaxy was picked at
random.  In the end, we had defined 20 slits for 22 targets, of which ten had
spectroscopic information, six appropriate colors and six were just fillers. After
data reduction, the sample turned out to have eleven cluster members. Five galaxies
four of which are cluster members have no or very weak emission lines and are,
therefore, not suitable for a kinematic analysis of their gas velocity fields.

To each galaxy we assign a number and put a ``C'' or an ``F'' in front of it, 
indicating whether it is a cluster member or a field galaxy.  In Table
\ref{centdist}, we give their redshifts and for the cluster members we give the
projected distance from the center, together with some information taken from the
literature (morphological type and redshift).  For C11, F7 and F11 we could not
determine the redshift from our data.  F8 and C11 were observed within the same
slit.  They are so close together that their spectra can not be distinguished. 
The redshift that is calculated using the emission lines in the composite spectrum
is $z=0.4443$.  \citet{EYAMC98} shows that C11 is an elliptical cluster member and
that the emission we measure comes from F8.  In the spectra of F7 and F11, one
strong emission line is visible and no other feature can be identified.  In both
cases, different possibilities for identification of the line rule out that these
two galaxies are cluster members.  Therefore they are flagged with an ``F''
showing that they are field galaxies.  In Table \ref{centdist} we give a redshift of galaxy F7
and F11 from \citet{M08}.  The redshift of galaxy F11 is photometric.  Since
photometric redshifts have big uncertainties, we do not use it in our analysis. 
All cluster members are within 2 Mpc from the center and most of them are closer
than 1 Mpc, i.e. they are well inside one virial radius (or $R_{200}$) of the
cluster.

\begin{figure}
   \includegraphics[width=\columnwidth]{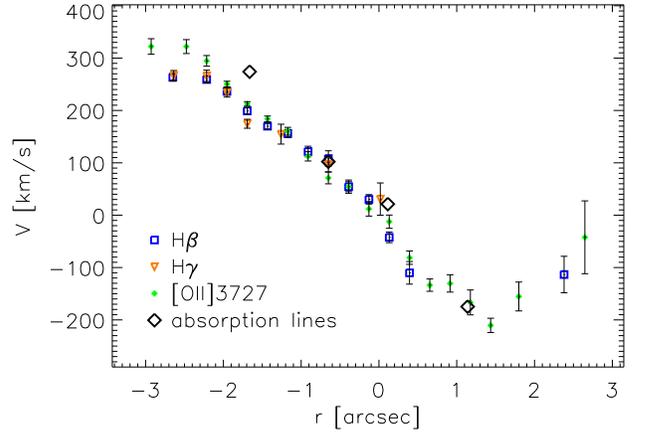}
 \caption{The observed gas and stellar rotation curves of a cluster member
 (galaxy C8) extracted along the central slit with no correction for inclination and
 seeing. Both components have their principal motions along the same direction and
 with similar amplitude.
            }
         \label{rcgs}
   \end{figure}

Our spectra cover the wavelength range 5000\,\AA\ $\leq \lambda_{\rm obs}\leq$
8000\,\AA.  Several lines can be observed within this interval depending on the
redshift.  In case of visible emission lines excited by different mechanisms we can
use them to investigate the physical state of different gas clouds.  We checked
whether the kinematics derived from different lines are consistent with each other or
whether there are prominent discrepancies that may indicate ongoing or recent
interaction processes.  If strong absorption lines are present, we also compare the
stellar rotation with the gas kinematics (see Sect.2.5  and, e.g.
Fig.\ref{rcgs}).

\begin{figure}
   \includegraphics[width=\columnwidth]{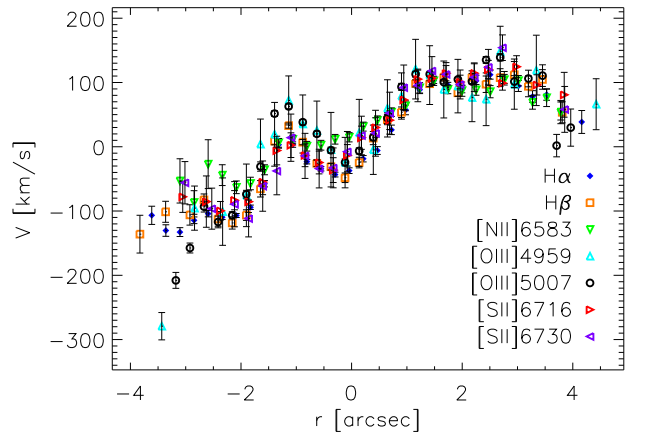}
 \caption{Rotation curves of a foreground galaxy at $z=0.1573$
(galaxy F5) extracted along the central slit using the prominent emission lines
in its spectrum (no correction for inclination and seeing is applied).
            }
         \label{pv02}
   \end{figure}

In case of cluster spirals, the visible emission lines are [OII]3727, $H\gamma$, $H\beta$,
[OIII]4959 and [OIII]5007, of which [OII]3727 is usually the strongest.  The rotation
curves extracted along the central slit using prominent emission lines of a cluster member
(C8) are shown together with its stellar rotation curve in Fig.\ref{rcgs}.  The redshift
range of field galaxies is $0.1\leq z \leq 0.9$.  For the highest redshift in this range, the
only visible line in the spectrum is [OII]3727 doublet while for small redshifts several
lines are visible.  For example the emission lines covered in the spectrum of F5 at
$z=0.1573$ are $H\gamma$, $H\beta$, [OIII]4959, [OIII]5007, OI, $H\alpha$, [NII]6583,
[SII]6716 and [SII]6730.  There are three foreground galaxies for which the $H\alpha$ line is
visible (F4, F5 and F6) and it is the strongest line in all cases.  In Fig.\ref{pv02} we show
the rotation curves extracted along the central slit using the prominent lines of F5. 

\begin{table}[h]
\caption{Basic galaxy information}
\label{centdist}
{\tiny
\begin{center}
\begin{tabular*}{\columnwidth}{lccccc}
\hline\hline
ID &      z   &  d & NED Name & Type, ref & z, ref 	\\
$(1)$ &	$(2)$ &	$(3)$ & $(4)$ &	$(5)$ &	$(6)$ \\
\hline
C1 &  0.5421  &  1.7  &  PPP 002414  & Sb-Sc (1)	  &	0.54160 (1)\\
 &    &    &    &  Sab (2)	  &	 0.54240 (2)\\
C2 &  0.5486  &  0.7  &  PPP 000802  & E (1)	  &	0.54956 (1) \\
C3 &  0.5465  &  0.7  &  PPP 001123  & S (2)	  &	0.54747 (2) \\
C4 &  0.5324  &  0.9  &  PPP 001795  & S (2)	  &	0.53345 (2)	\\
C5 &  0.5312  &  1.1  &  PPP 001787  & Sc-Irr (1)  &	0.53156 (1) \\
 &    &    &    & S0 (2) &	 \\
C6 &  0.5305  &  1.1  &  PPP 001802  & Sb-Sc (1)  & 0.53094 (1) \\
 &    &    &    &  S (2) &  0.53007 (2)\\
C7 &  0.5277  &  0.8  &  PPP 001696  & S (2)	  &	0.52700 (3) \\
C8 &  0.5325  &  0.7  &  PPP 001482  & Sc-Irr (1) &	0.53257 (1)  \\
 &    &    &    &  S(2)  &	 0.53215 (2) \\
C9 &  0.5246  &  0.5  &  [SED2002] 111  & E/S0 (4)    &	0.52420 (2)  \\
 &    &    &    & Irr (2)  &	  \\
C10 &  0.5312 &  0.8  &  -	     &   -	  &	-	\\
C11 &  -      &  0.1    &  PPP 001264  &   E (1)	  &	 0.53070 (1)\\
F1 &  0.9009 &  -    &  -	    &	-	 &     0.90011 (2)     \\
F2 &  0.5795 &  -    &  PPP 001818  &	Sb-Sc (1)    &   0.57924 (1)  \\
 &   &      &    &	    &   0.57918 (2) \\
F3 &  0.5667 &  -    &  PPP 002130  &	Sc-Irr (1)   &	0.56686 (1)\\
 &   &      &    &	 S (2)  &	\\
F4 &  0.1867 &  -    &  PPP 001358  &	Sc-Irr (1)  &   0.18727 (1)\\
 &   &     &    &	S (2)  &  \\
F5 &  0.1573 &  -    &  PPP 001403  &	Irr (4)   &   0.15679 (2) \\
 &   &      &    &	 S (2)  &    \\
F6 &  0.0982 &  -    &  PPP 001542  &	S (2)	 &    0.09809 (2)      \\
F7 &  -      &  -    &  -  	     &   Merger (2)   &	 0.91251  (2)	 \\
F8 &  0.4443 &  -    &  PPP 001259  &	Sc-Irr (1)   &  0.44392 (1)\\
F9 &  0.3259 &  -    &  PPP 000798  &   -	  &	 0.32561 (2) 	\\
F10 &  0.4947 &  -    &  -	     &   Irr (2)  &	 0.648 (2)(ph.) \\
F11 &  -      &  -    &  PPP 001478  &   -	  &	 0.319 (2)(ph.) \\

\hline
\end{tabular*}
\end{center}
}
Column (1) Object ID. \\
Column (2) Redshift. \\
Column (3) The projected distance from the cluster center in Mpc. \\
Column (4) The name of the galaxy in Nasa Extragalactic Database (NED). \\ 
Column (5) The morphological type of the galaxy and its reference. \\
Column (6) Redshift of the galaxy and its reference.  Photometric redshifts are indicated with "(ph.)".\\

{\it References: (1): \citep{EYAMC98},  (2): \citep{M08}, (3): \citep{MMTES07},
(4): \citep{SEDHP02}.  All galaxy names given in Column 4 begin with ``MS
0451.6-0305:''.  [SED2002]=Stanford+Eisenhardt+Dickinson.  For C11, F7 and F11,
we couldn't measure the redshifts using our data as explained in the text.  The
distance of C11 to the cluster center is calculated using its redshift from the
literature.} 

\end{table}

%-------------------------------------------------------------
%
%_____________________________________________________________

%-------------------------------------------------------------
%
%_____________________________________________________________

In the Appendix, we give some information about each galaxy.  In case the
galaxy has emission, we present: \\ \textit{a-} the HST-ACS image of the
galaxy in the I band; \\ \textit{b-} rotation curves of different
emission lines (and for some cases based on the absorption lines) extracted
along the central slit without correction for inclination and seeing; \\
\textit{c-} position angles of kinematic and photometric axes as a function of
radius; \\ \textit{d-} rotation curves extracted along the central
slit and the kinematic major axis; \\ \textit{e-} velocity field obtained
using the strongest line in the spectrum; \\ \textit{f-} velocity map
reconstructed using 6 harmonic terms;  \\  \textit{g-} residual of the
velocity map and the reconstructed map;   \\   \textit{h-} normalized flux map
of the line used for constructing the velocity field; \\ \textit{i-} simple
rotation map constructed for position angle and ellipticity fixed to their
global values;  \\  \textit{j-} residual of the velocity map and the simple
rotation model;  \\ \textit{k-} position angle and flattening as a function of
radius;  \\ \textit{l-} $k_{3}/k_{1}$ and $k_{5}/k_{1}$ (from the analysis
where position angle and ellipticity are fixed to their global values) as a
function of radius.

There are some emission-line galaxies for which we did not analyze the velocity
fields, and therefore no figures are given related to their velocity fields:
galaxies F1, F8, C5 and C6.  Galaxy F8 lies very close to galaxy C11, and their
FORS spectra could not be separated.  Galaxy F1 was observed with only 1 slit
position.  Galaxies C5 and C6 are interacting, and have extremely irregular
velocity fields.

\subsection{Kinematic analysis methods}

\begin{figure*}
   \includegraphics[width=170mm]{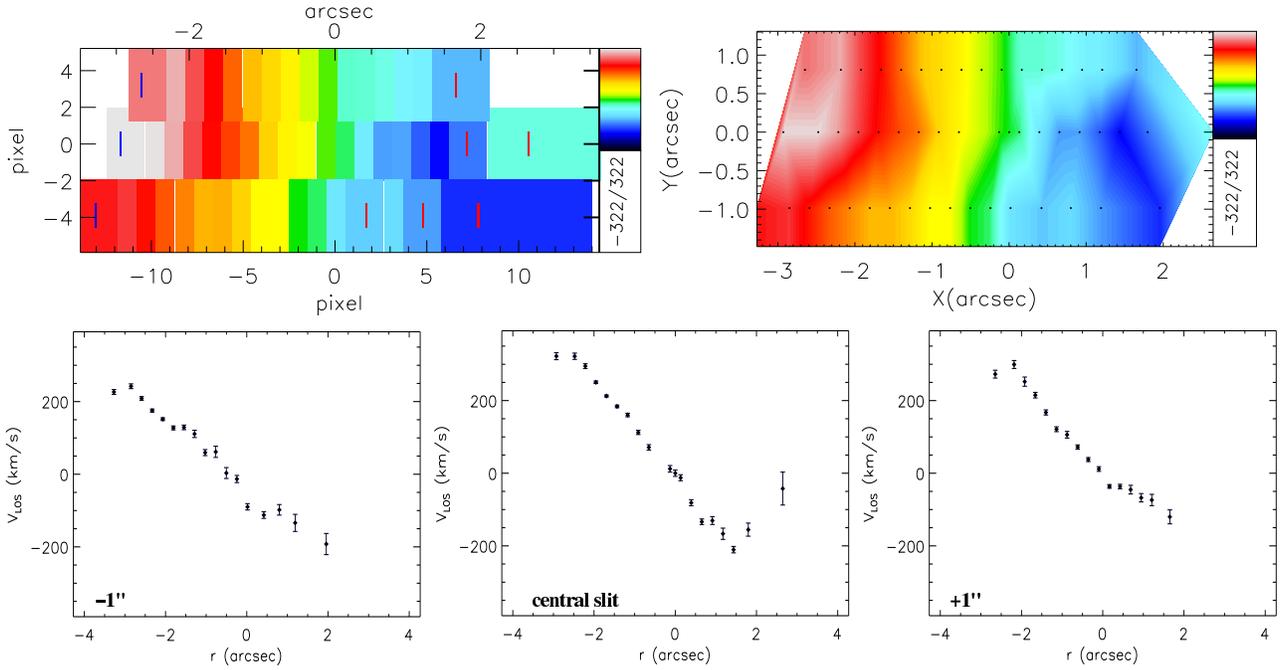}
   \centering
   \caption { 
   \textit{Top Left:} Pixel map of the galaxy in Fig.\ref{grid} (galaxy C8),
    based on [O\,\textsc{ii}]3727 emission line measurements.  In case that the
    pixels are binned, the flux weighted position is indicated with a red/blue
    line on the map.  The intersection between the pixel rows is caused by the
    spatial offsets between the masks mentioned in Sect.2.3. Velocities are
    given in km s$^{-1}$.
   \textit{Top Right:} The velocity map visualizing the data on
   the left. Velocities are interpolated linearly.
   \textit{Bottom Left:} Rotation curve extracted along the slit
$1\arcsec$ below the central one.
    \textit{Bottom Middle:} Rotation curve extracted along the
central slit.
    \textit{Bottom Right:} Rotation curve extracted along the
slit $1\arcsec$ above the central one.
              }
         \label{vfobs} \label{pv1} \label{pv2} \label{pv3}
   \end{figure*}

The redshifts were calculated using all visible emission lines and a few prominent
absorption lines in the integrated 1D-spectra of the central slit.  Then, for each
galaxy and for each emission line, the photometric center of the galaxy was
determined by fitting a Gaussian to the continuum light distribution in the
spatial direction averaged over 160\,\AA\  excluding the emission line itself. 
For each emission line separately, line centers and widths were measured along the
spatial axis by Gaussian fits row by row, after the underlying continuum had been
subtracted.  The measurements were done by starting at the photometric center and
continuing outwards on both sides.  In regions of low signal, rows were first
binned until a minimum flux in the emission line was reached.  This means that in
the outer regions of a galaxy, not every pixel of the grid shown in Fig.\ref{grid}
corresponds to an independent data point for the velocity field (Fig.\ref{vfobs},
top left).  Finally, rotation curves were constructed.  This was done
for the spectra at all three positions (Fig.\ref{vfobs}, bottom row).

%-------------------------------------------------------------
%
%_____________________________________________________________

To construct the 3D-velocity fields from the three rotation curves,
all measurements were transformed to a common coordinate system defined by (the
continuum center of) the central slit.  The spatial offsets of the three masks
relative to each other were determined using spectra of stars observed with 2
slits oriented perpendicular to each other so that the shift along both x and y
axes could be determined.  Because of the shift, the slit offsets along the
minor axis of the galaxies between the three setups were not exactly
$\pm1\arcsec$ as intended, leading to a slightly different coverage of each
galaxy from the ideal version of the grid shown in Fig.\ref{grid}.  The maximum
absolute difference that we measured between the position of each mask is
$\sim$1.5 pixels.  Taking this into account, the observed velocity map was
constructed using linear interpolation (Fig.\ref{vfobs}, top right).  We also
made emission line strength maps (see the Appendix).  Here we will explain how
we analyzed the velocity maps, and we derived parameters such as kinematic
position angle ($\Gamma$) and kinematic inclination ($i$) \citep{FVPEBCZ05}.

To analyze the velocity fields, we used the kinemetry package of
\citet{KCZC06} that was extensively tested and used for SAURON IFU observations
of local galaxies.  For a rotating disk the line of sight velocity is given by:

\begin{equation}  
V_{los}(R) = V_{sys} + V_{rot}(R) \cos {\psi} \sin {i}  
\label{los}
\end{equation}

\noindent 
where $V_{los}$ is the observed line of sight velocity, $V_{\rm sys}$ is the systemic
velocity, $V_{\rm rot}$ is the circular velocity and (R,$\psi$) are polar coordinates
in the plane of the galaxy with $\psi$ measured from the apparent major axis. 
According to Eq.\ref{los}, the observed velocity along a certain orbit can be
expressed as a simple cosine form when plotted versus azimuthal angle.  In general,
since the motion of gas in galaxies can often be explained by rotation plus a smaller
perturbation,  one can generalise Eq.\ref{los} to

\begin{equation}  
V(R,\psi)=A_{0}(R) + \sum_{n=1}^N A_{n}(R) \sin (n\psi)+ B_{n}(R) \cos (n\psi).
\label{fourier}
\end{equation}

\noindent

A grid of position angle -- flattening combinations covering the whole parameter
space is constructed for each radius, where flattening is defined as
q=b/a=1-ellipticity.  Fourier analysis is performed on the velocity profiles
extracted along each ellipse.  The best sampling ellipse is determined by
minimizing \citep[see][]{KCZC06}:

\begin{equation}  
\chi^2=A_{1}^2 + A_{3}^2 + B_{3}^2.
\label{fterm}
\end{equation}

%-------------------------------------------------------------
   \begin{figure*}
   \includegraphics[width=170mm]{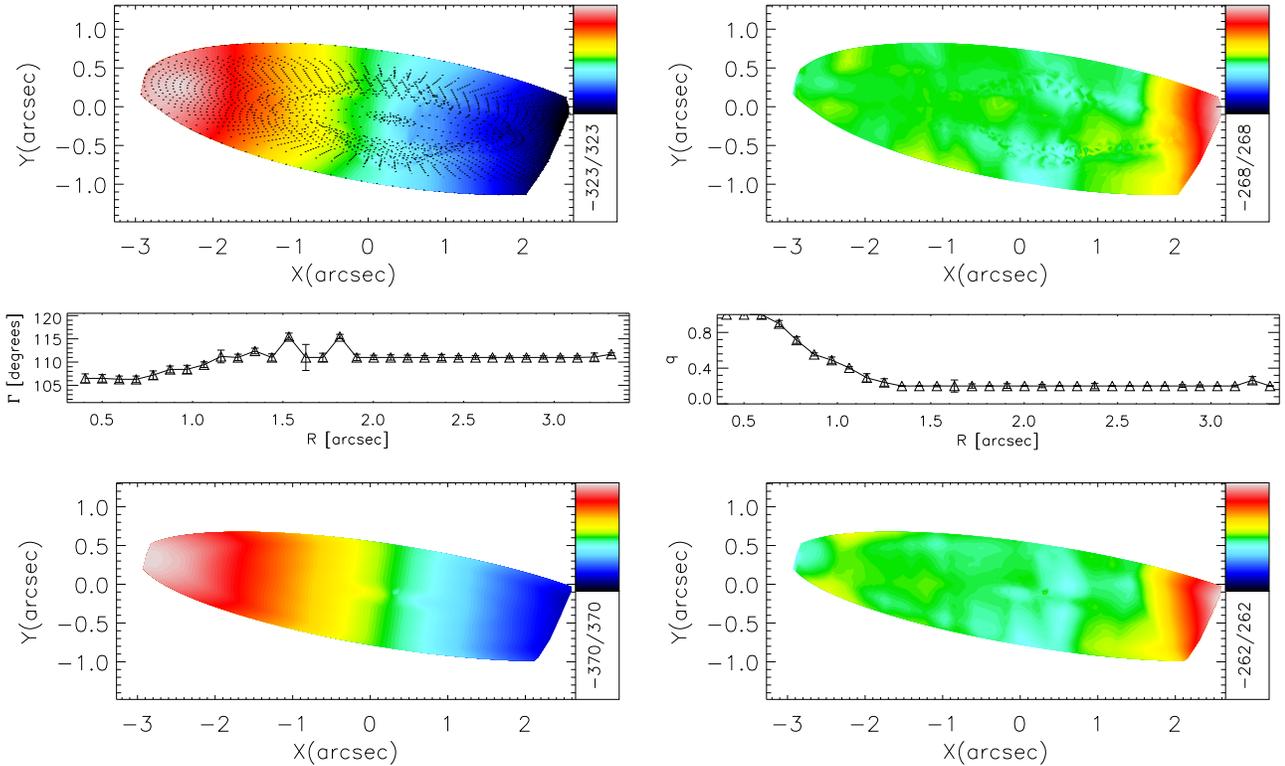}
 \caption{
\textit{Top Left:} Reconstructed velocity field of the galaxy in 
Fig.\ref{grid} (galaxy C8) constructed using 6 harmonic terms of the
Fourier analysis performed by kinemetry.
\textit{Top Right:} Residual of the observed velocity map and the 
reconstructed map given on the left.
\textit{Middle:} Kinematic position angle $\Gamma$ and flattening $q$ 
as a function of distance to the kinematic center.
\textit{Bottom Left:} Velocity field of the best fitting rotating 
disk ($B_{1}\cos {\psi}$) constructed fixing the position angle $\Gamma$
and flattening $q$ to their global values.
\textit{Bottom Right:} Residual of the observed velocity map and the rotating
disk velocity field given on the left.
}
         \label{kinpar}
   \end{figure*}
%
%_____________________________________________________________

\noindent

\begin{figure}
   \includegraphics[width=\columnwidth]{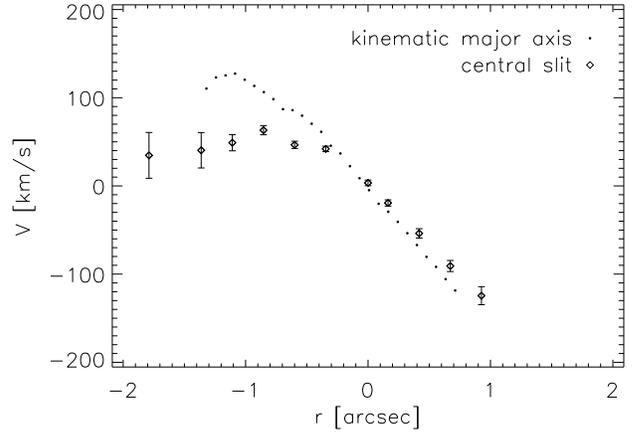}
 \caption{Rotation curves of a field galaxy at $z=0.5667$ (galaxy
 F3) extracted along the central slit which is 3$^{\circ}$ misaligned w.r.t.
 the photometric major axis, and the kinematic major axis. 
            }
         \label{rccomp}
   \end{figure}

For an infinitesimally thin disk the flattening of the isovelocity contours is
related to the inclination  by $q~=~b/a~=~\cos i$.  For every galaxy we determined
a global position angle and inclination by taking the median of the values at
several radii outside half the FWHM (full width at half maximum) of the seeing.  In
Fig.\ref{rccomp} we show, for a field galaxy in our sample, the comparison between
the rotation curves extracted along the central slit and along the
kinematic major axis.  This reveals a very important advantage of measuring the
full velocity field instead of just one rotation curve.  The rotation
curve along the kinematic major axis clearly reaches a higher value of $V_{\rm
max}$ than the photometric major axis rotation curve,  which is mostly
used for the Tully--Fisher analysis.

%-------------------------------------------------------------
%
%_____________________________________________________________

To determine the deviations from simple rotation, we repeated the Fourier analysis
of the velocity maps fixing the position angle and the inclination of the orbits
to these global values.  By subtracting the model of a regular rotation disk $B_{1}~\cos
\psi$ from the observed velocity field, non-circular motions and perturbations can
be seen in the residual image (Fig.\ref{kinpar}).  We can quantify them using
the Fourier terms A$_n$ and B$_n$.  We use the amplitude of the n$^{th}$ order
Fourier term, defined in \citet{KCZC06}:

\begin{equation}  
k_{n} = \sqrt{A_{n}^2 + B_{n}^2}.
\label{kcoeff}
\end{equation}

\noindent
The coefficients $k_3$ and $k_5$  (Eq.\ref{kcoeff}) represent deviations
from simple rotation.  We do not fit the even Fourier terms, since they should
vanish if the velocity fields are point-anti-symmetric.  To measure the total deviation,
we also define an additional coefficient:

\begin{equation}  
k_{3,5} = \sqrt{A_{3}^2 + B_{3}^2+A_{5}^2 + B_{5}^2}.
\label{kcoeff2}
\end{equation}

Before running kinemetry the kinematic center has to be known.  To determine it  we use
the fact that when the velocity field is regular the velocity gradient is the highest
at the kinematic center \citep{AMLB97}.  The peak of the squared sum of the velocity
gradient in perpendicular directions is taken to be the kinematic center.  In case that
there are shocks  in the gas, this method possibly does not work, but then the
kinematic center can generally be found as a secondary peak in the velocity gradient
map.  For instance for galaxy F6 (Appendix, Fig.A13), a galaxy that has a kinematically
decoupled core (KDC),  we took the peak of the gradient in the KDC as the kinematic
center although the gradient in the transition region was much higher.

\subsection{Stellar Kinematics}

In addition to the analysis of the gas velocity fields, we can also investigate
the kinematics of the stellar component, if absorption lines are present in the
galaxy spectra with sufficient $S/N$.  This prerequisite is mostly fulfilled only
for the central slit spectrum, where the surface brightness is large enough.  In
these cases, we determine the stellar rotation and velocity dispersion from
several strong absorption lines using the pPXF software \citep{CE04}.  This program
fits different linear combinations of stellar templates to a wavelength region in
pixel space.  As templates, we use SSP (Single age and metallicity Stellar
Population) models created with the MILES library \citep{SPJCC06}.  In
Fig.\ref{rcgs}, we compare, as an example, the gas and stellar rotation curves of
a cluster member, both extracted along the central slit.  

In our sample, there are 6 galaxies for which we could extract the stellar rotation curve.  For those, we have compared the 
rotation of stars and gas (see Appendix, Fig.\ref{gal15fig}.b, Fig.\ref{gal9fig}.b,
Fig.\ref{gal6fig}.b, Fig.\ref{gal5fig}.b, Fig.\ref{gal22fig}.b,
Fig.\ref{gal8fig}.b).  In the case of galaxy C8 and F2 they show a similar
behaviour.  For galaxy C5 and galaxy C9, there is a discrepancy between the two
curves.  Galaxy C5 is interacting with its companion (Fig.\ref{gal9fig}). 
Galaxy C9 has a big misalignment between its kinematic and photometric axes.  It
appears that the gas component in both galaxies is disturbed.  In case of  galaxy
C7, which has an irregular gas velocity field, while the $H\beta$ and H$\gamma$
rotation curves do not show a big discrepancy with the stellar one, the
[O\,\textsc{ii}]3727 curve does show a somewhat larger discrepancy.  
Galaxy C1 has very weak [O\,\textsc{ii}]3727
emission, so a comparison in this case would not give reliable information.

In all cases, the stellar rotation curve is regular.  The cross sections for gas clouds in
spiral disks are large enough that disturbances in the gas velocities are triggered rather
easily, whereas stellar orbits are collisionless \citep{SFDBBCZEFKKMP06}.  Therefore, gas
velocity fields are expected to be more distorted and richly structured than their stellar
counterparts \citep[e.g.][]{DMEN07}.  However, comparing with local spiral galaxies
\citep[see][]{FBBCD06} one sees that in the inner regions of galaxies such discrepancies
between stars and gas are rare in the local Universe.  For the local cluster members, the
situation might be different.  From H$\alpha$ morphologies of galaxies in the Virgo cluster
\citet{KK04} find that about half of their Virgo spiral galaxies have truncated H$\alpha$
disks, while several spiral galaxies have asymmetric H$\alpha$ enhancements at the outer edge
of their disks, and at least two highly inclined spiral galaxies have extraplanar
concentrations of H{\sc II} regions.  \citet{CBCCA06} investigated $H\alpha$ velocity fields
of local cluster spirals, and showed that they exhibit typical kinematical perturbations like
streaming motions along spiral arms, twist of the major axis, Z shape of velocities due to
the presence of a bar, decoupled nuclear spiral or misalignment between photometric and
kinematic major axes.  In general, however, these perturbations are minor: for  most galaxies
the kinematics of the gas can be considered a perturbation to the  stellar kinematics.

\section{Photometry}

\subsection{Groundbased Photometry}

To study the photometry of our galaxies we used data from the HST and  ground-based
telescopes.  To determine luminosities and colours of our target galaxies we make
use of photometry obtained at ground-based telescopes.  For the cluster MS0451,
we have taken direct images with FORS2  at the VLT (ESO PID 66.A--0547).  Frames in
the $V$, $R_C$ and $I_C$ filter were reduced in the standard manner  (bias and
flatfield correction, registering, astrometric calibration using IRAF\footnote{IRAF
is distributed by the National Optical Astronomy Observatories, which is operated
by the Association of Universities for Research in Astronomy, Inc., under
cooperative agreement with the National Science Foundation.} tasks) and combined to
yield total exposure times of 3300, 1800 and 2400\,s, respectively.  Photometric
calibration was achieved via standard stars from \citet{Lando92}.  For
determination of colours, the final combined $V$, $R$ and $I$ images were convolved
with a 2D-Gaussian filter to a common seeing of $0\farcs9$.  Photometry was
performed using the SExtractor package \citep{BA96}.  While \textsf{MAG\_BEST}
magnitudes are taken as total magnitudes, apertures with diameter $2\farcs0$ were
used to compute colours.  All magnitudes were corrected for Galactic extinction by
$A_V=0.110$, $A_R=0.088$ and $A_I=0.064$ given for the cluster coordinates in NED
(NASA/IPAC Extragalactic Database), which are based on $E(B-V) = 0.033$
\citep{SFD98}.  The three filter magnitudes were transformed to rest-frame
Johnson-$B$ using the k-correction algorithm by \citet{BR07}.  The absolute total
luminosities of the galaxies are the second parameter entering the Tully--Fisher
relation.  All derived magnitudes of the objects for which we have spectra are
listed in Table \ref{tabrun2}. The tabulated $M_B$ magnitudes of the spiral
galaxies are not corrected for internal dust (inclination).

\subsection{HST imaging and structural analysis}

For the cluster MS0451 presented here, we exploit existing imaging in the ACS/$F814W$
filter from the ST--ECF archive.  We basically used the pipeline reduced images
but with an additional cosmic ray rejection routine. 

Structural parameters of the galaxy disks such as scale length, inclination,
ellipticity and position angle were determined by decomposing the 3D-surface
brightness distribution of the target objects into an exponential disk and a
S\'{e}rsic bulge using the GALFIT algorithm \citep{PHIR02} (Table \ref{tab4}).

\begin{figure}
   \includegraphics[width=\columnwidth]{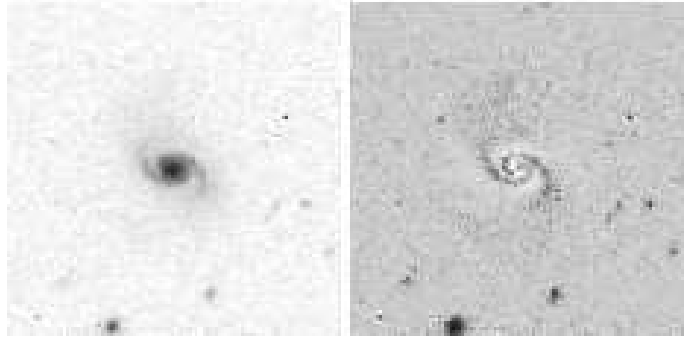}
 \caption{\textit{Left} The HST-ACS I band image of a cluster member 
 (Galaxy C2)
    \textit{Right} The residual of the exponential disk + S\'{e}rsic bulge fit.
    }
         \label{photometry}
   \end{figure}

The model-subtracted residual images reveal additional structural features like
spiral arms (Fig.\ref{photometry}) and bars, or signatures of ongoing or recent
interaction processes such as tidal tails that can be compared to the kinematic
tracers. 

%-------------------------------------------------------------
%
%_____________________________________________________________

To further investigate the structure of these galaxies we also determined the
asymmetry and concentration index of the galaxies \citep[e.g.][]{AVYB94}.  To
calculate the concentration parameter, we use the ellipticity, position angle and
central coordinates (derived with SExtractor on the second order moments of the
images) to define an elliptical area that includes all the flux above $2\sigma$ of
the sky level and an elliptical area for which the semi-major axis is 30\% of the
outer one (Eq.\,\ref{conc}).
                                        
\begin{equation}
C=\frac{\sum_{}\sum_{i,j\in E(0.3)}I_{ij}}{\sum_{}\sum_{i,j\in E(1)}I_{ij}}.
\label{conc}
\end{equation}

\begin{figure}
   \includegraphics[width=\columnwidth]{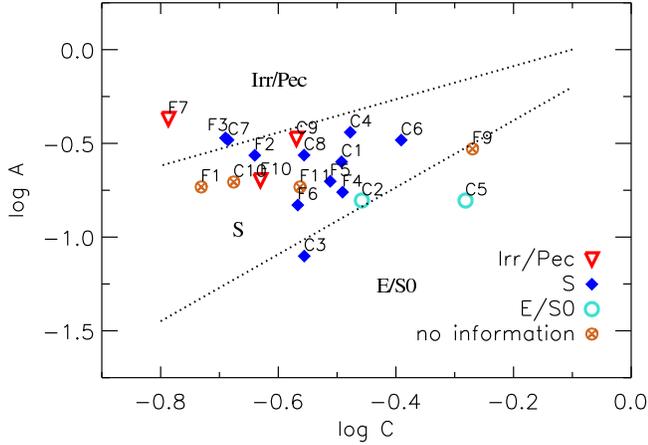}
 \caption{Distribution of our galaxies in the A-C plane.  Symbols are used to indicate
 morphologies of these galaxies. The references for this information are given in Table
 \ref{centdist}.  In case inconsistent morphology information was given in different
 references, we used the morphologies from \citet{M08}.  The dashed lines are the
 selection limits taken from \citet{MFMBMBI06}. 
            }
         \label{asymconc}
   \end{figure}

For the asymmetry index, we applied the definition and centering method of
\citet{CBJ00}.  The sky subtracted galaxy image and its counterpart rotated by
180\,degrees are compared with each other in an elliptical area including all flux
above  the $1.5\sigma$ sky level.  A $3\times3$ grid was defined around the
initial center to search for the coordinates that give the smallest asymmetry. 
The distance from the central point to the eight surrounding points was set to
0\farcs1.  The grid was reconstructed around the coordinates that gave the
smallest asymmetry in the previous grid.  That was done until the smallest value
was measured at the center.  Similarly, the asymmetry of a blank region was
calculated in the sky subtracted image using the same elliptical mask that was
used for the galaxy.  To correct for the contribution of the noise, this ``blank
asymmetry'' was then subtracted from the asymmetry of the galaxy:

\begin{equation}
A=min\left(\frac{\sum_{}|I_{0}-I_{180}|}{\sum_{}|I_{0}|}\right)-\left(\frac{\sum_{}|B_{0}-B_{180}|}{\sum_{}|I_{0}|}\right)
\label{asym}
\end{equation}

\noindent where $I$ represents the image pixel values and $B$ represents the blank
region pixel values.  The asymmetry and concentration parameters of our galaxies are
given in Table ~\ref{tabrun2} in the Appendix.  This information can be used for
morphological classification.  We present our galaxies on the A-C plane together with
their morphologies from the literature (Table \ref{centdist}) and the selection limits
from \citet{MFMBMBI06} in Fig.\ref{asymconc}. 

%-------------------------------------------------------------
%
%_____________________________________________________________

\section{Analysis}
Here we want to classify our galaxies using a number of parameters that are
expected to be good indicators of the irregularity of velocity
fields: $k_{3,5}/k_{1}$, $\sigma_{PA}$ and $\Delta \phi$.  The deviation from
simple rotation was measured using the $k_{3,5}/k_{1}$ parameter, described in
Sect.2.4.  It is the squared sum of the higher order Fourier terms (normalized
by the rotation velocity) obtained for a fit with a fixed position angle and
ellipticity. For each galaxy we obtained an average $k_{3,5}/k_{1}$ value.  
$\sigma_{PA}$ is the standard deviation of the
kinematic position angle within a galaxy and $\Delta \phi$ is the absolute value
of the average difference between the kinematic and photometric position angle at
several radii.  The radial extent over which these parameters have been calculated
depends on the distance of the galaxy, its surface brightness and its ionized gas
contents.  Since not at every radius enough gas is present we could not fit the
velocity fields at all radii for all galaxies.  The maximum radius for which we
could measure these parameters for each galaxy ($R_{max}$) is given in Table
\ref{tab2}.  $\sigma_{PA}$, $\Delta \phi$ and $k_{3,5}/k_{1}$ are calculated
between $r=0\farcs39$ and the $R_{max}$ (Table \ref{tab2}).  The errors in these
parameters are the variance of the parameters in the range of the observations.

\begin{table}[h]
\caption{Parameters quantifying the irregularity of the gas kinematics measured
for our sample}
\label{tab2}
\begin{center}
\begin{tabular*}{0.35\textwidth}{@{\extracolsep{\fill}}c@{\hspace{0.07in}}c@{\hspace{0.1in}}c@{\hspace{0.18in}}c@{\hspace{0.07in}}c@{\hspace{0.17in}}c@{\hspace{0.01in}}c@{\hspace{0.005in}}c}
\hline\hline
ID &  $R_{max}$ &  $\sigma_{PA}$    &   $\Delta \phi$  &   &  $ k_{3,5}/k_{1}$   &\\
$(1)$ & $(2)$ & $(3)$  &  $(4)$ & & $(5)$  &  \\
\hline
\end{tabular*}
\begin{tabular*}{0.35\textwidth}{@{\extracolsep{\fill}}lllr@{\hspace{0.07in}}@{$\pm $}@{\hspace{0.07in}}lr@{\hspace{0.07in}}@{$\pm $}@{\hspace{0.07in}}l}
 C7 &  13.8	     &  23 &  68&26   & 0.32 & 0.20   \\
 C8 &  20.9	     &   2 &   9& 6   & 0.06 & 0.05   \\
 C9 &	9.2	     &  19 &  66&20   & 0.10 & 0.03   \\ 
 C10 &  11.7	     &   9 &  14& 9   & 0.26 & 0.08   \\
 F2 &  10.5	     &   2 &  35&37   & 0.08 & 0.02   \\
 F3 &	10.0	     &   7 &  39& 9   & 0.07 & 0.05   \\
 F4 &	5.5	     &  21 &  18& 20  & 0.30 & 0.19   \\
 F5 &  11.3	     &   5 &  46& 9   & 0.08 & 0.03   \\ 
 F6 &	4.4	     &  29 &  57&44   & 0.27 & 0.18   \\ 
 F7 &  14.1  	     &   3 &   0&11   & 0.05 & 0.02   \\
 F10 &	8.4	     &   8 &   1& 7   & 0.25 & 0.05   \\
 F11 &  1\farcs7 &   5 &  18& 5   & 0.05 & 0.02   \\
\hline
\end{tabular*}
\end{center}

Column (1) Object ID. \\
Column (2) Maximum radius for which kinematic parameters could be calculated.  The conversion from
arcsecond into kpc was done as explained in \citet{W06}. \\
Column (3) Standard deviation of the kinematic position angle ($\sigma_{PA}$).\\
Column (4) Mean misalignment between the kinematic and photometric position angles
($\Delta \phi$). \\
Column (5) Mean $k_{3,5}/k_{1}$ of the analysis done
while fixing the position angle and the ellipticity to their global values. \\

{\it For galaxy F11, only the photometric redshift is available.  Since
photometric redshifts have big uncertainties, we give $R_{max}$ of this galaxy in
arcseconds.  $k_{3,5}/k_{1}$ of galaxy $F5$, $\Delta \phi$ of the galaxies that
have $\epsilon \leq 0.1$ (galaxy F2) and all parameters for galaxy F10 are
rather meaningless as explained in Sect.4.1.  Therefore they are
excluded from the analysis.}
\end{table}

\begin{figure}
   \includegraphics[width=\columnwidth]{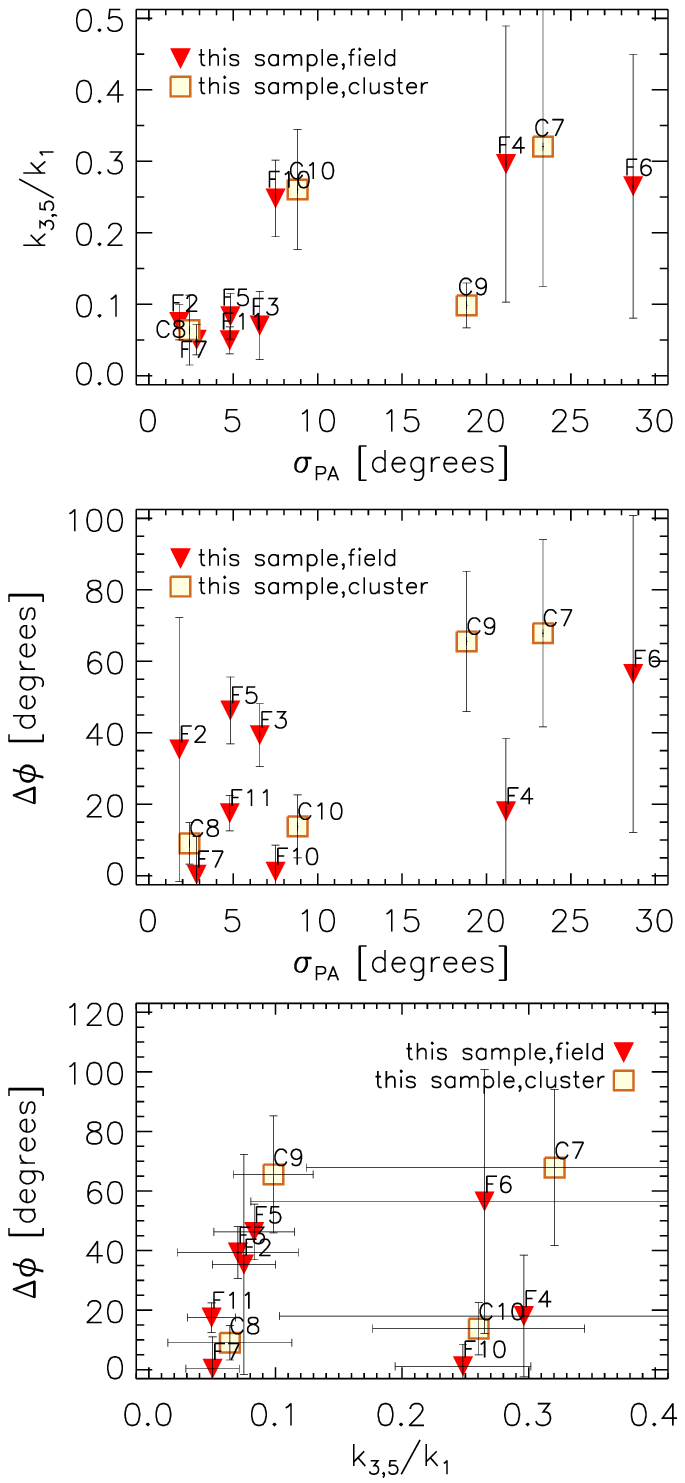}
 \caption{
\textit{Top:} 
Standard deviation of the kinematic position angle ($\sigma_{PA}$) versus mean
$k_{3,5}/k_{1}$. 
\textit{Middle:} Mean misalignment between kinematic and 
photometric axes ($\Delta \phi$) 
versus standard deviation of kinematic position angle ($\sigma_{PA}$).
\textit{Bottom:} Mean $k_{3,5}/k_{1}$ versus mean misalignment between 
kinematic and photometric axes ($\Delta \phi$). }
\label{cop2}
\end{figure}

\begin{figure}
   \includegraphics[width=\columnwidth]{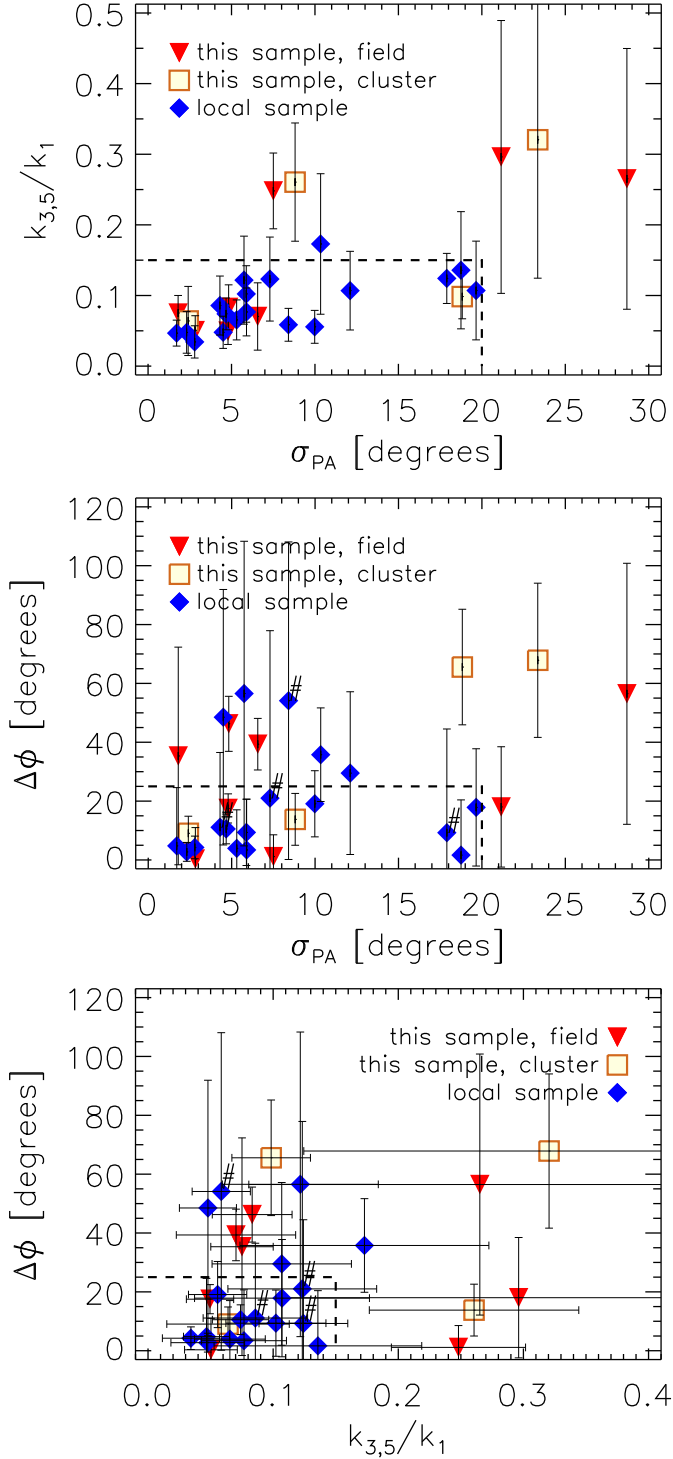}
 \caption{
The same plots as in Fig.\ref{cop2}, now together with the local sample.  
The boundaries inside which we call the gas kinematics regular are indicated with dashed lines
for each parameter. The galaxies that are indicated
with \# symbol are the galaxies for which the $\Delta \phi$ is doubtful as
explained in Sect.4.1.  Their $\Delta \phi$ measurements are excluded form the
analysis.
 }
\label{loc}
\end{figure}

In Fig.\ref{cop2}, we show how the galaxies are distributed in the parameter space
of $k_{3,5}/k_{1}$, $\Delta \phi$ and
$\sigma_{PA}$.  To be able to decide whether these velocity fields are regular we
should determine a regularity treshold for each parameter.  To do that, we measured
the parameters for the $H\alpha$ velocity fields of the SINGS local sample
\citep{DCAHCBK06} in which most galaxies have regular kinematics.  Among the
velocity fields we were provided with (all but NGC925, NGC2403, UGC5423, NGC3198,
NGC4321, NGC6946) we excluded from the analysis the galaxies that have magnitudes
which are very different from the ones in our sample (m81dwb, NGC2915) and the
velocity fields that are very noisy (IC4710 and NGC5398).  For every galaxy we ran
kinemetry to determine the parameters $k_{3,5}/k_{1}$, $\Delta \phi$ and
$\sigma_{PA}$.  To be able to calculate $\Delta \phi$, the photometric position
angle should be known as a function of radius.  To measure that, we used the red
Palomar DSS images ($6450\AA$), apart from NGC3521, NGC3621 and NGC5713, for which
we used blue UK-Schmidt images ($4680\AA$), also from the DSS.  All parameters are
given in Table \ref{tabloc}.  It is pointed out in \citet{DCAHCBK06}, Table 1 that
some of these galaxies are peculiar.  These galaxies are indicated next to their ID
in Table \ref{tabloc}.  We plotted the parameters against each other for both the
local sample and our sample together in Fig.\ref{loc}.  Using the location of the
local galaxies that are not peculiar and do not belong to the Virgo cluster, we
defined a limit for each parameter below which we call the gas kinematics
``regular''. These regularity criteria are: $\Delta \phi \leq 25$, $\sigma_{PA}
\leq 20$ and $k_{3,5} / k_{1} \leq 0.15$. 

%-------------------------------------------------------------
%_____________________________________________________________
%-------------------------------------------------------------

\begin{table}[h]
\caption{Parameters quantifying the irregularity of the gas kinematics measured
for the local sample}
\label{tabloc}
\begin{center}
\begin{tabular*}{0.37\textwidth}{@{\extracolsep{\fill}}c@{\hspace{0.52in}}cc@{\hspace{0.11in}}@{$ $}@{\hspace{0.1in}}lr@{\hspace{0.1in}}c@{\hspace{0.02in}}l}
\hline\hline
ID &  & $\sigma_{PA}$      & $\Delta \phi$ &  &   $ k_{3,5}/k_{1}$  &      \\
$(1)$ & & $(2)$ &  $(3)$ &  & $(4)$ &      \\
\hline
\end{tabular*}
\begin{tabular*}{0.37\textwidth}{@{\extracolsep{\fill}}llr@{\hspace{0.07in}}@{$\pm $}@{\hspace{0.07in}}lr@{\hspace{0.07in}}@{$\pm $}@{\hspace{0.07in}}l}
NGC0628 	&      18      &    9   &    35    &  0.12   &  0.04	 \\
NGC2976 (pec)   &    	6      &    3   &    17    &  0.08   &  0.03	 \\
NGC3031 	&   	19     &    2   &    19    &  0.14   &  0.08	 \\
NGC3049 	&      20      &   18   &    20    &  0.11   &  0.07	 \\ 
NGC3184 	&       7      &   21   &    57    &  0.12   &  0.06	 \\ 
NGC3521 	&       2      &    3   &     3    &  0.05   &  0.03	 \\ 
NGC3621 	&       2      &    5   &    20    &  0.05   &  0.02	 \\
NGC3938 	&       4      &   11   &    25    &  0.09   &  0.04	 \\
NGC4236 	&       5      &   11   &     5    &  0.07   &  0.02	 \\
NGC4536 	&       5      &    4   &    13    &  0.07   &  0.03	 \\
NGC4569 (Virgo)	&   	10     &   19   &    11    &  0.06   &  0.02	 \\
NGC4579 (Virgo)	&   	10     &   36   &    16    &  0.17   &  0.10	 \\
NGC4625 (pec) 	&       6      &   57   &    52    &  0.12   &  0.06	 \\
NGC4725 (pec) 	&      12      &   30   &    28    &  0.11   &  0.06	 \\
NGC5055 	&       3      &    4   &     4    &  0.03   &  0.02	 \\
NGC5194 (pec) 	&       5      &   49   &    43    &  0.05   &  0.02	 \\ 
NGC5713 (pec) 	&       8      &   54   &    54    &  0.06   &  0.02	 \\ 
NGC7331 	&       6      &    9   &    11    &  0.10   &  0.04	 \\

\hline			   
\end{tabular*}		   
\end{center}	

Column (1) Object ID. \\
Column (2) Standard deviation of the kinematic position angle ($\sigma_{PA}$).\\
Column (3) Mean misalignment between the kinematic and photometric position angles
($\Delta \phi$). \\
Column (4) Mean $k_{3,5}/k_{1}$ of the analysis done
while fixing the position angle and the ellipticity to their global values. \\

{\it Peculiar galaxies and members of Virgo Cluster are indicated on the
first column with ``(pec)/(Virgo)'' next to their
name.  $\Delta \phi$ of the galaxies that have $\epsilon \leq 0.1$ (NGC 628,
NGC 3184, NGC 3938, NGC 5713) have large errors as explained
in Sect.4.1, and are therefore excluded from the analysis.} 

\end{table}

%_____________________________________________________________
%-------------------------------------------------------------

To check whether the 3 parameters $k_{3,5}/k_{1}$, $\Delta \phi$ and $\sigma_{PA}$ 
are indicators of the same phenomena, and correlate with each other, we calculated
the uncertainty weighted linear Pearson correlations separately for the two
samples.  The correlation coefficients for each relation are given in the first two
rows of Table \ref{tab3}.  $\sigma_{PA}$ and $k_{3,5}/k_{1}$ seem to correlate with
each other while the correlation between $\Delta \phi$ and $\sigma_{PA}$ holds only
for the higher redshift sample and the correlation between $\Delta \phi$ and
$k_{3,5}/k_{1}$ holds only for the local sample.  We will discuss this result in
more detail in Sect.5.

\begin{table}[h]
\caption{Linear Pearson correlation coefficients}
\label{tab3}
\begin{center}
\begin{tabular}{lccc}

\hline\hline
&$\sigma_{PA}$-$k_{3,5}/k_{1}$ & $\Delta \phi$-$\sigma_{PA}$ & $k_{3,5}/k_{1}$-$\Delta \phi$   \\
\hline
this sample &   0.5			   & 0.7		       & 0.2			         \\
local sample &  0.7			   & 0.1		       & 0.7			         \\
\hline
& $R_{d}$-$k_{3,5}/k_{1}$ & $R_{d}$-$\sigma_{PA}$ & $R_{d}$-$\Delta \phi$		 \\
\hline
this sample &  -0.3		 & -0.4 	       & -0.3				     \\
local sample &  0.0		 & 0.1	 	       & 0.0				     \\
\hline
&   $M_{B}$-$k_{3,5}/k_{1}$	 &  $M_{B}$-$\sigma_{PA}$   &  $M_{B}$-$\Delta \phi$		 \\
\hline
this sample &  0.5		     &	     0.4 		  &	  0.4  		    \\
local sample & 0.1		     &	     -0.2		  &	   0.1  		    \\
\hline
& A-$k_{3,5}/k_{1}$ &  A-$\sigma_{PA}$ & A-$\Delta \phi$		 \\
\hline
this sample & 0.0 	     & -0.3		 & 0.0  		     \\
\hline
& C-$k_{3,5}/k_{1}$ & C-$\sigma_{PA}$  & 	C-$\Delta \phi$	 \\
\hline
this sample & 0.1	  & 0.2 	    & 0.1	    		      \\
\hline
 & $F_{em}$-$k_{3,5}/k_{1}$ & $F_{em}$-$\sigma_{PA}$ & 	$F_{em}$-$\Delta \phi$\\
this sample & -0.5 	& -0.4		& 	-0.7	\\
\hline
 & $F_{em}/L_{B}$-$k_{3,5}/k_{1}$ & $F_{em}/L_{B}$-$\sigma_{PA}$ & $F_{em}/L_{B}$-$\Delta \phi$\\
this sample & -0.4 	& -0.5		& -0.7		\\
\hline
 & z-$k_{3,5}/k_{1}$ & z-$\sigma_{PA}$ & 	z-$\Delta \phi$\\
 both samples & -0.1 & -0.2 & 0.4 \\
\hline
\end{tabular}
\end{center}

{\it $\Delta \phi$ of the galaxies that have $\epsilon \leq 0.1$ (galaxy F2 in
``this sample'', NGC628, NGC3184, NGC3938 and NGC5713 in the local sample),
$k_{3,5}/k_{1}$ of galaxy $F5$ (this sample) and all parameters for galaxy F10
(this sample) are doubtful as explained in Sect.4.1.  Therefore they are
excluded while calculating the correlation coefficients.  For the calculation
of the correlations with the redshift, only field galaxies were used. So the
results do not have the bias of the environment.  Including the cluster
galaxies does not change the result.  }	\\

\end{table}

\begin{figure}
   \includegraphics[width=\columnwidth]{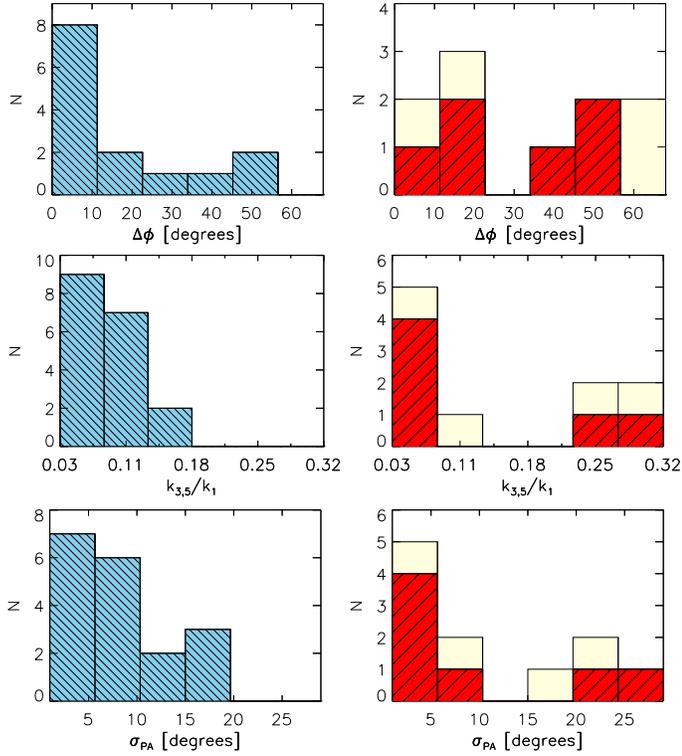}
 \caption{ Histograms of the mean misalignment between kinematic and photometric
major axes ($\Delta \phi$), mean $k_{3,5}/k_{1}$ and standard deviation of
kinematic position angle ($\sigma_{PA}$).  The local sample is given in blue
(the column on the left, fine-shaded at $-45^{\circ}$), our sample (the column
on the right): field galaxies in red (coarse-shaded at $45^{\circ}$) and cluster
galaxies in light yellow (not shaded).  $\Delta \phi$ of the galaxies that have
$\epsilon \leq 0.1$ (galaxy F2 in ``this sample'', NGC628, NGC3184, NGC3938 and
NGC5713 in the local sample), $k_{3,5}$ of galaxy $F5$ (this sample) and all
parameters for galaxy F10 (this sample) are doubtful as explained in Sect.4.1. 
Therefore they are excluded from the histograms.  The cluster galaxies are
distributed within similar ranges as the field galaxies.}
         \label{cop1}
   \end{figure}

We compared the distributions of $k_{3,5}/k_{1}$, $\Delta \phi$ and $\sigma_{PA}$
for the local and higher redshift galaxies (Fig.\ref{cop1}).  While some galaxies
in our sample have $k_{3,5}/k_{1}$ values that are far above the region where the
local galaxies are distributed, this is not the case for the $\sigma_{PA}$ and
$\Delta \phi$ parameters.  Galaxies in both samples are distributed within similar
ranges.  $\sigma_{PA}$ of galaxy F6 in our sample is $9^{\circ}$ above the highest
value among the local galaxies.  This galaxy has a kinematically decoupled core and
that causes the $\sigma_{PA}$ to be large.  All local galaxies that have a large
$\Delta \phi$ value are either peculiar or a member of the Virgo Cluster. 

Using the distributions for the cluster and field galaxies in our sample
(Fig.\ref{cop1}), we investigate the effect of the environment on the gas kinematics. 
All 4 cluster galaxies are distributed in a similar range as the field galaxies.  The
fraction of galaxies that have irregular gas kinematics in the whole sample, among the
field and cluster galaxies separately, is given in Table \ref{tab5}.  Without
distinguishing between different criteria we find that $75 \pm 22$\% of the cluster
members, $50 \pm 18$\% of the field galaxies and $58 \pm 14$\% of the whole sample have
irregular gas kinematics.  These numbers are underestimates because the spatial resolution
of the $z\approx0.5$ observations is low.  In Appendix A we show that the fractions
given above are lower limits to the real number of irregular galaxies.

%-------------------------------------------------------------
%
%_____________________________________________________________

\begin{table}[h]
\caption{Irregularity fraction}
\label{tab5}
\begin{center}
\begin{tabular}{lccc}

\hline\hline
& $frac_{\sigma_{PA}}$ & $frac_{\Delta \phi}$ & $frac_{k_{3,5}/k_{1}}$	\\
& $(1)$ &$(2)$ &$(3)$	\\
\hline
field \& cluster & 27 $\pm$ $13\%$ & 50 $\pm$ $16\%$ & 40 $\pm$ $15\%$  \\
only field & 29 $\pm$ $17\%$ & 50 $\pm$ $20\%$ & 33 $\pm$ $19\%$  \\
only cluster & 25 $\pm$ $22\%$ & 50 $\pm$ $25\%$ & 50 $\pm$ $25\%$  \\
\hline
\end{tabular}
\end{center}

Column (1): Fraction of irregular velocity fields according to $\sigma_{PA}$ criterion and its Poisson error \\
Column (2): Fraction of irregular velocity fields according to $\Delta \phi$ criterion and its Poisson error  \\
Column (3): Fraction of irregular velocity fields according to $k_{3,5}/k_{1}$ criterion and its Poisson error \\

Note: {\it $k_{3,5}/k_{1}$ of galaxy $F5$, $\Delta \phi$ of the galaxies that have
$\epsilon \leq 0.1$ (galaxy F2) and all parameters for galaxy F10 have been excluded
from the calculations as explained in Sect.4.1}	\\

\end{table}

\begin{figure*}
   \includegraphics[width=170mm]{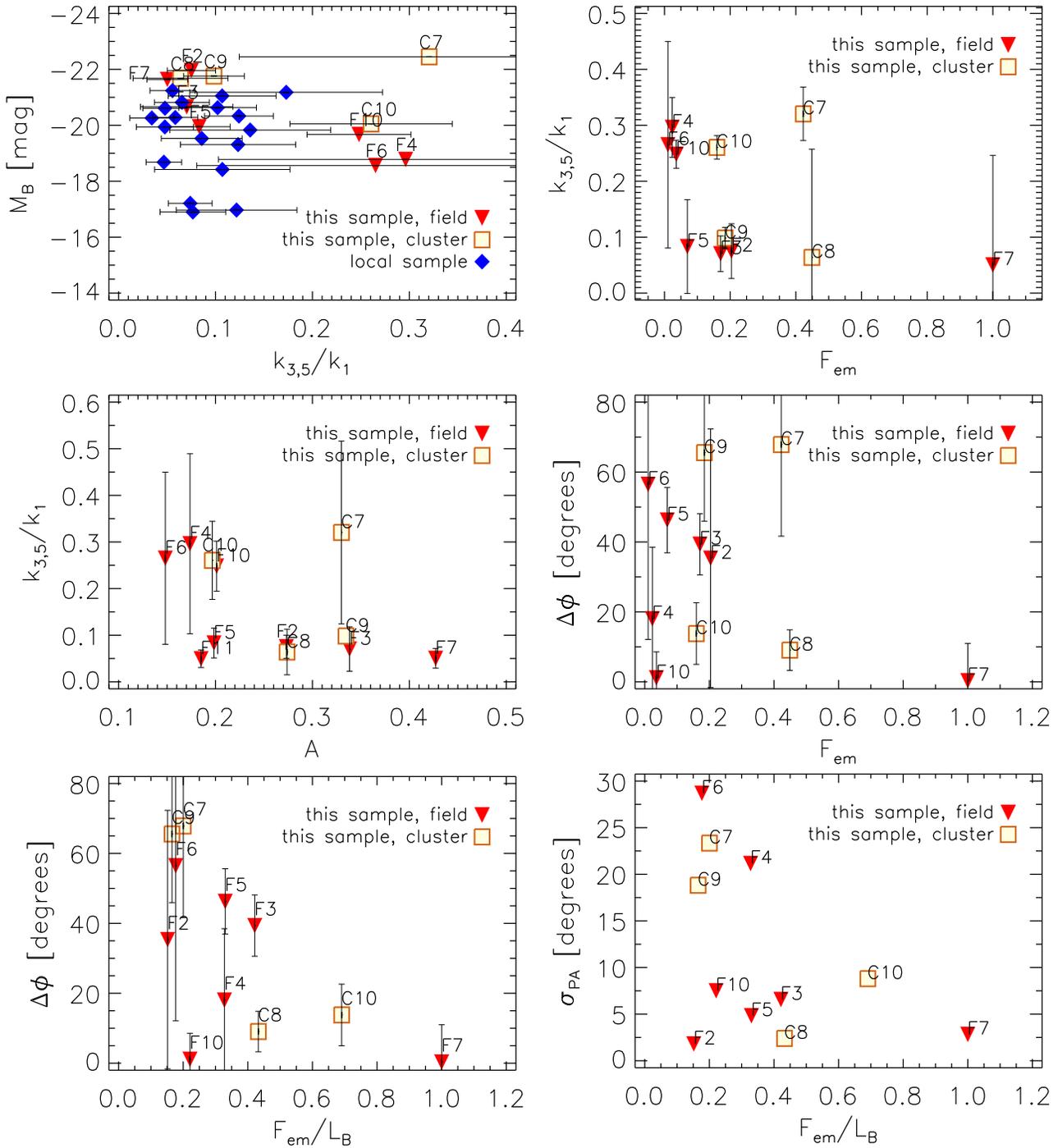}
 \caption{
\textit{Top Left:} $k_{3,5}/k_{1}$ versus absolute magnitude in the B band
($M_{B}$).
\textit{Middle Left:} Asymmetry index (A) versus $k_{3,5}/k_{1}$.
\textit{Bottom Left:} Relative [OII]3727 flux per unit B band luminosity
($F_{em}/L_{B}$) versus the mean misalignment between the the photometric and the
kinematic position angle ($\Delta \phi$).
\textit{Top Right:} $k_{3,5}/k_{1}$ versus the relative [OII]3727 flux
($F_{em}$).
\textit{Middle Right:} Mean misalignment between the the photometric and the
kinematic position angle ($\Delta \phi$) versus the relative [OII]3727 flux
($F_{em}$).
\textit{Bottom Right:} Standard deviation of the kinematic position angle
($\sigma_{PA}$) versus relative [OII]3727 flux per unit B band luminosity
($F_{em}/L_{B}$).
}
         \label{cop3}
   \end{figure*}

We now checked whether there is a correlation between the irregularity in the gas kinematics
and some other parameters derived from the photometry: the absolute B magnitude, disk scale
length, asymmetry and concentration index (see Sect.3.2).  We did this for our sample and
the local sample separately.  For the local sample, we do not have all these parameters, so we
give the correlations only for the absolute magnitude in the B band ($M_{B}$) and the disk
scale length.  We calculated their disk scale length assuming $R_{d}$ $\sim$ 0.25 $R_{25}$
\citep[$R_{25}$ values are taken from][]{DCAHCBK06}.  We find a correlation between $M_{B}$
and $k_{3,5}/k_{1}$ only for our sample (Fig.\ref{cop3}).  The distribution of the local
sample galaxies in the plot is showing that galaxies with very different $M_{B}$ values can
have very similar $k_{3,5}/k_{1}$ parameters.  This correlation is therefore most probably
spurious. 

%------------------------------------------------------------
%
%_____________________________________________________________

We do not find a correlation between the irregularity in the gas kinematics and either
the concentration index C, which is an indicator of the morphological type of the
galaxy, or the photometric asymmetry (Table \ref{tab3}).

We have investigated whether the irregularity in the gas kinematics of the galaxy
correlates with its ionized gas contents.  This is calculated by measuring the
integrated emission line flux and correcting it for the distance of the galaxy
(multiplying it by $d^{2}$).  Since the redshift interval of the sample is large
($0.1\leq z \leq 0.9$) we do not have the same lines for every galaxy.  The strongest
emission line that we have for all cluster galaxies is [OII]3727.  Using the mean
emission line flux ratios derived from our spectra ($H\alpha / [OII]3727=6.60$,
$H\beta / [OII]3727=0.82$, $[OIII]5007 / [OII]3727=1.77$), we calibrated the flux
from other lines to the [OII]3727 flux.  The emission line that was used for each
galaxy is given in Table \ref{gasamount}.  We then normalized the ionized gas fluxes
to the value of galaxy F7, which has the highest flux (Table \ref{gasamount}).  The
correlation coefficients of the relations between the amount of gas ($F_{em}$) and the
irregularity in the gas kinematics are given in Table \ref{tab3}.  $\Delta \phi$ and
$k_{3,5}/k_{1}$ have an anti-correlation with $F_{em}$.  These parameters are
plotted against each other in Fig.\ref{cop3}. 

\begin{table}[h]
\caption{Relative emission line fluxes}
\label{gasamount}
\begin{center}
\begin{tabular}{lccc}

\hline\hline
name &  $F_{em}$ &$F_{em}$/$L_{B}$	& line		\\
$(1)$ & $(2)$ &   $(3)$ 		& (4)		\\
\hline
 C7 &  0.42 	&  0.20  & [OII]3727       \\
 C8 &  0.45 	&  0.43  & [OII]3727       \\ 
 C9 &  0.18 	&  0.17  & [OII]3727       \\
 C10 & 0.16 	&  0.69  & [OII]3727       \\
 F2 &  0.20 	&  0.15  & [OII]3727       \\
 F3 &  0.17 	&  0.42  & $H\beta$	 \\
 F4 &  0.02 	&  0.33  & $H\beta$	 \\   
 F5 &  0.07 	&  0.33  & $H\beta$	 \\
 F6 &  0.01 	&  0.18  & $H\alpha$     \\
 F7 &  1.00 	&  1.00  & [OII]3727      \\
 F10 &  0.04	&  0.22  & [OIII]5007      \\
\hline
\end{tabular}
\end{center}
Column (1): Galaxy ID	\\
Column (2): Relative [OII]3727 flux   \\
Column (3): Relative value of [OII]3727 flux per unit B-band luminosity     \\
Column (4): Emission line used to calculate the [OII]3727 flux for each galaxy      \\
\end{table}

Using the concentration index, we find no correlation between the galaxy type and
the irregularity in the gas kinematics.  The amount of gas per unit luminosity is
also an indicator of the galaxy type such that later type spirals have more gas and
they are less luminous. We find an anti-correlation between this parameter and
both $\Delta \phi$ and $\sigma_{PA}$ (Table \ref{tab3}, Fig.\ref{cop3}). In all correlations
with $F_{em}$ and $F_{em}/L_{B}$, one sees that the galaxies with large emission line fluxes
generally have very regular velocity fields (Fig.\ref{cop3}).
Since the number of objects that have large
$F_{em}$ and $F_{em}/L_{B}$ values is very small, the correlations we find
for these two parameters are most probably spurious.

Galaxies in our sample have redshifts between 0.1 and 0.9.  Using both our
sample and the local sample together, we checked whether the regularity of the
gas kinematics is a function of redshift.  In order not to have the bias of the
environment on this analysis, we used only the field galaxies.  No correlation
was found.  Including the cluster galaxies in the correlation does not change
the result.  Since the number of galaxies in our sample is small, this issue
should be revisited in the future.

\subsection{Special cases}

In our sample, some galaxies need some special mention: Galaxies F2,  F5 and F10.
 Galaxy F2 has a large $\Delta \phi$, even though it has regular kinematics
according to the other criteria.  It is nearly face-on $(\epsilon=0.1)$
(Fig.\ref{gal8fig}.a), which causes the photometric position angle
measurements to be unreliable.  Therefore, $\Delta \phi$ is not meaningful
in this case. 

For galaxy F5, only the inner part of the galaxy is included (Appendix, Fig.
A.12.i) for the parameter $k_{3,5}/k_{1}$ due to the large misalignment of the
kinematic and photometric axis, so that the range for which it is determined is
much smaller than for $\Delta \phi$ and $\sigma_{PA}$.  There are a few more
galaxies that have large misalignments (e.g. galaxy F3 and galaxy C9).  In those
cases, however, the redshifts are much higher and the radial extent of the 
observed velocity fields is rather small, so that the measurements of all
parameters is limited to the same region.

The data of galaxy F10 are very noisy, so the high order Fourier terms are
strongly affected (Fig.\ref{gal13fig}.e).  In our sample we exclude
$k_{3,5}/k_{1}$ of galaxy F5, $\Delta \phi$ of the galaxies that have $\epsilon
\leq 0.1$ (which is the case only for galaxy F2) and all parameters for galaxy
F10 from the analysis.

In the local sample, we excluded the $\Delta \phi$ of the galaxies that have
$\epsilon \leq 0.1$ (NGC628, NGC3184, NGC3938, NGC5713).

\section{Discussion}

Field spirals that have an unbarred morphology and no nearby neighbors are expected to
have regular velocity fields \citep{RWK99}.  However, about 70 percent of the spiral
galaxies are expected to have bars \citep{EFPQDDHKRSTT00,KSP00}.  A bar and spiral arms
may cause large scale deviations from circular rotation.  This is discussed in
\citet{FADKMSZ05} together with different possible reasons for small scale anomalies
(e.g. nuclear activity, gas outflow from a star formation burst region, a shock
excitation mechanism, etc.).  See also \citet{WBB04} and \citet{FVPEF05} where
noncircular motions such as bar streaming, inflow and a warp are distinguished from one
another using a harmonic decomposition of the velocity fields.  However, in the case of
distant galaxies, such small scale perturbations would be smeared out in  observations,
mainly due to seeing \citep{KKSZ07}.

Interactions between galaxies may cause big deviations from circular rotation
\citep{RPABBR05}.  In clusters of galaxies, several interaction mechanisms can take
place in addition to the galaxy-galaxy interactions \citep{P04} such as ram pressure
stripping \citep{GG72,QMB00,AMB99,RB07,KKFUS08}, gas starvation \citep{BCS02,LTC80},
harassment \citep{MLKDO96,MLK98} and viscous stripping \citep{N82}. \citet{METSRS07}
give an estimate of where in MS0451 different interaction processes are expected to
be effective.  These phenomena may produce large scale perturbations in the velocity
field, similar to the effect of galaxy-galaxy interactions \citep{RWK99}.  Therefore
the fraction of irregular velocity fields is expected to be higher for cluster
galaxies than for field galaxies.

We use three indicators of irregular gas kinematics: $\Delta \phi$, $\sigma_{PA}$ and
$k_{3,5}/k_{1}$ (see Sect.4). $\sigma_{PA}$ and $k_{3,5}/k_{1}$ seem to correlate with
each other, while this is not the case for all parameters.  It is also not expected
that all phenomena that may cause kinematical irregularities would increase each
parameter.  In case of interaction processes, the disturbance in the velocity field
might cause all parameters to increase.  A secondary component in the velocity field
would increase the $k_{3,5}/k_{1}$ term while the components do not have to be
kinematically misaligned.  The information from all parameters together could be used
to interpret what kind of a phenomenon might be responsible from the irregularities in
the gas kinematics.  However, the resolution effects should be taken into account
while doing that (see Appendix A).

We have measured these parameters for the galaxies in our sample and in the SINGS
local sample.  All local galaxies apart from one cluster member have regular gas
kinematics according to the $k_{3,5}/k_{1}$ criterion while all galaxies are regular
according to the $\sigma_{PA}$ criterion.  There are four galaxies that do not meet
the $\Delta \phi$ criterion (excluding the face-on cases as mentioned in Sect.4.1). 
Each galaxy in the local sample that has a large $\Delta \phi$ value is either
peculiar or is a cluster member.  While some galaxies in our high z sample have
$k_{3,5}/k_{1}$ values that are far above the region where the local galaxies are
distributed, this is not the case for the $\Delta \phi$ and $\sigma_{PA}$ parameters
(the highest $\sigma_{PA}$ value in our sample belongs to galaxy F6 which has a
kinematically decoupled core).  Galaxies in both samples are distributed within
similar $\sigma_{PA}$ and $\Delta \phi$ ranges.  \citet{GMABG05} reported for their
local sample that, on average, the misalignment between the kinematic and the
photometric position angle is $15^\circ \pm 19^\circ$ and the maximum value they
measured is $75^\circ$.  Also in our sample, the fraction of galaxies that are
irregular according to the $\Delta \phi$ criterion is the highest.  This kind of
information might give clues to the dominant phenomena that are responsible for the
irregularities or the disturbances that take longer to recover in the gas kinematics. 
That would only be reliable, however, in case of high resolution data.  At
low spatial resolution, some combinations of a large $k_{3,5}/k_{1}$ parameter and
sampling can cause the measured values of $\sigma_{PA}$ and $\Delta \phi$ to be
boosted (see Appendix A).

The percentage of galaxies in our sample that have irregular gas kinematics are: $27
\pm 13$\% (according to $\sigma_{PA}$ criterion), $50 \pm 16$\% ( $\Delta \phi$
criterion), $40 \pm 15$\% ($k_{3,5} / k_{1}$ criterion) and $58 \pm 14$\% (any
criterion).  Among these galaxies, there are 4 cluster members and they all are
closer than 1 Mpc to the center.  Since this is well inside the virial radius of the
cluster, we expect the ICM density to be rather high and, therefore, able to distort
the ISM of the galaxy disks.  On the other hand, relative velocities are quite high so
that direct encounters would have rather negligible effects.  When we analyze the
distributions of each of the three parameters (Fig.\ref{cop1}), we see that cluster
and field galaxies have similar behaviour.

\citet{FHPAB06} made a kinematical classification based on the alignment  of the
photometric and kinematic axes and the agreement between the peak of their $\sigma$
map and the kinematic center.  They found that $35\%$ of the disks are rotating and
as such regular, for a sample of 35 galaxies at $0.4<z<0.75$.  For the galaxies in
our sample that are in the same redshift interval as Flores et al.'s sample, (7
galaxies, but we excluded the special cases mentioned in Sect.4.1 from the analysis)
the fraction of objects that have regular gas kinematics is: $40 \pm 22$\% (according
to $\Delta \phi$ criterion ), $83 \pm 15$\% ($\sigma_{PA}$ criterion) and $67 \pm
19$\% ($k_{3,5}/k_{1}$ criterion).  So, using the $\Delta \phi$ criterion, our result
is comparable with their result.  \citet{FGBVESSDL06} investigated galaxy kinematics
at a much higher redshift.  They obtained $H\alpha$ maps of 14 massive star forming
galaxies at $z\approx2-3$.  Four of the six best spatially resolved cases match the
expectations for a simple rotating disk.  \citet{SGSTB08} introduce a method to
distinguish merging and non-merging systems.  Using kinemetry, they measure the
deviation from the ideal case for both the gas velocity field and the stellar
continuum intensity map of the galaxies.  They use these parameters of 29 merging and
regular template galaxies to define a criterion that discerns between the two.  They
then apply this method to 11 $z\approx2$ galaxies and find that $>50$\% of these
systems are consistent with a single rotating disk interpretation. Considering this
result together with high (up to $100 M_{\sun} yr^{-1}$) star formation rates of these
systems, they argue that the smooth accretion mechanism can play an important role in
the early stages of the massive-galaxy evolution.

Using H$\alpha$ velocity fields of local galaxies, \citet{GMABG05} found that
the asymmetry of the rotation curves (or lopsidedness) is correlated with the
morphological type, the luminosity, the (B-V) colour and the maximal rotational
velocity of galaxies.  They find that brighter, more massive and redder galaxies
have smaller deviations from regular kinematics.  Here we do not use a parameter
that corresponds to kinematic lopsidedness.  Although we know that lopsidedness
would cause $k_{3,5}/k_{1}$ to increase, large  $k_{3,5}/k_{1}$ values are not
necessarily due to lopsidedness (e.g. secondary kinematical  component can also
do the job).  \citet{GMABG05} also measured the misalignment between the
kinematic and photometric position angles of their galaxies and found no
correlation with morphological type.  We investigated this effect  using two
indicators of galaxy type: the concentration index and the amount of gas per
unit luminosity.  We do not find a correlation with the concentration index
while we do find a negative correlation using $F_{em}$ / $L_{B}$ which means
that later type galaxies have smaller misalignments.  This correlation could be
spurious as explained in Sect.4.1.  

We do not find a correlation between the photometric asymmetry and irregularity in
the gas kinematics.  The asymmetry index indicates the inhomogenity in the I-band
light distribution.  This parameter is a function of the galaxy type.  For the
E/S0s it is the smallest, it becomes higher for spirals and it is the highest for
irregular galaxies.  Apart from the intrinsic asymmetry of a galaxy, interaction
processes or mergers can also cause the asymmetry index to increase.  In case of
an interaction process that is strong enough to affect the distribution of the
stars, it is expected that the gas component is also disturbed.  On the other
hand, the phenomena that cause irregularities in the gas kinematics do not
necessarily affect the stellar component since it is gravitationally more
stable.   There is one case in our sample for which both the photometry and the
kinematics are irregular (galaxy C9).  It has $\Delta \phi = 66 \pm 20$, and it is
regular according to the other criteria ($\sigma_{PA}$ and $k_{3,5}/k_{1}$).

\section{Summary and Conclusions}

In this paper we present the kinematic and photometric analysis of 22 galaxies,
11 of which are members of MS0451 at $z=0.54$.  All cluster members are within 2
Mpc from the center and most of them are closer then 1 Mpc.  Among the cluster
objects, there are two in interaction with each other and 5 with weak or no
emission.  For 8 field and 6 cluster galaxies we were able to obtain the gas
velocity fields.  We used kinemetry to characterise these velocity fields using
Fourier coefficients.  The maximum radius for which we could measure the
kinematic parameters (position angle, ellipticity, etc.) of each galaxy is given
in Table \ref{tab2}. 

We use 3 parameters to quantify the regularity of the velocity fields:
$k_{3,5}/k_{1}$, $\sigma_{PA}$ and $\Delta \phi$ (Sect.4) and introduce a
criterion for each parameter to distinguish between regular and irregular gas
kinematics.  To do that, we also measured each parameter for nearby galaxies in the
SINGS survey and investigated the distribution of these galaxies in the parameter
space.  The regularity criteria we define are: $\Delta \phi \leq 25$, $\sigma PA
\leq 20$ and $k_{3,5} / k_{1} \leq 0.15$. 

We calculated the fraction of galaxies in our sample that have irregular gas
kinematics according to each criterion (without subdividing as field galaxies
and cluster members).  $27 \pm 13\%$ (according to $\sigma_{PA}$ criterion), $50
\pm 16\%$ ($\Delta \phi$ criterion), $40 \pm 15\%$ ($k_{3,5} / k_{1}$
criterion) and $58 \pm 14\%$ (any criterion).  Among these objects, there are four cluster galaxies for which we
were able to measure these parameters.  We find that field and cluster galaxies
span similar range of parameter values (Fig.\ref{cop1}).  In particular, we find
irregular cases in both environmental classes: $29 \pm 17\% / 25 \pm 22\% $
(according to $\sigma_{PA}$ criterion), $50 \pm 20\% / 50 \pm 25\%$ ($\Delta
\phi$ criterion), $33 \pm 19\% / 50 \pm 25\%$ ($k_{3,5} / k_{1}$ criterion) and $50
\pm 18\% / 75 \pm 22\%$ (any criterion). 
However, we note that the number of galaxies we have analyzed is not
statistically sufficient to give robust results.  In a future paper we will
analyse a much larger sample in the same way, increasing the statistics
considerably.

We investigated the spatial resolution effects on our kinematic analysis and
found that the fraction of irregular velocity fields in a sample determined from low
spatial resolution data is an underestimate of the real fraction.  We also found out
that for some combinations of a large $k_{3,5}/k_{1}$ and sampling, the measured
values of $\Delta\phi$ and $\sigma_{PA}$ will be boosted, so that the different types
of irregularity become degenerate in the rebinned data. Hence, identifying irregular
galaxies without distinguishing between the three types of irregularity gives more
reliable results in case of low spatial resolution.

6 galaxies (5 of them are cluster members) have strong absorption lines (mainly
CaII H,K and Balmer lines) which enabled us to derive their stellar
rotation curves. For three of them (C5, C7 and C9), stellar and gas
curves show different behaviours.  These are galaxies that are classified to
have irregular gas kinematics (C5 is interacting with its companion).  C7 and C9
also have relatively large photometric asymmetries. 

We have performed a surface photometry analysis of each galaxy (Table \ref{tab4}),
and calculated their rest-frame absolute B band magnitude (Table \ref{tabrun2}). We
also measured some other photometric properties (e.g. asymmetry and concentration
index) and investigated the existence of a correlation of these parameters with the
kinematic regularity. Although we do not find any significant correlation, these should
be re-investigated with a larger sample.

\begin{acknowledgements}
We are thankful to the authors of \citet{DCAHCBK06} for kindly providing us with the $H\alpha$
velocity fields of the galaxies in the SINGS local sample, Laurent Chemin for the $H\alpha$
velocity field of NGC4254 and Davor Krajnovi{\'c} for the Kinemetry software.  We thank the
anonymous referee for helpful comments and Isabel P\'erez Martin for fruitful discussion,  We
appreciate the efficient support of ESO and the Paranal staff. This work has been financially
supported by the  \emph{Volkswagen Foundation} (I/76\,520), the \emph{Deut\-sche
For\-schungs\-ge\-mein\-schaft, DFG\/}  (project number ZI\,663/6) within the Priority Program
1177, the Kapteyn Astronomical Institute of the University of Groningen and the German Space
Agency \emph{DLR\/} (project numbers 50\,OR\,0602\,\&\,50\,OR\,0404,\&\,50\,OR\,0301).  This
research has made use of the NASA/IPAC Extragalactic Database (NED) which is operated by the Jet
Propulsion Laboratory, California Institute of Technology, under contract with the National
Aeronautics and Space Administration.  HST/ACS images were retrieved from the ST--ECF archive.  
\end{acknowledgements}

\bibliographystyle{aa}
%\bibliography{/net/bellevue/data/users/kutdemir/documents/paper/velpap/}
\bibliography{abb,all}

\clearpage

\appendix
\section{Resolution Effects on the Kinematic Analysis}

\begin{figure*}
   \includegraphics[width=170mm]{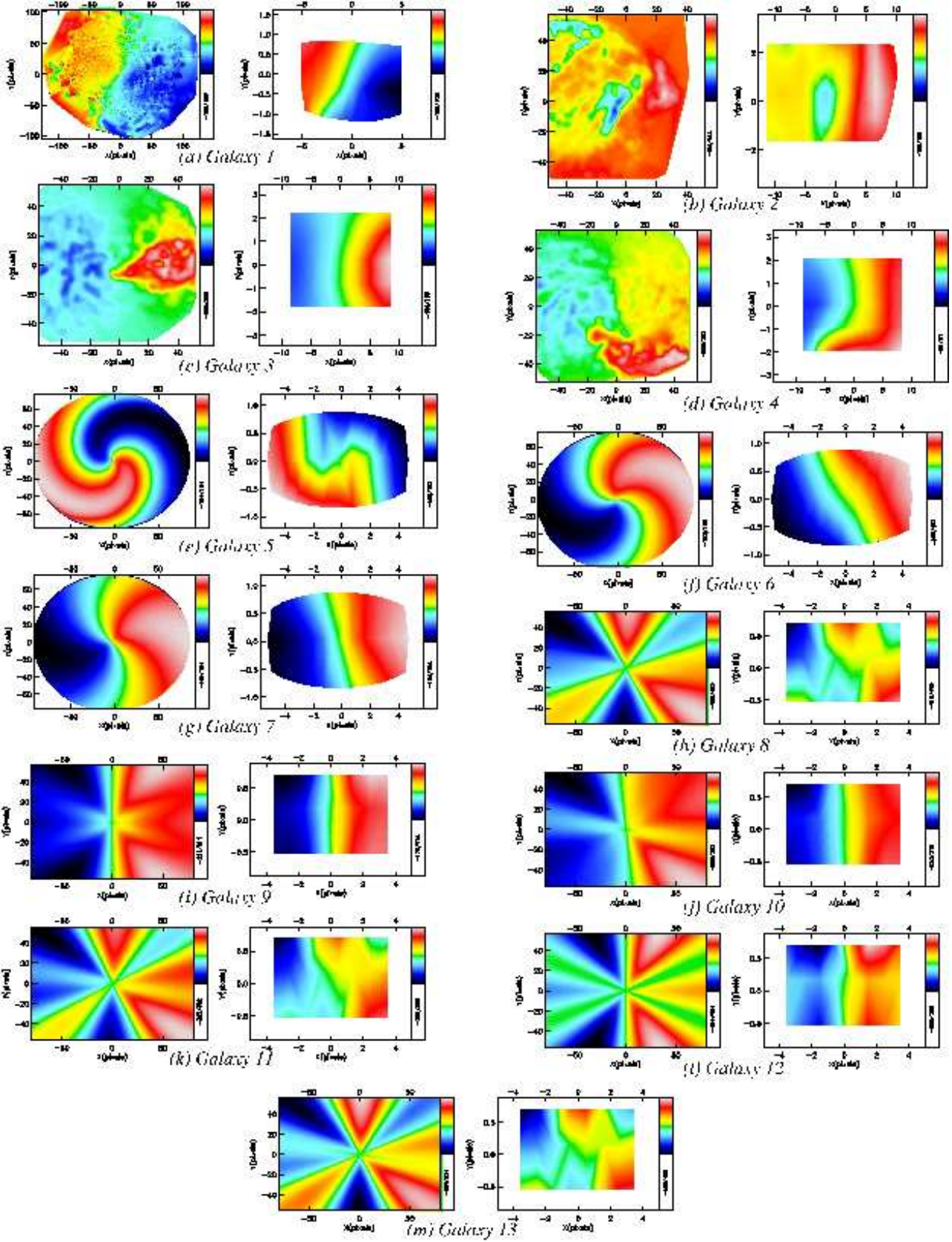}
   \centering
   \caption { 
   \textit{a:} H$\alpha$ velocity field of NGC4254 (galaxy 1) at distance d with
pixel size of $m \times m$ and at distance 25.6d with pixel size of $16m \times 64m$
respectively.
   \textit{b,c and d:} Simulated irregular velocity fields: galaxy 2, galaxy 3 and
   galaxy 4, each at distance d with pixel size of $m \times m$ and at distance 8d
   with pixel size of $4m \times 16m$ respectively.
   \textit{e,f and g:} Warped disk models with simple rotation along each orbit:
galaxy 5, galaxy 6 and galaxy 7 at distance d with pixel size of $m \times m$ and at
distance 25.6d with pixel size of $16m \times 64m$ respectively.
   \textit{h,i,j,k,l and m:} 3rd and/or 5th order Fourier terms with different
amplitudes added to a simple rotating disk model: galaxy 8, galaxy 9 and galaxy 10,
galaxy 11, galaxy 12 and galaxy 13 at distance d with pixel size of $m \times m$ and
at distance 25.6d with pixel size of $16m \times 64m$ respectively.   
              }
         \label{vfs}
   \end{figure*}

More than half of our galaxies are at redshifts above 0.5 and were observed with pixels
of $1\arcsec\times0\farcs25$ and a seeing of $0\farcs76$ (FWHM).  Here we test how much
our kinematical analysis depends on these observing conditions.  For that, we
simulate the combined effect of seeing and spatial resolution on the observed velocity
field, as a function of distance to the observed galaxy.  We start with the velocity
field of a galaxy at distance $d$ with a pixel size of $m\times4m$.  To represent the
observed velocity field of this galaxy when it is N times further away, we first
convolve it with a Gaussian psf:

\begin{equation}  
{
FWHM_{new}=\sqrt{{(N\times FWHM_{current})}^2-{FWHM_{current}}^2}
\label{psf}
}
\end{equation}

\noindent where $FWHM_{current}$ is the FWHM of the point spread function of the
velocity field at the distance $d$.  After the convolution process, we rebin the
velocity field to have a pixel size of $Nm \times 4Nm$.  To calculate the velocity
corresponding to the new pixel, we measure the intensity weighted velocity within its
borders.  We applied this method to several velocity fields that have different types
and levels of irregularity.  These include a regular H$\alpha$ velocity field
(Fig.\ref{vfs},a), simulated velocity fields of interacting galaxies
\citep{KKSZ07}(Fig.\ref{vfs},b-d), warped models that have simple rotation along each
orbit (Fig.\ref{vfs},e-g) and some models made by adding 3rd and 5th order Fourier
terms with different amplitudes to the velocity field of a simple rotating disk
(Fig.\ref{vfs},h-m).  The warped velocity fields were made to measure the resolution
effects on the $\sigma_{PA}$ and $\Delta\phi$ parameters while the last group of
models were made to evaluate the resolution effects on the $k_{3,5}/k_{1}$ term. 

\begin{figure*}
   \includegraphics[width=\textwidth]{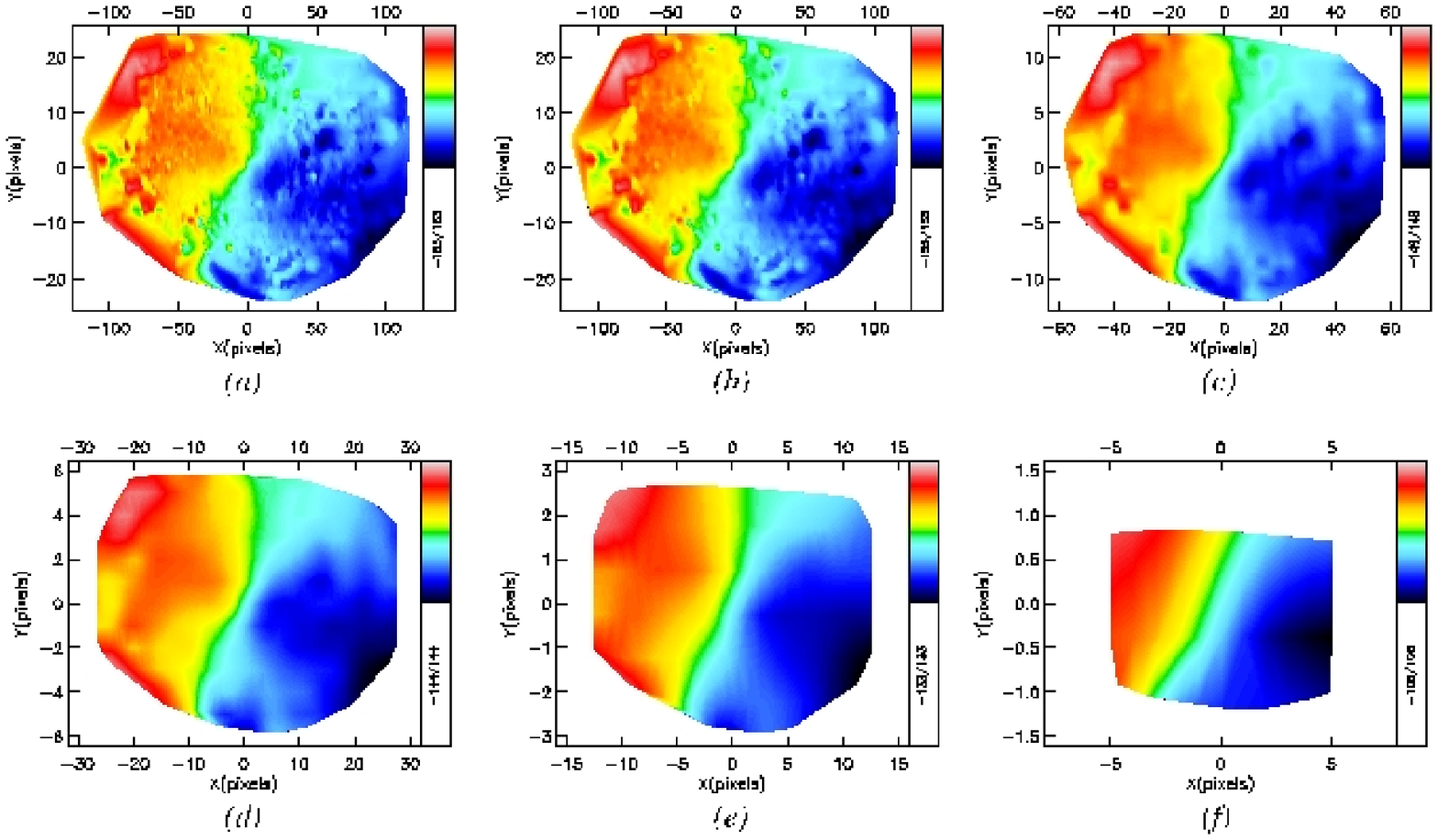}
   \centering
   \caption { Galaxy1 
   \textit{a:} At distance $d$ with $m \times 4m$ pixel size.
   \textit{b to f:} At distance $1.6d$, $3.2d$, $6.4d$, $12.8d$ and $25.6d$ with pixel
   sizes of $m \times 4m$, $2m \times 8m$, $4m \times 16m$, $8m \times 32m$, $16m \times
   64m$ respectively.
              }
         \label{galngc4254}
   \end{figure*}

For the intensity weighting of galaxy 1, we used its H$\alpha$ image.  For galaxies
2-4, we used their stellar mass density maps assuming that they have constant mass
to light ratios.  For the rest of the models, we used an intensity map that has the
same orientation and ellipticity as the corresponding velocity field and is in the
form of 

\begin{equation}  
{
I(r)=\frac{A}{(1+\frac{r}{B})^2},
\label{intmap}
}
\end{equation}

\noindent where $A$ and $B$ are constants that are chosen to make an intensity profile that is
similar to the H$\alpha$ intensity profile of galaxy 1.  $r$ is the distance to the
center of the map.  The velocity field of galaxy 1 calculated at several distances is
shown in Fig.\ref{galngc4254}.  For the other galaxies, we show only the velocity field
at the largest simulated distance, in Fig.\ref{vfs}.  The velocity maps in these
figures show the velocity corresponding to each pixel at its center.  The values in
between the pixel centers were visualized using linear interpolation. 
%-------------------------------------------------------------
%
%_____________________________________________________________

%-------------------------------------------------------------
%
%_____________________________________________________________

The velocity fields that are used for this exercise do not all have the same initial
distance and resolution.  However, they all have similar grid sizes with our observations
at their largest simulated distance.  This better indicates how comparable the simulations
are with our observations than the actual resolution.  Because our sample is not homogenius
in size and redshift, but all galaxies were observed with 3 pixels along their minor axis. 
For the redshift of our cluster galaxies ($z=0.54$), the physical size of our pixels is
$\approx 1.6\times6.4kpc$.  For a magnification of $25.6$, the range of our simulations,
the pixel size in the local universe would be $63 \times 250 pc.$ This would correspond to
an angular resolution of $\approx 0\farcs85 \times 3\farcs4$ at the distance of Virgo,
somewhat larger than can be obtained with current ground-based telescopes in the optical. 
For the velocity fields of galaxies 2-4, the intrinsic resolution was so low, that we could
only degrade them by a factor 8.  For the rest the maximum factor was 25.

\begin{figure*}
   \includegraphics[width=\textwidth]{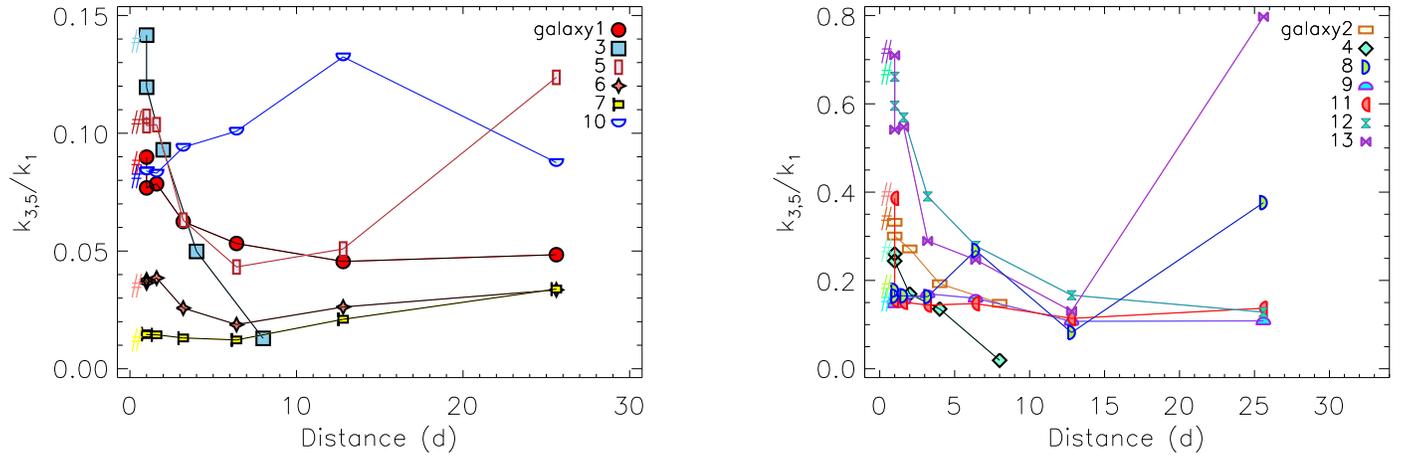}
   \centering
   \caption {$k_{3,5}/k_{1}$ as a function of the distance. The measurements for the original velocity fields are also given in these
   plots. They are indicated with \# symbol to be able to distinguish them from the
   velocity field at the same distance with the same pixel size along the x axis and 4
   times as large pixel size along the y axis.
   \textit{Left:} galaxies that have $k_{3,5}/k_{1}$ less than 0.15.
   \textit{Right:} galaxies that have $k_{3,5}/k_{1}$ greater than 0.15.
              }
         \label{comp2}
   \end{figure*}

For each velocity field simulated at several resolutions, we measured the
(ir)regularity parameters ($\sigma_{PA}$, $\Delta\phi$ and $k_{3,5}/k_{1}$).  In
Fig.\ref{comp2} we plot $k_{3,5}/k_{1}$ as a function of distance for galaxies that
have $k_{3,5}/k_{1}$ greater and less than 0.15, to avoid overcrowding of the plots. 
For galaxies that have a large intrinsic value of $k_{3,5}/k_{1}$, the  value that is
recovered at low resolution is always smaller except in a few cases.  For some
galaxies, a decrease in the resolution causes an increase in this parameter as a result
of a different gradient pattern introduced by intensity weighted average velocities
within larger pixels.  This is the case for galaxies 8 and 13.  For a few galaxies,
$k_{3,5}/k_{1}$ decreases substantially with decreasing resolution, e.g. galaxy 4. 
Regular velocity fields show a regular, slow decrease, which is accompanied by a small
deviation in some cases.  Overall, small values of $k_{3,5}/k_{1}$ measured for
galaxies at large distances do not provide information about their intrinsic values,
while large measured values indicate large intrinsic values.  There are three cases
where the measured $k_{3,5}/k_{1}$ at the largest distance of simulations is above the
irregularity treshold (galaxies 2,8 and 13) and they all have intrinsic values that are
higher than the treshold.  However, only $43\pm19\%$ of the velocity fields that have
intrinsic values of $k_{3,5}/k_{1}$ above the irregularity threshold would be
identified as such at the lowest resolution.  This means that using $k_{3,5}/k_{1}$ as
an indicator, a lower limit is found to the number of irregular galaxies.

\begin{figure*}
   \includegraphics[width=\textwidth]{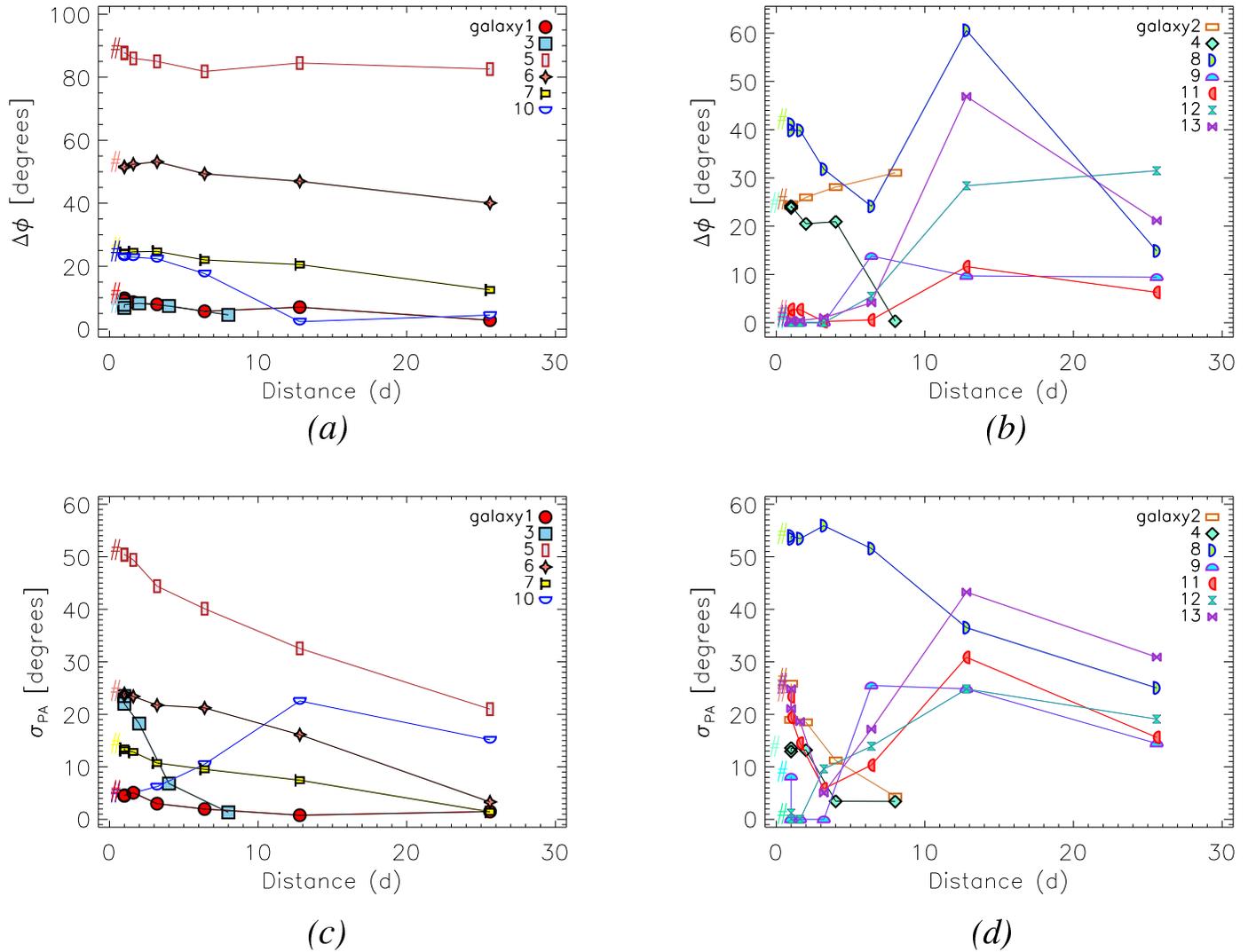}
   \centering
   \caption {$\Delta \phi$ and $\sigma_{PA}$ as a function of the distance.The measurements for the original
   velocity fields are also given in these plots. They are indicated with \# symbol to be able to distinguish
   them from the velocity field at the same distance with the same pixel size along the x axis and 4 times as
   large pixel size along the y axis.
   \textit{a:} $\Delta \phi$ as a function of the distance for galaxies that have $k_{3,5}/k_{1}$ less than 0.15.
   \textit{b:} $\Delta \phi$ as a function of the distance for galaxies that have $k_{3,5}/k_{1}$ greater than 0.15.
   \textit{c:} $\sigma_{PA}$ as a function of the distance for galaxies that have $k_{3,5}/k_{1}$ less than 0.15.
   \textit{d:} $\sigma_{PA}$ as a function of the distance for galaxies that have $k_{3,5}/k_{1}$ greater than 0.15.
              }
         \label{comp}
   \end{figure*}

\begin{figure*}
   \includegraphics[width=\textwidth]{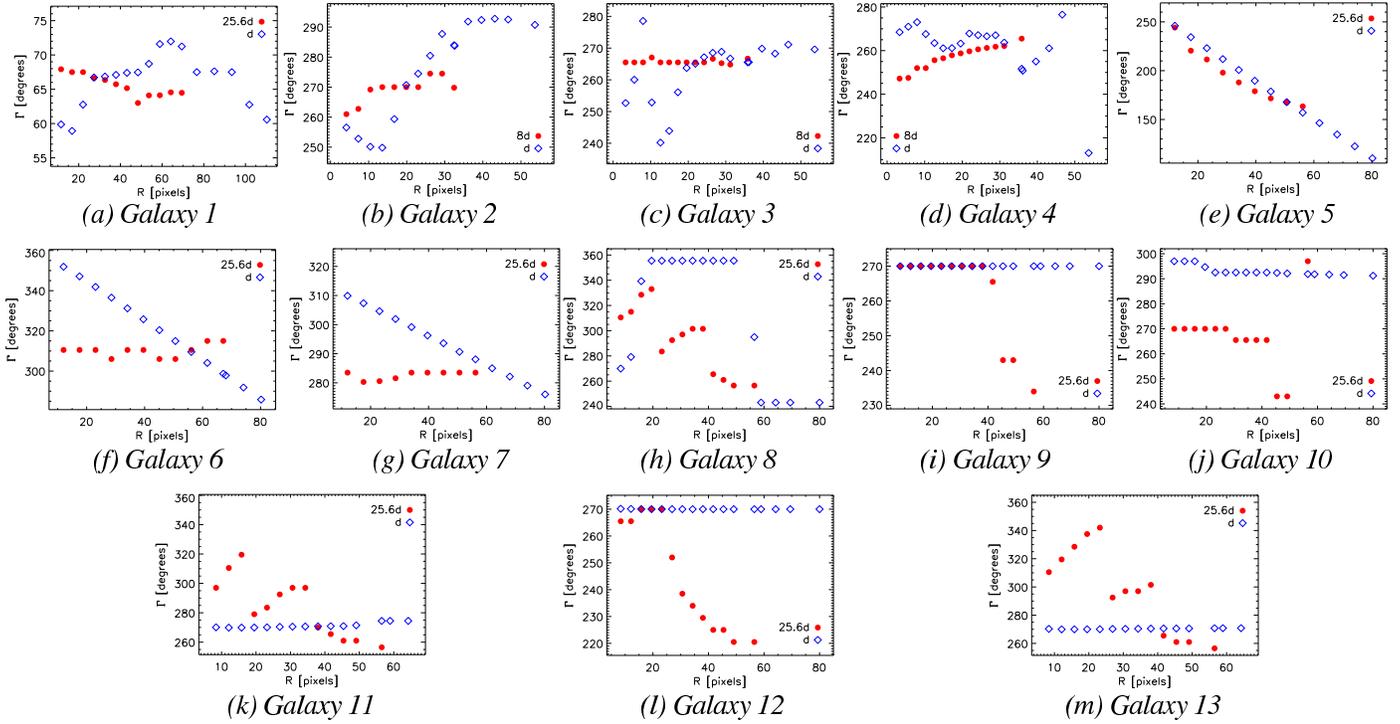}
   \centering
   \caption { Kinematic position angle as a function of radius for the original velocity
   field (blue) together with the velocity field at the largest distance (red). The
   radius unit is the pixels of the original velocity field.
               }
         \label{prof}
   \end{figure*}

In Fig.\ref{comp}, we plot $\Delta\phi$ and $\sigma_{PA}$ as a function of distance for
the same groups of galaxies as in Fig.\ref{comp2} separately.  We also give the
position angle as a function of radius for both the original velocity field and the
velocity field at the largest distance in Fig.\ref{prof}.  While most galaxies show a
continuous decrease of these two parameters with increasing distance, some of the
galaxies that have a large $k_{3,5}/k_{1}$ parameter show a strongly varying behavior. 
The reason for this is the same fact causing the deviation in $k_{3,5}/k_{1}$ parameter
itself as explained above.  The increase in the $\Delta\phi$ and $\sigma_{PA}$ terms at
lower resolution does not only depend on the $k_{3,5}/k_{1}$ parameter, but also on the
gradient structure of the velocity field and how the pixels are positioned on it.  At
low resolution, $\Delta\phi$ values of galaxy 12 and 13 are measured to be larger than
the irregularity treshold although their intrinsic values are around zero.
$\sigma_{PA}$ of galaxies 9 and 10 also go above the irregularity treshold at low
resolution although their intrinsic values are below 10.  Among these galaxies, all but
galaxy 10 have large $k_{3,5}/k_{1}$ parameters (for galaxy 10, $k_{3,5}/k_{1}=$0.08).
This means that in the case of low spatial resolution, the combined effect of a large
$k_{3,5}/k_{1}$ and the pixel positions will cause $\Delta\phi$ and $\sigma_{PA}$ to be
larger than their intrinsic values.  For galaxies that have a large intrinsic value of
$\Delta\phi$ and $\sigma_{PA}$, the fraction of information that is lost with
decreasing resolution is generally higher.  In case of galaxies that do not have large
$k_{3,5}/k_{1}$ values, $\Delta\phi$ does not change much with resolution.
$75\%\pm22\%$ of the velocity fields that are irregular according to the $\Delta\phi$ 
criterion could be identified at the lowest resolution case.  There are
7 velocity fields that are irregular according to the $\sigma_{PA}$ criterion and
$43\%\pm19\%$ of them could be identified at the lowest resolution case.

The results of this test show that the fraction of irregular galaxies in a sample
determined from low spatial resolution data will be underestimated.  A combination of a
high $k_{3,5}/k_{1}$ and pixel positioning can cause the measured values of
$\Delta\phi$ and $\sigma_{PA}$ to be boosted, so that the different types of
irregularity become degenerate in the rebinned data.  Hence, identifying irregular
galaxies without distinguishing between the three types of irregularity gives more
reliable results in case of low spatial resolution.  At the lowest resolution,
$60\%\pm15\%$ of all irregular galaxies is classified as such according to at least one
of our irregularity criteria, while all three regular cases were correctly
identified.  $90\%\pm10\%$ of the cases that are identified to be irregular at the
lowest resolution were intrinsically irregular.

\begin{figure*}
   \includegraphics[width=\textwidth]{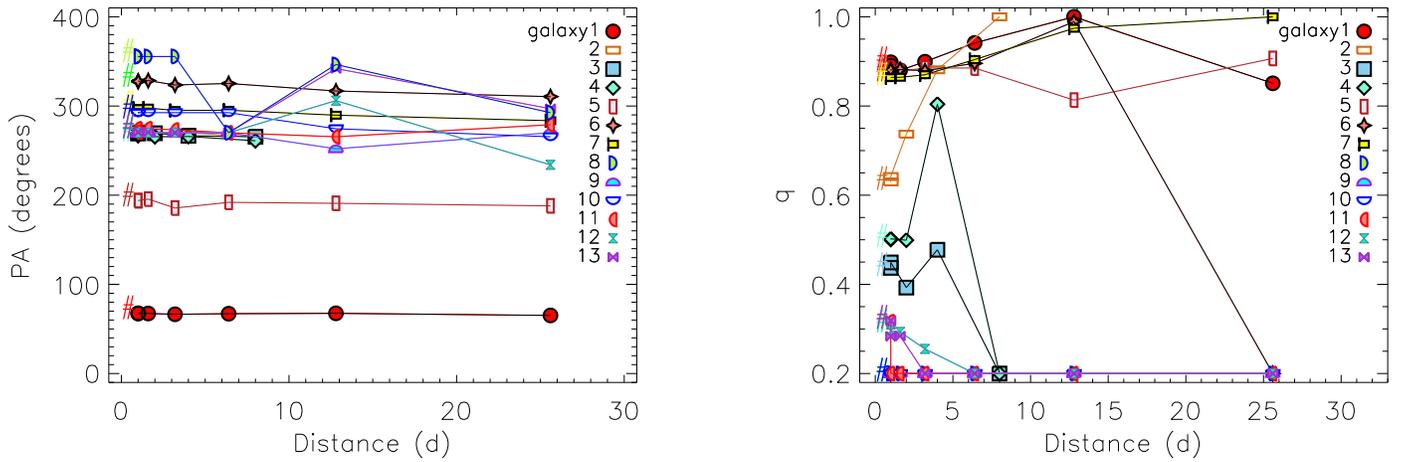}
   \centering
   \caption { 
   \textit{Left:} The global kinematic position angle as a function of the distance.
   \textit{Right:} The global kinematic flattening as a function of the distance.
              }
         \label{pa_q}
   \end{figure*}

We also checked how the global kinematic position angle and flattening change as a
function of the distance.  This is given in Fig.\ref{pa_q}.  The global position angle
is very robust while this is much less the case for the ellipticity.

%-------------------------------------------------------------
%

%-------------------------------------------------------------

%-------------------------------------------------------------
%

%-------------------------------------------------------------
%
%_____________________________________________________________

%-------------------------------------------------------------
%
%_____________________________________________________________
\clearpage

\section{Individual Galaxies}

Here we give some information on each object in our sample.  In the figures, the HST
image and the velocity field of each galaxy have the same orientation (the slit position
is parallel to the x-axis).  The velocities and positions are given with respect to the
continuum center of the galaxies. 

\subsection{Cluster Galaxies}

\textbf{Galaxy C1:}

\begin{figure}
   \includegraphics[height=\columnwidth, angle=270]{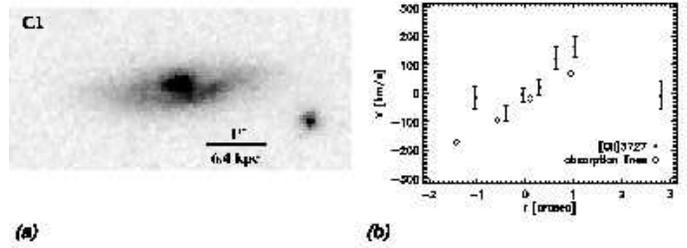}
 \caption{
 \textit{a)} HST-ACS image of the galaxy in the I band.
 \textit{b)} Rotation curves of the [OII]3727 emission line and several
absorption lines extracted along the central slit.
	}
         \label{gal15fig}
   \end{figure}

Has very weak OII emission that is not strong enough to provide a velocity field. 
Stellar and gas rotation curves obtained along the central slit are
given in Fig.\ref{gal15fig}.b.

\textbf{Galaxy C2:}

\begin{figure}
   \includegraphics[height=\columnwidth, angle=270]{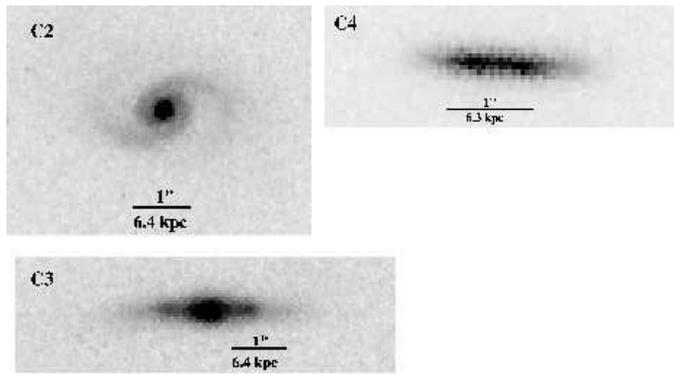}
 \caption{I band HST-ACS images of galaxies that do not have emission. The corresponding galaxy name
 (C2, C3 and C4) is indicated on each image.
 }
         \label{noemission}
   \end{figure}

This is a spiral galaxy that has no emission (Fig.\ref{noemission}).

\textbf{Galaxy C3:}

Has very weak OII emission that is not strong enough to provide a velocity field
(Fig.\ref{noemission}).

\textbf{Galaxy C4:}

This is a galaxy that has no emission.  It was observed with only one slit position
(Fig.\ref{noemission}). 

\textbf{Galaxies C5 and C6:}

\begin{figure*}
   \includegraphics[height=\textwidth, angle=270]{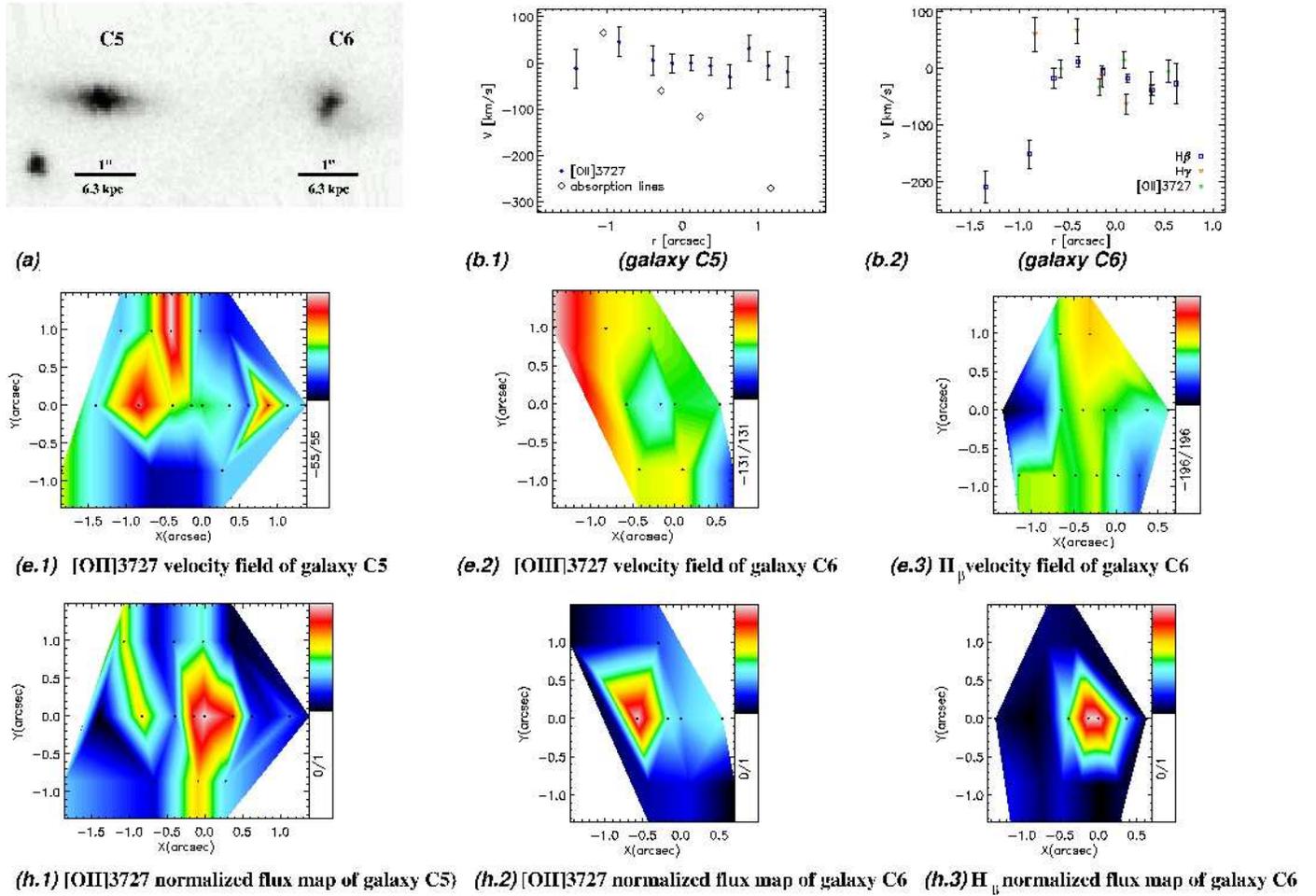}
 \caption{
 \textit{a)} HST-ACS image of the galaxies in the I band.
 \textit{b.1)}  Rotation curves of the [OII] emission line and several
absorption lines extracted along the central slit for galaxy C5.
 \textit{b.2)}  Rotation curves of different emission lines extracted along the
central slit for galaxy C6.
 \textit{e.1)} [OII]3727 velocity field of galaxy $C5$.
\textit{e.2)}  [OII]3727 velocity field of galaxy $C6$.
\textit{e.3)} $H\beta$ velocity field of galaxy $C6$.
 \textit{h.1)} Normalized [OII]3727 flux map of galaxy $C5$.
 \textit{h.2)} Normalized [OII]3727 flux map of galaxy $C6$.
\textit{h.3)} Normalized $H\beta$ flux map of galaxy $C6$.
	}
         \label{gal9fig}
   \end{figure*}

They were observed with the same slit (Fig.\ref{gal9fig}.a).  Velocity fields of
both are very irregular, which means that they are probably interacting
(Fig.\ref{gal9fig}.e).  The stellar and gas rotation curves of galaxy
C5 plotted together also show that the gas is disturbed
(Fig.\ref{gal9fig}.b).

Galaxy C5 is an example of the fact that a galaxy with very complex kinematics
can have a very regular surface brightness profile.  Using its asymmetry and
concentration indexes, it is classified as an E/S0 galaxy (Sect.3.2, Fig.\ref{asymconc}).

\textbf{Galaxy C7:}

\begin{figure*}
   \includegraphics[height=\textwidth, angle=270]{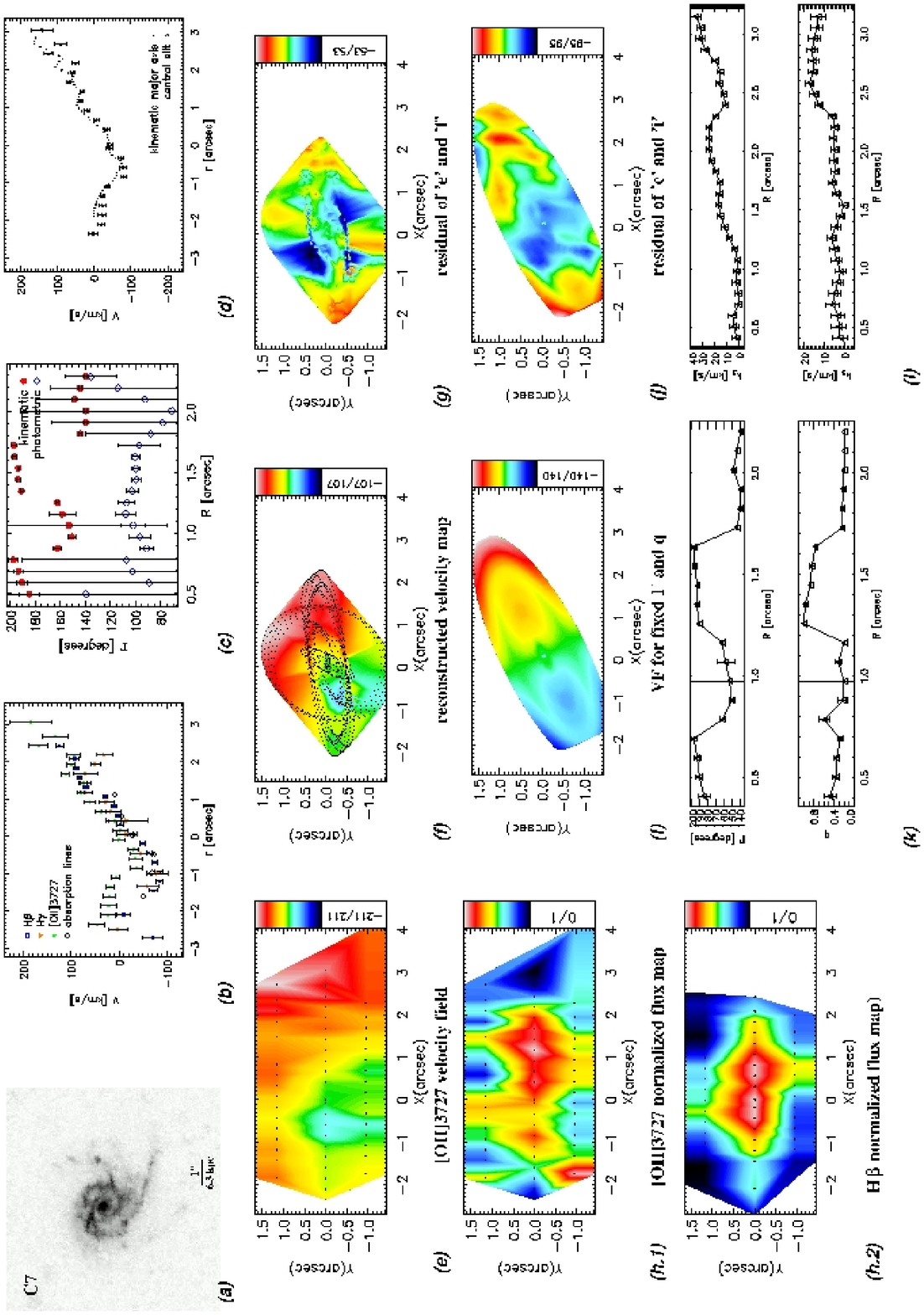}
 \caption{
 \textit{a)} HST-ACS image of the galaxy in the I band.
 \textit{b)} Rotation curves of different emission lines
and several absorption lines extracted along the central slit.
\textit{c)} Position angles of kinematic and photometric axes as a
	function of radius.
\textit{d)} Rotation curves extracted along the central
	slit and the kinematic major axis.
\textit{e)} [OII]3727 velocity field.
 \textit{f)} Velocity map reconstructed using 6 harmonic terms.
\textit{g)} Residual of the velocity map and the reconstructed map.
\textit{h.1)} Normalized [OII]3727 flux map.
\textit{h.2)} Normalized $H\beta$ flux map.	 	
\textit{i)} Simple rotation map constructed for position angle and
	ellipticity fixed to their global values.
\textit{j)} Residual of the velocity map and the simple rotation map.
\textit{k)} Position angle and flattening as a function of radius.
\textit{l)} $k_{3}/k_{1}$ and $k_{5}/k_{1}$ (from the analysis where position angle and
	ellipticity are fixed to their global values) as a function of
	radius.
	}
         \label{gal6fig}
   \end{figure*}

The mean misalignment between its kinematic and photometric axes is large ($\Delta
\phi=68^{\circ}\pm26^{\circ}$) and the standard deviation of its kinematic position angle
is big ($23^{\circ}$).  So this galaxy has irregular gas kinematics (Fig.\ref{gal6fig}). 
It is classified as an irregular galaxy according to its photometric asymmetry and
concentration measurements (Sect.3.2, Fig.\ref{asymconc}).  The stellar and gas velocity
fields of this galaxy have some discrepancies. 

\textbf{Galaxy C8:}

\begin{figure*}
   \includegraphics[height=\textwidth, angle=270]{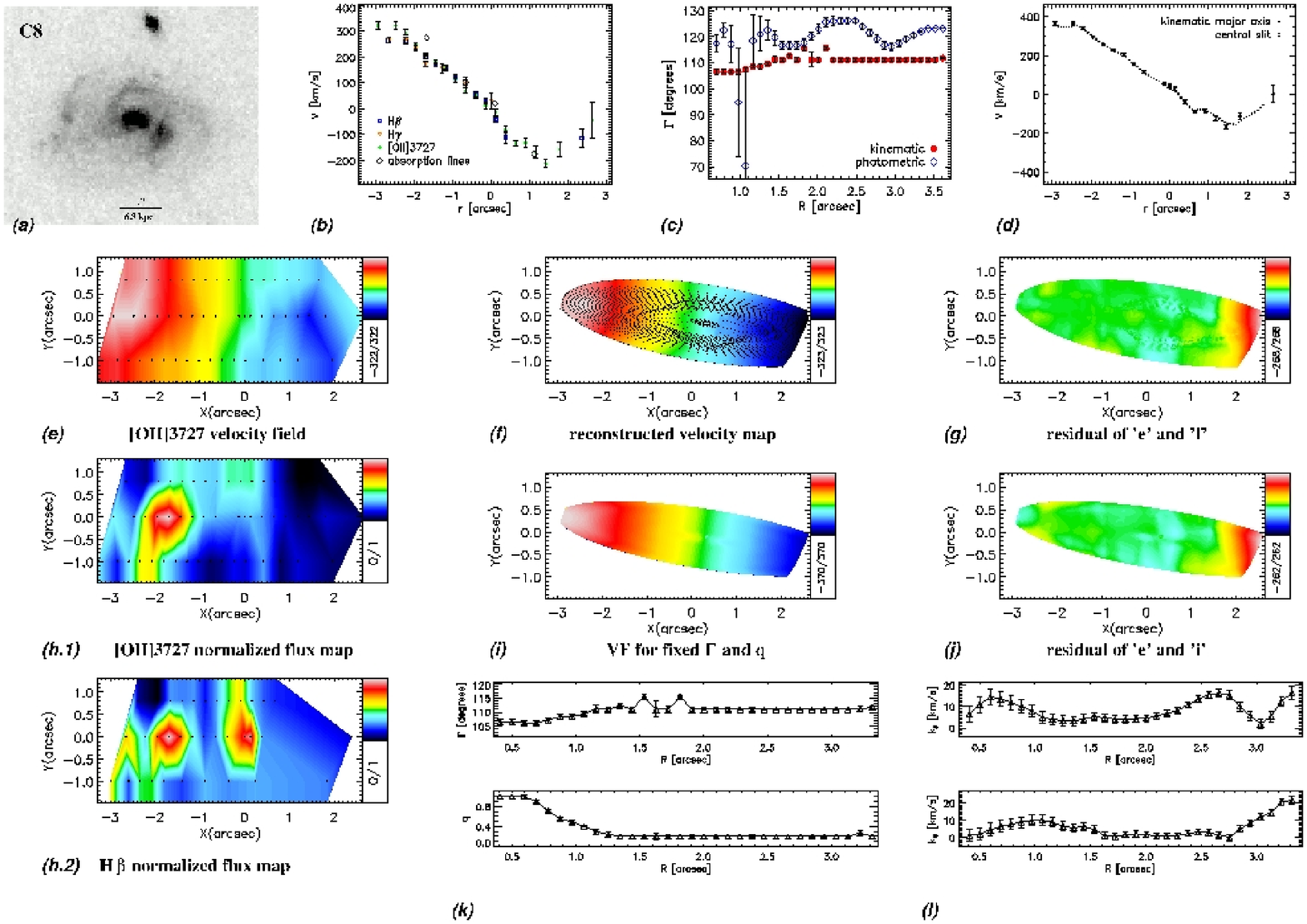}
 \caption{
 \textit{a)} HST-ACS image of the galaxy in the I band.
 \textit{b)} Rotation curves of different emission
lines and several absorption lines extracted along the central
slit.
\textit{c)} Position angles of kinematic and photometric axes as a
	function of radius.
\textit{d)} Rotation curves extracted along the central
	slit and the kinematic major axis.
\textit{e)} [OII]3727 velocity field.
 \textit{f)} Velocity map reconstructed using 6 harmonic terms.
\textit{g)} Residual of the velocity map and the reconstructed map
\textit{h.1)} Normalized [OII]3727 flux map.
\textit{h.2)} Normalized $H\beta$ flux map.	
\textit{i)} Simple rotation map constructed for position angle and
	ellipticity fixed to their global values.
\textit{j)} Residual of the velocity map and the simple rotation map.
\textit{k)} Position angle and flattening as a function of radius.
\textit{l)} $k_{3}/k_{1}$ and $k_{5}/k_{1}$ (from the analysis where position angle and
	ellipticity are fixed to their global values) as a function of
	radius. 
	}
         \label{gal5fig}
   \end{figure*}

Even though it is very close to the cluster center (0.7 Mpc) in projection, this
galaxy has regular gas kinematics according to all criteria we defined
($\sigma_{PA}$, $\Delta \phi$ and $k_{3,5} / k_{1}$) (Fig.\ref{gal5fig}).

\textbf{Galaxy C9:}

\begin{figure*}
   \includegraphics[height=\textwidth, angle=270]{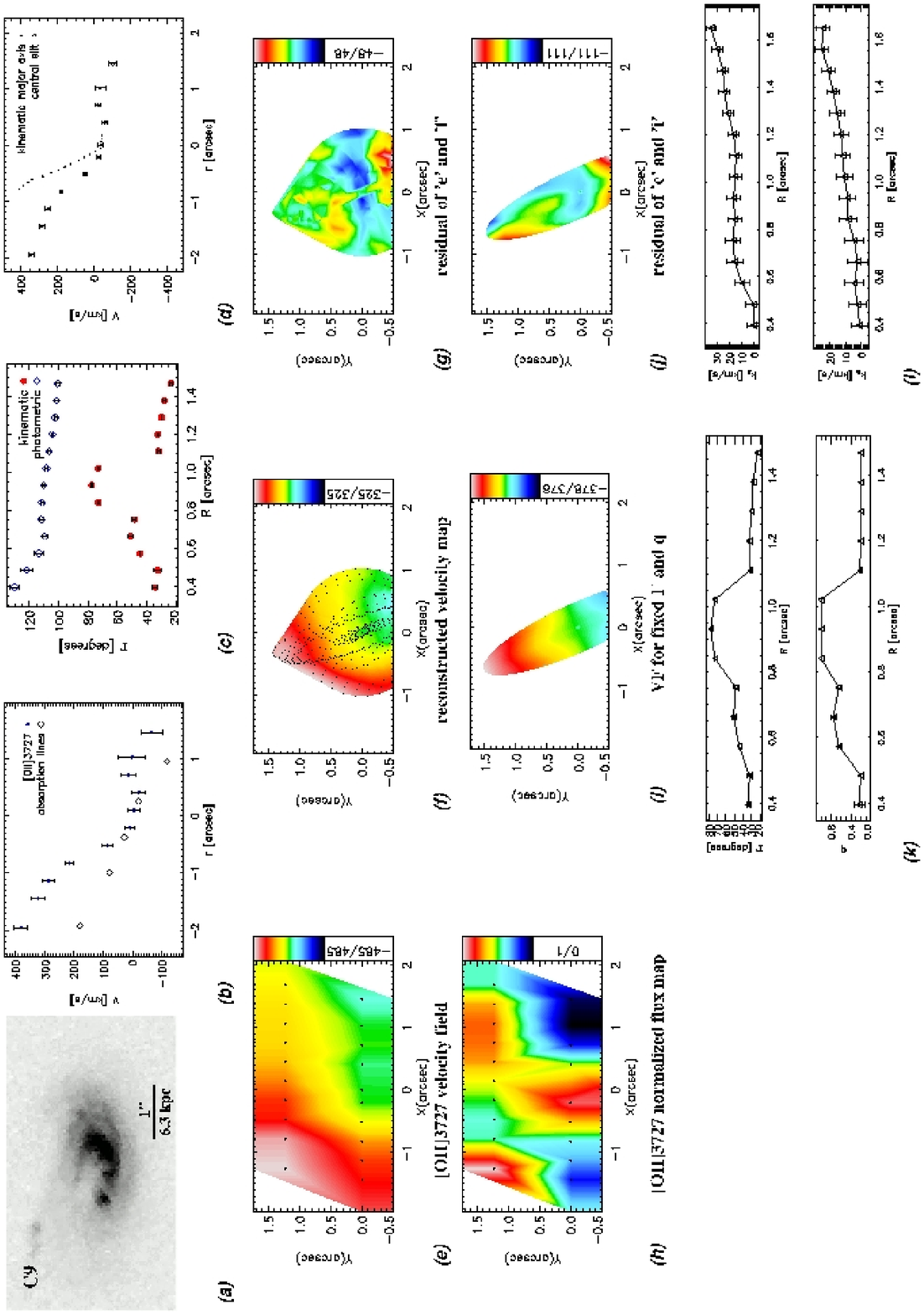}
\caption{
\textit{a)} HST-ACS image of the galaxy in the I band.
\textit{b)} Rotation curves of OII emission line and several
absorption lines extracted along the central slit.
\textit{c)} Position angles of kinematic and photometric axes as a
	function of radius.
\textit{d)} Rotation curves extracted along the central
	slit and the kinematic major axis.
\textit{e)} [OII]3727 velocity field.
\textit{f)} Velocity map reconstructed using 6 harmonic terms.
\textit{g)} Residual of the velocity map and the reconstructed map.
\textit{h)} Normalized [OII]3727 flux map.	
\textit{i)} Simple rotation map constructed for position angle and
	ellipticity fixed to their global values.
\textit{j)} Residual of the velocity map and the simple rotation map.
\textit{k)} Position angle and flattening as a function of radius.
\textit{l)} $k_{3}/k_{1}$ and $k_{5}/k_{1}$ (from the analysis where position angle and
	ellipticity are fixed to their global values) as a function of
	radius.
	}
         \label{gal22fig}
   \end{figure*}

Since the signal in one of the slits is not enough to extract a rotation curve,
data from just two slits were used to construct its OII velocity field
(Fig.\ref{gal22fig}.e).  The average misalignment between its kinematic and
photometric axes is very large $(\Delta \phi=66^{\circ}\pm20^{\circ})$.  There is also
a big discrepancy between its stellar and gas rotation curves.

\textbf{Galaxy C10:}

\begin{figure*}
   \includegraphics[height=\textwidth, angle=270]{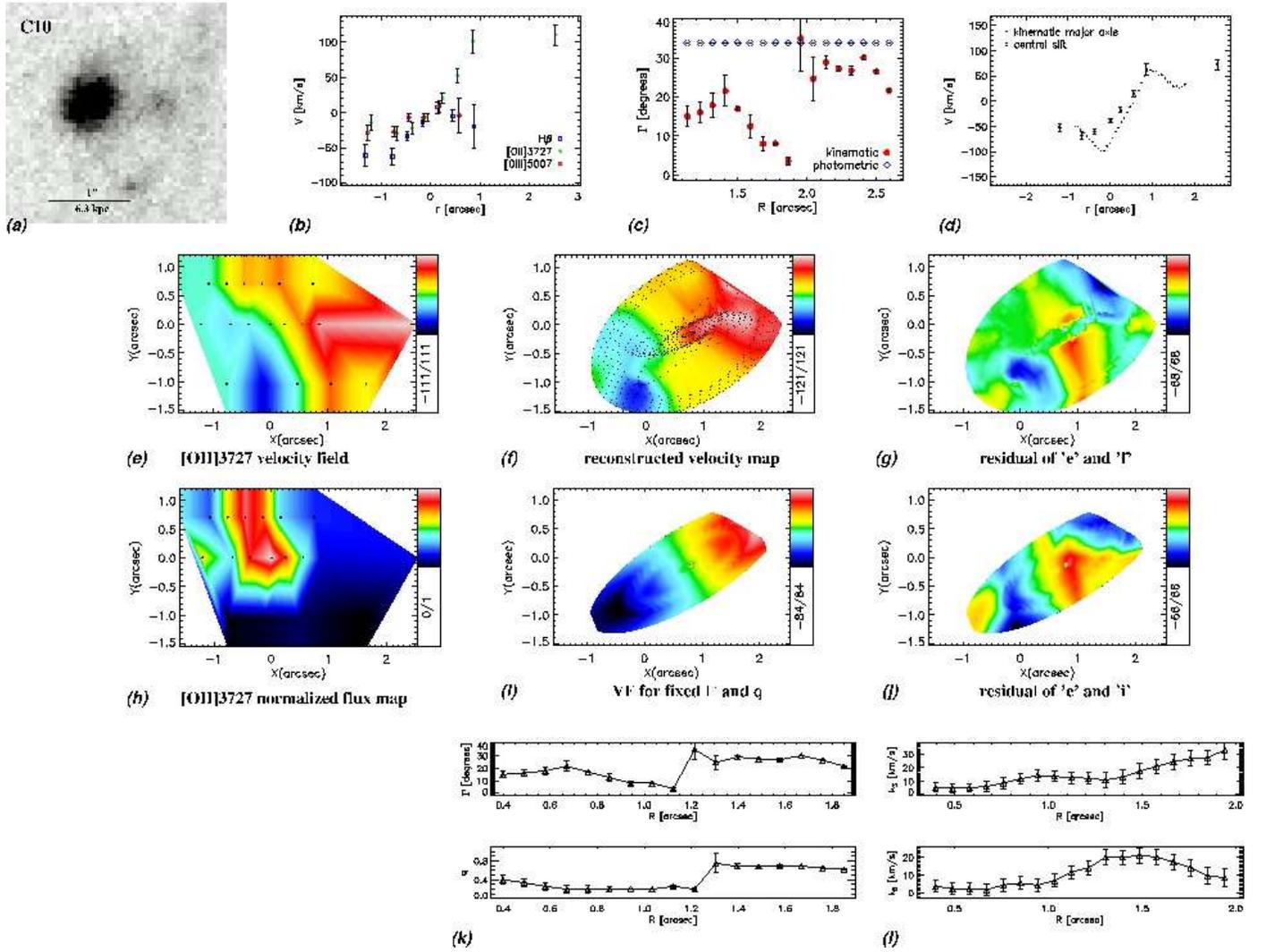}
 \caption{
\textit{a)} HST-ACS image of the galaxy in the I band.
\textit{b)} Rotation curves of different emission lines
extracted along the central slit.
\textit{c)} Position angles of kinematic and photometric axes as a
	function of radius.
\textit{d)} Rotation curves extracted along the central
	slit and the kinematic major axis.
\textit{e)} [OII]3727 velocity field.
\textit{f)} Velocity map reconstructed using 6 harmonic terms.
\textit{g)} Residual of the velocity map and the reconstructed map.
\textit{h)} Normalized [OII]3727 flux map.	
\textit{i)} Simple rotation map constructed for position angle and
	ellipticity fixed to their global values.
\textit{j)} Residual of the velocity map and the simple rotation map.
\textit{k)} Position angle and flattening as a function of radius.
\textit{l)} $k_{3}/k_{1}$ and $k_{5}/k_{1}$ (from the analysis where position angle and
	ellipticity are fixed to their global values) as a function of
	radius.
	}
         \label{gal16fig}
   \end{figure*}

This galaxy is an example of a case where the rotation curves
extracted along the central slit looks more regular than the rotation
curve extracted along the kinematic major axis (Fig.\ref{gal16fig}).  It has
irregular gas kinematics according to the $k_{3,5}/k_{1}$ criterion.

\textbf{Galaxy C11:}
This is a galaxy that was classified as an elliptical in the literature (see Table
\ref{centdist}).  It was observed together with galaxy F8 within the same slit
(Fig.\ref{gal1fig}).  Their spectra are so close together that they could not be
distinguished.
%\clearpage
%------)))))))------------------------------------------------
%   \begin{landscape}
 %    \end{landscape}
%
%_____________________________________________________________
%------)))))))------------------------------------------------
%      \begin{landscape}  
%   \end{landscape}
%
%_____________________________________________________________
%\clearpage
%------)))))))------------------------------------------------
%   \begin{landscape}
%   \end{landscape}
%
%_____________________________________________________________
%\clearpage
%------)))))))------------------------------------------------
%   \begin{landscape}
%   \end{landscape}
%
%_____________________________________________________________
%\clearpage
%------)))))))------------------------------------------------
   % \begin{landscape}
     % \end{landscape}   
%
%_____________________________________________________________

%\clearpage
%------)))))))------------------------------------------------
     % \begin{landscape} 
      % \end{landscape}
%
%_____________________________________________________________
%\clearpage
%------)))))))------------------------------------------------
    % \begin{landscape} 
      % \end{landscape} 
%
%_____________________________________________________________
\clearpage

\subsection{Field Galaxies}

\textbf{Galaxy F1:}

\begin{figure}
   \includegraphics[height=\columnwidth, angle=270]{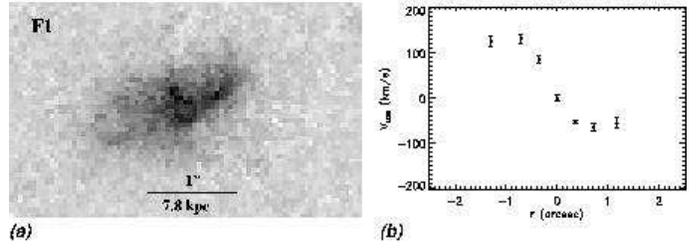}
 \caption{
 \textit{a)} HST-ACS image of the galaxy in the I band.
 \textit{b)} [OII]3727 rotation curve (this galaxy was observed 
	with only one slit position).
}
         \label{gal10fig}
   \end{figure}

It is a galaxy at $z=0.9009$ that was observed using only one slit position
(Fig.\ref{gal10fig}). 

\textbf{Galaxy F2:}

\begin{figure*}
   \includegraphics[height=\textwidth, angle=270]{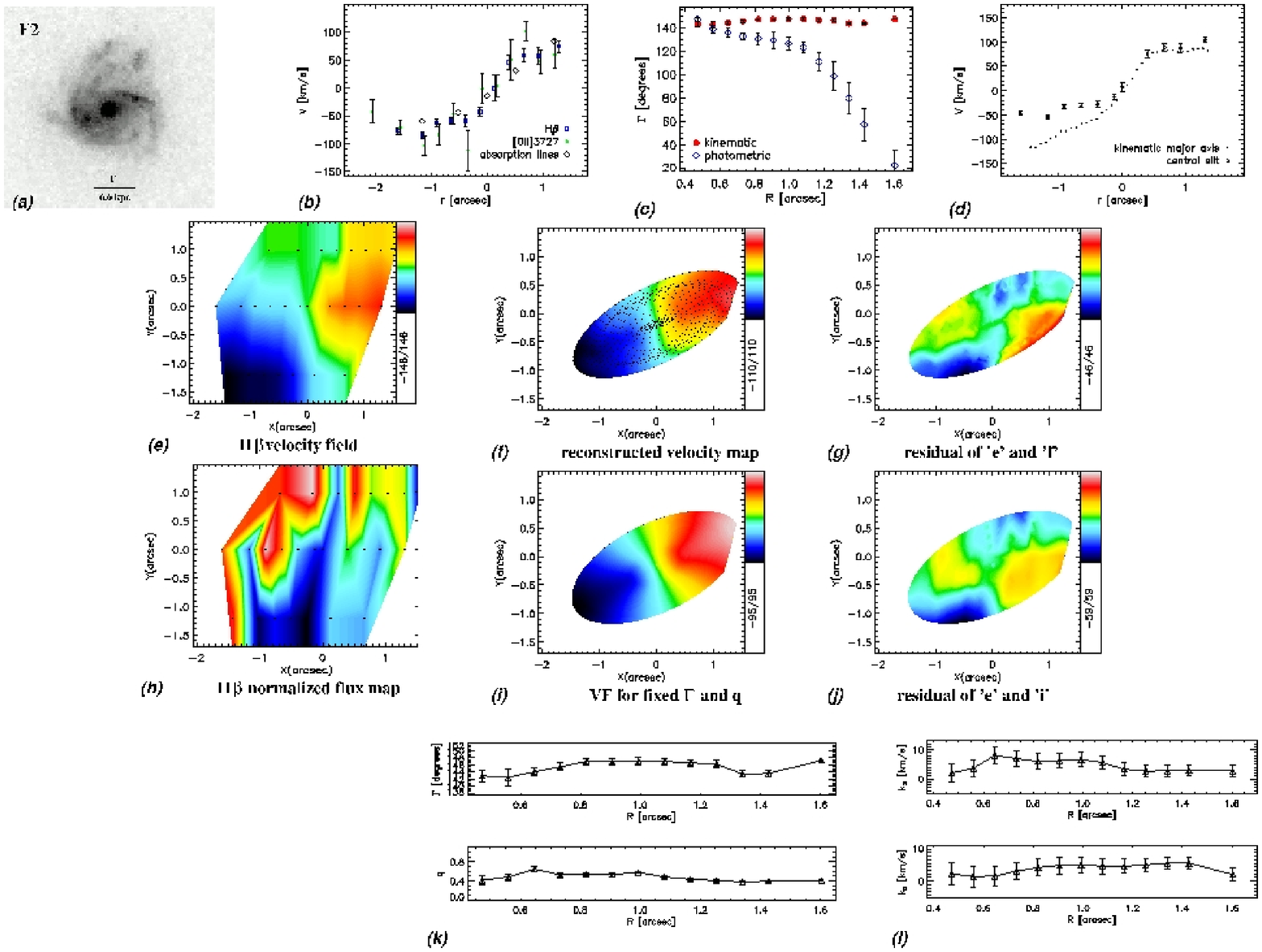}
 \caption{
 \textit{a)} HST-ACS image of the galaxy in the I band.
 \textit{b)} Rotation curves of different emission lines
and several absorption lines extracted along the central slit.
\textit{c)} Position angles of kinematic and photometric axes as a
	function of radius.
\textit{d)} Rotation curves extracted along the central
	slit and the kinematic major axis.
\textit{e)} $H\beta$ velocity field.
 \textit{f)} Velocity map reconstructed using 6 harmonic terms.
\textit{g)} Residual of the velocity map and the reconstructed map.
\textit{h)} Normalized $H\beta$ flux map.
\textit{i)} Simple rotation map constructed for position angle and
	ellipticity fixed to their global values.
\textit{j)} Residual of the velocity map and the simple rotation map.
\textit{k)} Position angle and flattening as a function of radius.
\textit{l)} $k_{3}/k_{1}$ and $k_{5}/k_{1}$ (from the analysis where position angle and
	ellipticity are fixed to their global values) as a function of
	radius.
	}
 
         \label{gal8fig}
   \end{figure*}

Background galaxy at $z=0.5795$ (Fig.\ref{gal8fig}).  It has regular gas
kinematics according to the $\sigma_{PA}$ and $k_{3,5} / k_{1}$ criteria.  Since
it is nearly face-on, its $\Delta \phi$ was excluded from the analysis.

\textbf{Galaxy F3:}

\begin{figure*}
   \includegraphics[height=\textwidth, angle=270]{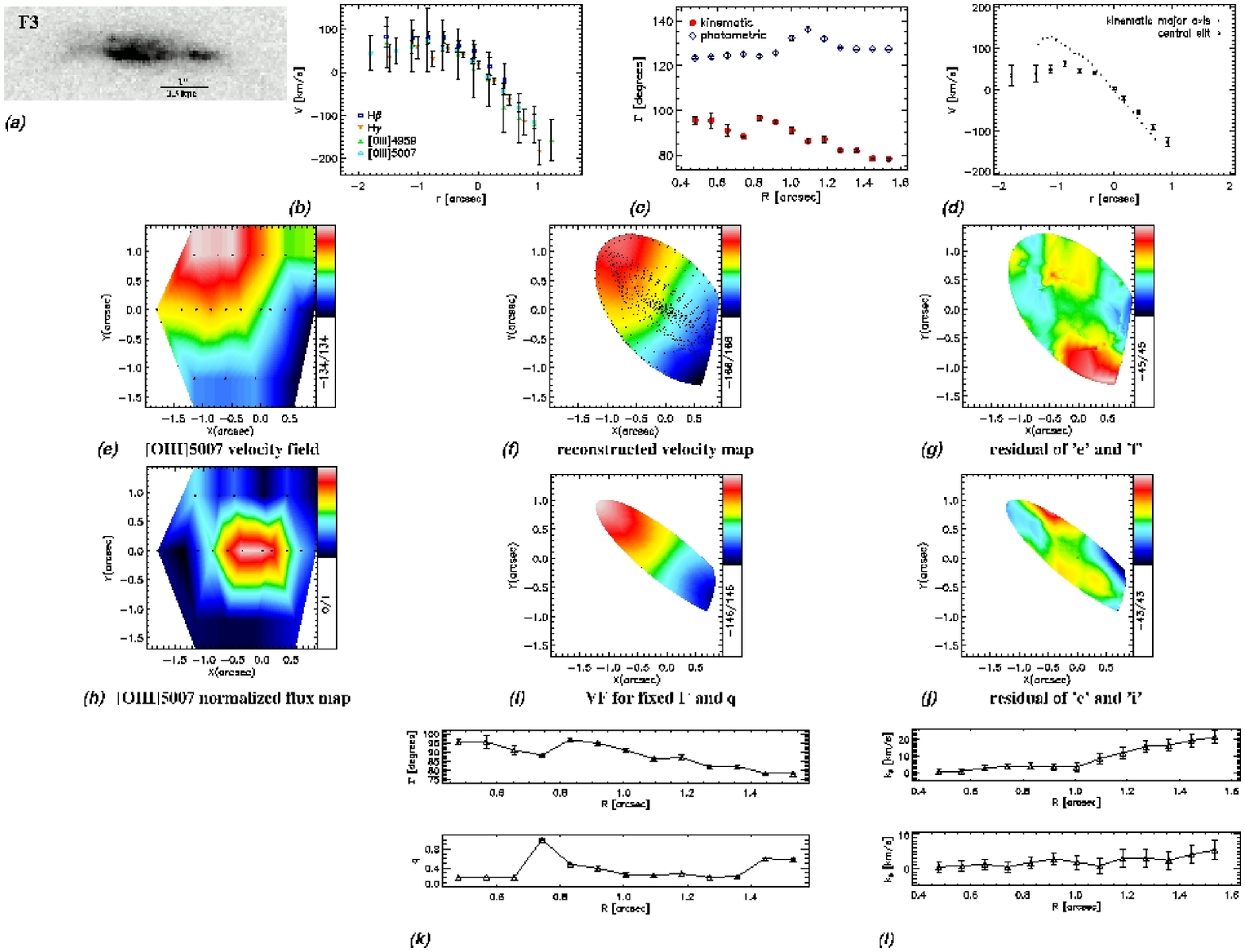}
 \caption{
 \textit{a)} HST-ACS image of the galaxy in the I band.
 \textit{b)} Rotation curves of different emission lines
extracted along the central slit.
\textit{c)} Position angles of kinematic and photometric axes as a
	function of radius.
\textit{d)} Rotation curves extracted along the central
	slit and the kinematic major axis.
\textit{e)} $[OIII]5007$ velocity field.
\textit{f)} Velocity map reconstructed using 6 harmonic terms.
\textit{g)} Residual of the velocity map and the reconstructed map.
\textit{h)} Normalized [OII]3727 flux map.	
\textit{i)} Simple rotation map constructed for position angle and
	ellipticity fixed to their global values.
\textit{j)} Residual of the velocity map and the simple rotation map.
\textit{k)} Position angle and flattening as a function of radius.
\textit{l)} $k_{3}/k_{1}$ and $k_{5}/k_{1}$ (from the analysis where position angle and
	ellipticity are fixed to their global values) as a function of
	radius.
	}
         \label{gal4fig}
   \end{figure*}

Background galaxy at $z=0.5667$ (Fig.\ref{gal4fig}).  It has irregular gas kinematics
according to the $\Delta \phi$ criterion $(\Delta \phi=39^{\circ}\pm9^{\circ})$. 
Classified as an irregular galaxy according to its photometric asymmetry and concentration
measurements (Sect.3.2, Fig.\ref{asymconc}).

\textbf{Galaxy F4:}

\begin{figure*}
   \includegraphics[height=\textwidth, angle=270]{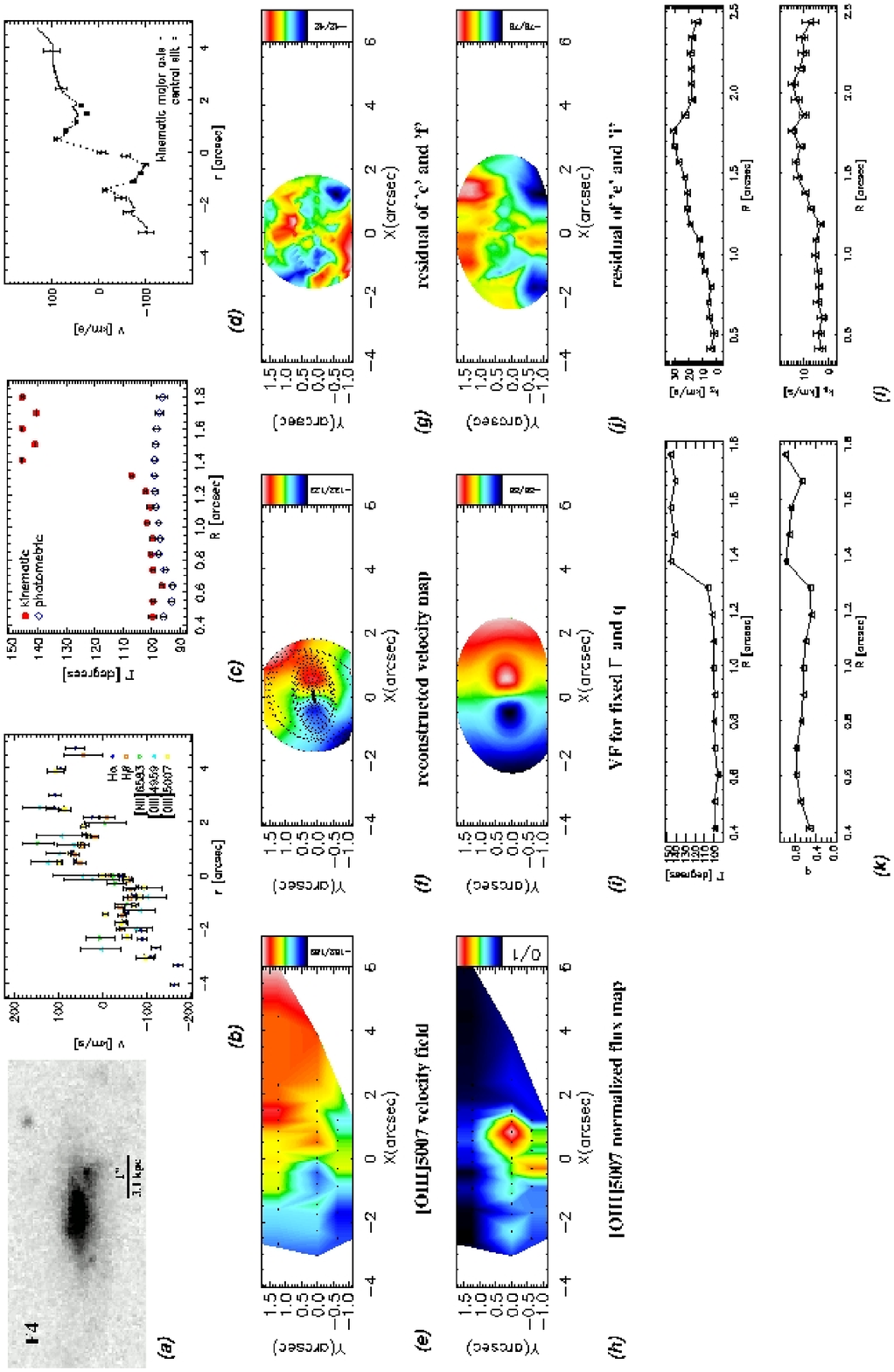}
 \caption{
 \textit{a)} HST-ACS image of the galaxy in the I band.
 \textit{b)} Rotation curves of different emission lines
extracted along the central slit.
\textit{c)} Position angles of kinematic and photometric axes as a
	function of radius.
\textit{d)} Rotation curves extracted along the central
	slit and the kinematic major axis.
\textit{e)} $[OIII]5007$ velocity field.
\textit{f)} Velocity map reconstructed using 6 harmonic terms.
\textit{g)} Residual of the velocity map and the reconstructed map.
\textit{h)} Normalized [OIII]5007 flux map.
\textit{i)} Simple rotation map constructed for position angle and
	ellipticity fixed to their global values.
\textit{j)} Residual of the velocity map and the simple rotation map.
\textit{k)} Position angle and flattening as a function of radius.
\textit{l)} $k_{3}/k_{1}$ and $k_{5}/k_{1}$ (from the analysis where position angle and
	ellipticity are fixed to their global values) as a function of
	radius.
	}
         \label{gal3fig}
   \end{figure*}

Foreground galaxy at $z=0.1867$ (Fig.\ref{gal3fig}).  It has irregular gas
kinematics according to the $\sigma_{PA}$ and $k_{3,5} / k_{1}$ criteria.  After R
$\sim$ 1\farcs3, there is a substantial change in its kinematic position angle
and ellipticity which may be the indicator of a secondary component. 

\textbf{Galaxy F5:}

Foreground galaxy at z=$0.1573$.  The spectral range of our observations cover
several emission lines at this redshift.  Rotation curves extracted
along the central slit using these lines are plotted together in Fig.A.12.b.  It
has irregular kinematics according to the $\Delta \phi$ criterion.  The $k_{3,5} /
k_{1}$ is not meaningful since the measurements cover only the inner part of the
velocity field in this particular case (see Fig.A.12.i).  The reason is that
$k_{3,5} / k_{1}$ is calculated while fixing the position angle to its global
value and the misalignment between this angle and the slit position angle is very
large.  Therefore $k_{3,5} /k_{1}$ of this galaxy is excluded from
our analysis.

\textbf{Galaxy F6:}

Foreground galaxy at $z=0.0982$. As the sudden change in its kinematic position
angle indicates (Fig.\ref{gal20fig}.k), and can be seen in its $H\alpha$
velocity field (Fig.\ref{gal20fig}.e), it has a counter rotating core.  Related
to this is the fact that it has irregular gas kinematics according to all the criteria we
defined:$\Delta \phi=57^{\circ}\pm44^{\circ}$, $\sigma PA=29^{\circ}$, $k_{3,5} /
k_{1}=0.27\pm0.18$.

\textbf{Galaxy F7:}

Only one emission line is visible in its spectrum and no other feature can be
identified.  Therefore we could not measure its redshift and used the value given in
the literature for our analysis \citep[z=0.9125,][]{M08}.  It has regular kinematics
according to all the criteria we defined ($\sigma_{PA}$, $k_{3,5} / k_{1}$ and $\Delta
\phi$) (Fig.\ref{gal12fig}).  Using its photometry (asymmetry and concentration
indexes), it is classified as an irregular/peculiar galaxy (Fig. \ref{asymconc}).

%-------------------------------------------------------------
%   % \begin{landscape}
%   % \end{landscape}
%
%_____________________________________________________________

%\clearpage

%-------)))))))-----------------------------------------------
      % \begin{landscape}
  % \end{landscape}
%
%_____________________________________________________________
%\clearpage
%-------)))))))-----------------------------------------------
   % \begin{landscape}
     % \end{landscape} 
%
%_____________________________________________________________
%
%\clearpage
%-------)))))))-----------------------------------------------
  % \begin{landscape} 
   % \end{landscape}
%
%_____________________________________________________________

%\clearpage
%-------)))))))-----------------------------------------------
% \begin{landscape}
   \begin{figure*}
   \begin{center} 
   \includegraphics[height=\textwidth, angle=270]{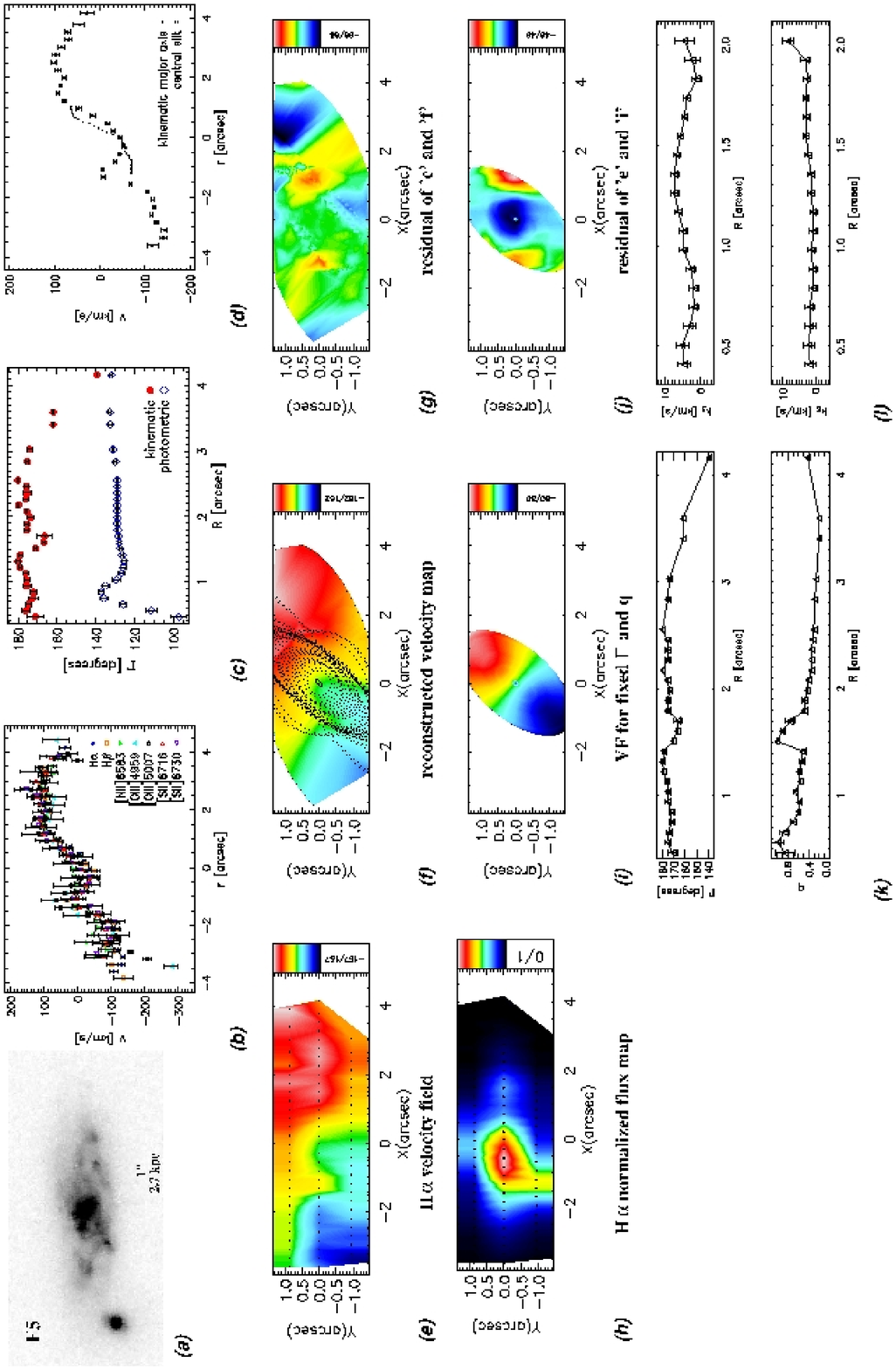}
 \caption{
 \textit{a)} HST-ACS image of the galaxy in the I band.
 \textit{b)} Rotation curves of different emission lines
extracted along the central slit.
\textit{c)} Position angles of kinematic and photometric axes as a
	function of radius.
\textit{d)} Rotation curves extracted along the central
	slit and the kinematic major axis.
\textit{e)} $H\alpha$ velocity field.
\textit{f)} Velocity map reconstructed using 6 harmonic terms.
\textit{g)} Residual of the velocity map and the reconstructed map.
\textit{h)} Normalized $H\alpha$ flux map.	
\textit{i)} Simple rotation map constructed for position angle and
	ellipticity fixed to their global values.
\textit{j)} Residual of the velocity map and the simple rotation map.
\textit{k)} Position angle and flattening as a function of radius.
\textit{l)} $k_{3}/k_{1}$ and $k_{5}/k_{1}$ (from the analysis where position angle and
	ellipticity are fixed to their global values) as a function of
	radius.
	}
	\end{center}
         \label{gal2fig}
   \end{figure*}
   	% \end{landscape}
%
%_____________________________________________________________

%\clearpage
%-------)))))))-----------------------------------------------
   % \begin{landscape}
   \begin{figure*}
   \includegraphics[height=\textwidth, angle=270]{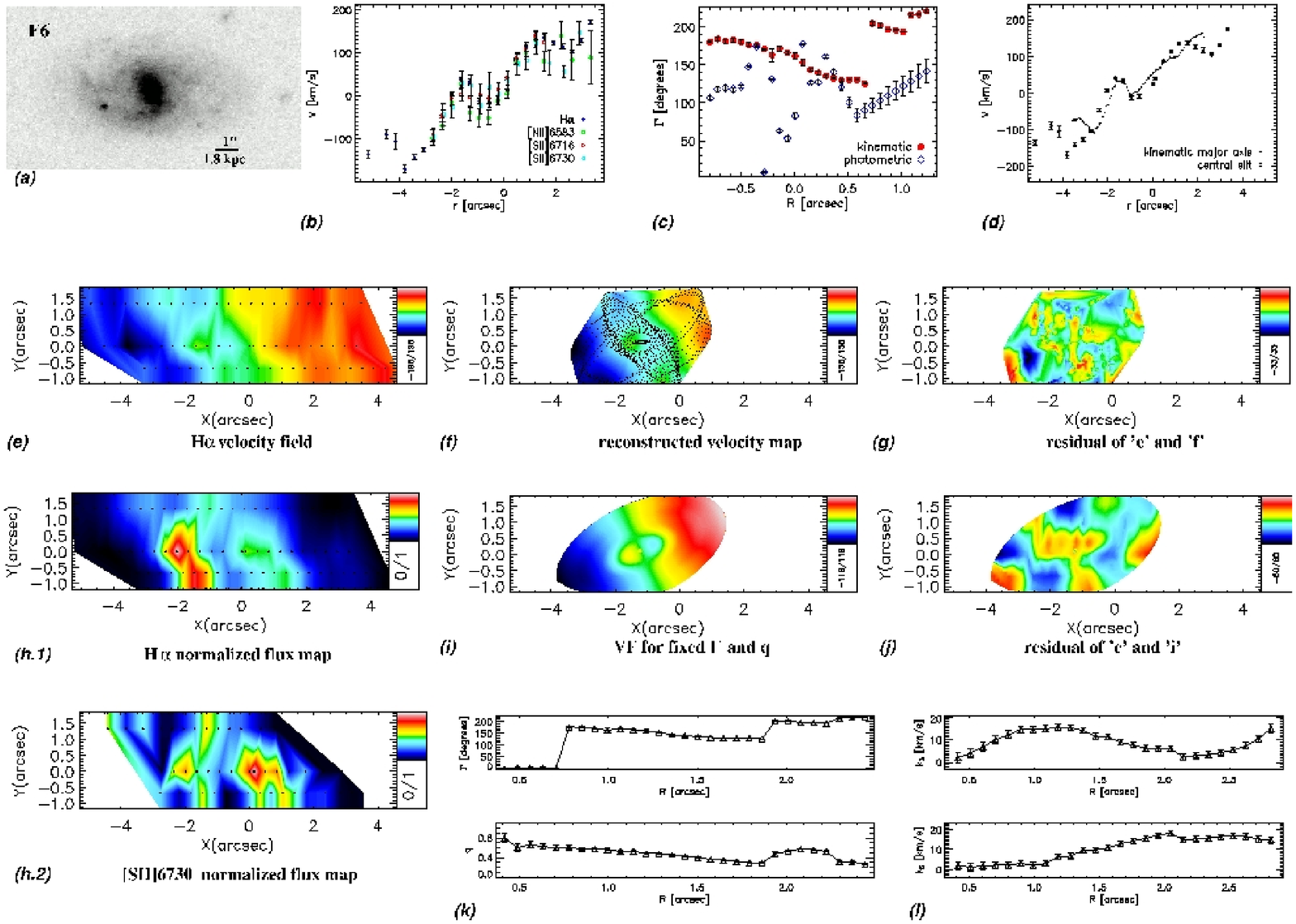}
 \caption{
\textit{a)} HST-ACS image of the galaxy in the I band.
\textit{b)} Rotation curves of different emission lines
extracted along the central slit.
\textit{c)} Position angles of kinematic and photometric axes as a
	function of radius.
\textit{d)} Rotation curves extracted along the central
slit and the kinematic major axis.
\textit{e)} $H\alpha$ velocity field.
\textit{f)} Velocity map reconstructed using 6 harmonic terms.
\textit{g)} Residual of the velocity map and the reconstructed map.
\textit{h.1)} Normalized $H\alpha$ flux map.	
\textit{h.2)} Normalized S[II]6730 flux map.
\textit{i)} Simple rotation map constructed for position angle and
ellipticity fixed to their global values.
\textit{j)} Residual of the velocity map and the simple rotation map.
\textit{k)} Position angle and flattening as a function of radius.
\textit{l)} $k_{3}/k_{1}$ and $k_{5}/k_{1}$ (from the analysis where position angle and
	ellipticity are fixed to their global values) as a function of
	radius.
	}
         \label{gal20fig}
   \end{figure*}
   % \end{landscape}
%
%_____________________________________________________________

%\clearpage
%-------)))))))-----------------------------------------------
   % \begin{landscape}
   \begin{figure*}
   \includegraphics[height=\textwidth, angle=270]{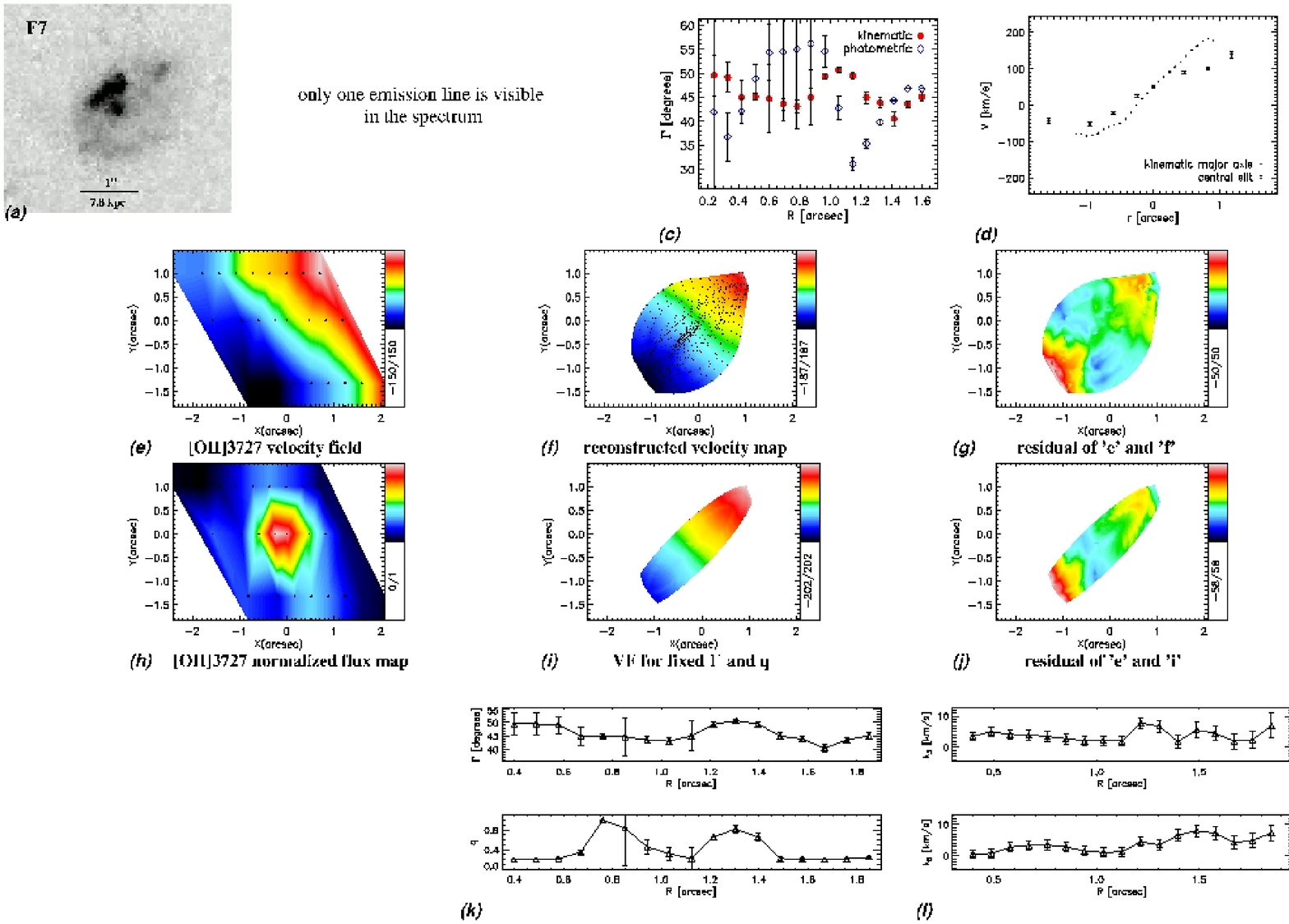}
 \caption{
 \textit{a)} HST-ACS image of the galaxy in the I band.
\textit{c)} Position angles of kinematic and photometric axes as a
	function of radius.
\textit{d)} Rotation curves extracted along the central
	slit and the kinematic major axis.		
\textit{e)} [OII]3727 velocity field.
 \textit{f)} Velocity map reconstructed using 6 harmonic terms.
\textit{g)} Residual of the velocity map and the reconstructed map.
\textit{h)} Normalized [OII]3727 flux map.
\textit{i)} Simple rotation map constructed for position angle and
	ellipticity fixed to their global values.
\textit{j)} Residual of the velocity map and the simple rotation map.
\textit{k)} Position angle and flattening as a function of radius.
\textit{l)} $k_{3}/k_{1}$ and $k_{5}/k_{1}$ (from the analysis where position angle and
	ellipticity are fixed to their global values) as a function of
	radius.
	}
         \label{gal12fig}
   \end{figure*}
    % \end{landscape}  
%
%_____________________________________________________________
\clearpage

\subsection{Galaxies difficult to treat}

Here we give information on the galaxies that were excluded from the analysis
because of the problems explained below for individual cases.  This section also
includes the galaxies for which the redshift is uncertain, since only one line
is visible in the spectrum.  In these cases, different possibilities for
identification of the line rule out that these galaxies are cluster members. 

\indent \textbf{Galaxy F8:}

\begin{figure}
   \includegraphics[height=\columnwidth, angle=270]{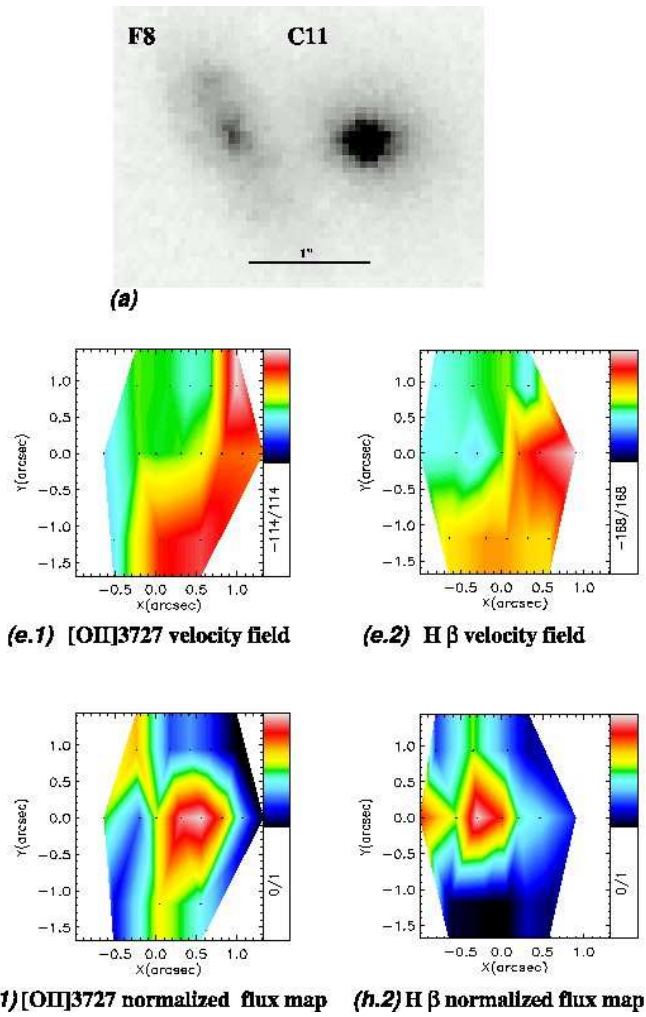}
 \caption{
\textit{a)} HST-ACS image of the galaxies in the I band.
\textit{e.1)} [OII]3727 velocity field constructed using composite
	spectra of the two objects.
\textit{e.2)} $H\beta$ velocity field constructed using composite
	spectra of the two objects.
\textit{h.1)} Normalized [OII]3727 flux map.
\textit{h.2)} Normalized $H\beta$ flux map.	 
	}
         \label{gal1fig}
   \end{figure}

This object is so close together to galaxy C11 that on the FORS-spectra, they
can not be distinguished (Fig.\ref{gal1fig}).  The redshift calculated using the
emission lines in the composite spectrum is $z=0.4443$.  Using the information
from the literature \citep{EYAMC98} we confirmed that the redshift we measure
belongs to galaxy F8.  Since the spectra can not be separated, the velocity fields
are very noisy and not reliable.  Therefore we excluded this object from our
analysis. 

\textbf{Galaxy F9:}

\begin{figure}
   \includegraphics[height=\columnwidth, angle=270]{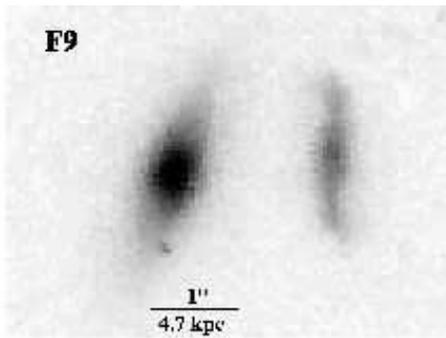}
 \caption{
 HST-ACS image of the galaxies in the I band.
 }
         \label{gal21fig}
   \end{figure}

Two objects were observed within the same slit with only one slit position
(Fig.\ref{gal21fig}).  A part of the spectrum of galaxy F9 is outside the frame.  It is
at z=0.3259 and has no emission.  This galaxy is classified as an E/S0 using its
asymmetry-concentration parameters (Sect.3.2, Fig.\ref{asymconc}).  It has an
exponential disk component according to our surface photometry results (Table
\ref{tab4}), so it could be classified as an S0 galaxy.  The other spectrum is so weak,
that no line can be identified in the noise. 

\textbf{Galaxy F10:}

\begin{figure*}
   \includegraphics[height=\textwidth, angle=270]{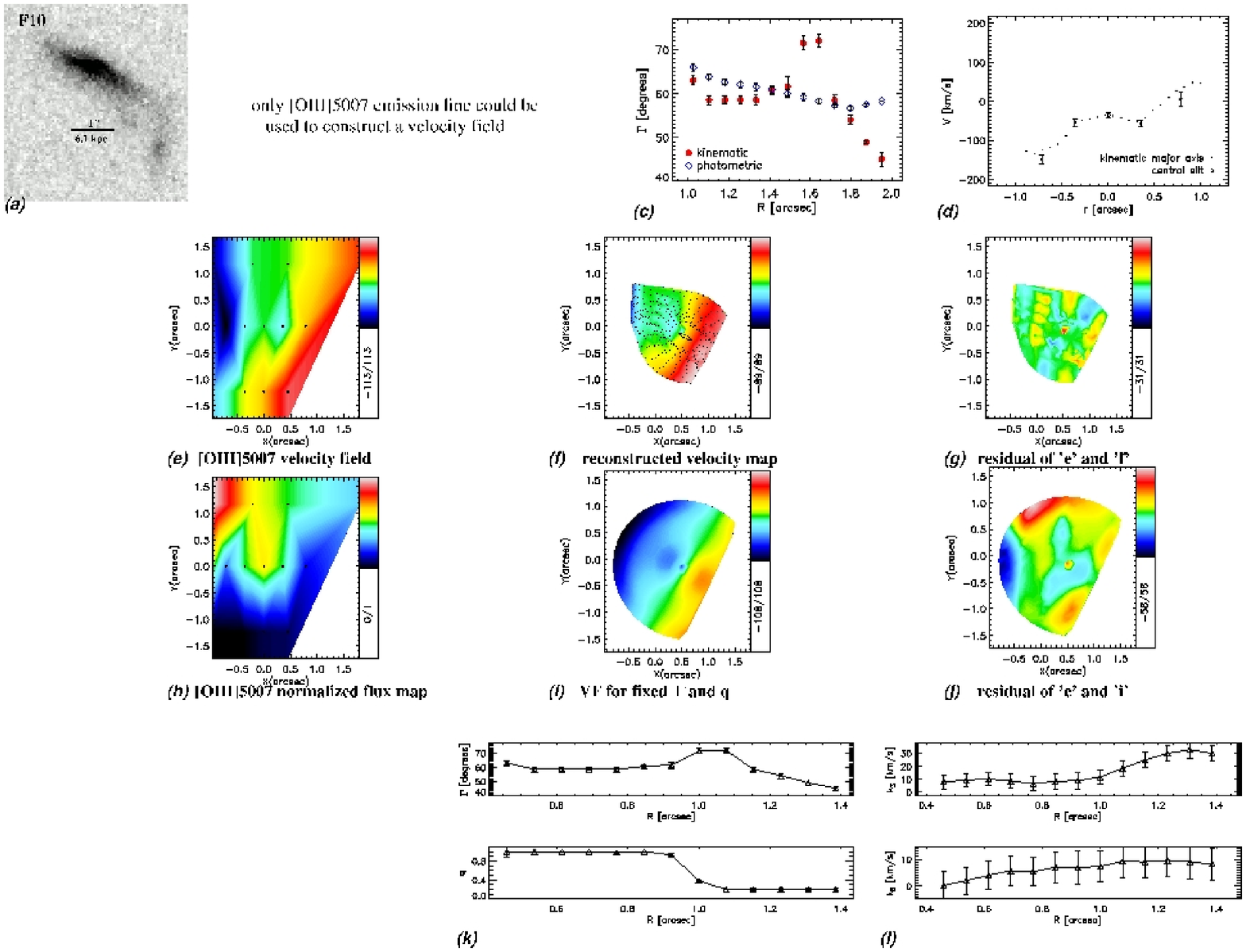}
 \caption{
 \textit{a)} HST-ACS image of the galaxy in the I band.
\textit{c)} Position angles of kinematic and photometric axes as a
	function of radius.
\textit{d)} Rotation curves extracted along the central
	slit and the kinematic major axis.		
\textit{e)} $[OIII]5007$ velocity field.
 \textit{f)} Velocity map reconstructed using 6 harmonic terms.
\textit{g)} Residual of the velocity map and the reconstructed map.
\textit{h)} Normalized [OIII]5007 flux map.
\textit{i)} Simple rotation map constructed for position angle and
	ellipticity fixed to their global values.
\textit{j)} Residual of the velocity map and the simple rotation map.
\textit{k)} Position angle and flattening as a function of radius.
\textit{l)} $k_{3}/k_{1}$ and $k_{5}/k_{1}$ (from the analysis where position angle and
	ellipticity are fixed to their global values) as a function of
	radius.
	}
         \label{gal13fig}
   \end{figure*}

Foreground galaxy at $z=0.4947$.  It has a tidal tail which indicates a
disturbance (Fig.\ref{gal13fig}).  The [OII]3727 emission line in its spectrum could not be used since
it is very noisy with the contribution from a strong sky line.  Its velocity field
constructed using the second prominent line ($[OIII]5007$) is quite noisy as well,
so this galaxy is excluded from our analysis.

\textbf{Galaxy F11:}

\begin{figure*}
   \includegraphics[height=\textwidth, angle=270]{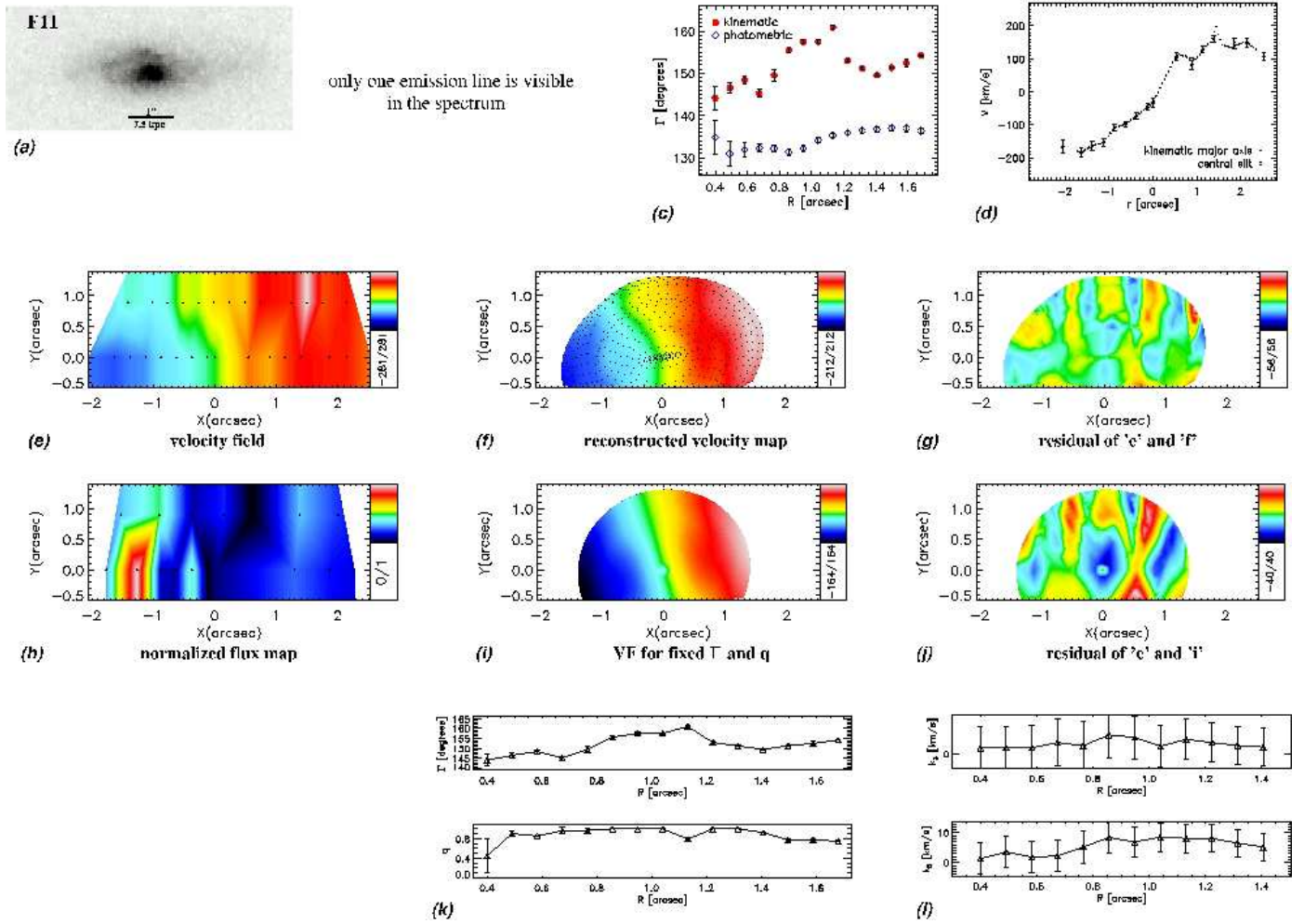}
 \caption{
\textit{a)} HST-ACS image of the galaxy in the I band.
\textit{c)} Position angles of kinematic and photometric axes as a
	function of radius.
\textit{d)} Rotation curves extracted along the central
	slit and the kinematic major axis.		
\textit{e)} Velocity field constructed using the emission line which could not
be identified.
\textit{f)} Velocity map reconstructed using 6 harmonic terms.
\textit{g)} Residual of the velocity map and the reconstructed map.
\textit{h)} Normalized flux map of the emission line which could not be
identified.
\textit{i)} Simple rotation map constructed for position angle and
	ellipticity fixed to their global values.
\textit{j)} Residual of the velocity map and the simple rotation map.
\textit{k)} Position angle and flattening as a function of radius.
\textit{l)} $k_{3}/k_{1}$ and $k_{5}/k_{1}$ (from the analysis where position angle and
	ellipticity are fixed to their global values) as a function of
	radius.
	}
         \label{gal19fig}
   \end{figure*}

Only one emission line is visible in its spectrum, and no other feature could be
identified.  Therefore its redshift is not certain.  Its photometric redshift is
available in the literature (see Table \ref{centdist}), but we did not use it in
our analysis.  Since the
signal in one of the slits is not enough to extract a rotation curve,
data from the two other slits were used to construct its velocity field
(Fig.\ref{gal19fig}).  It has regular kinematics according to all the criteria
we defined ($\sigma_{PA}$, $k_{3,5} / k_{1}$  and $\Delta \phi$).

%\clearpage
%------------------------------------------------------------
%   % \begin{landscape}
%   % \end{landscape}
%
%_____________________________________________________________

%-------)))))))-----------------------------------------------
%   % \begin{landscape}
%   % \end{landscape}
%
%_____________________________________________________________

\clearpage

%-------)))))))-----------------------------------------------
      % \begin{landscape}
   % \end{landscape}
%
%_____________________________________________________________

%\clearpage

%-------)))))))-----------------------------------------------
   % \begin{landscape}
   % \end{landscape}
%
%_____________________________________________________________
\clearpage

\section{Photometric tables for the selected objects}

\begin{table*}[h]
\caption{Photometric parameters for the MS0451 sample}
\label{tabrun2}
\begin{center}
\begin{tabular}{lllccccccccc}

\hline\hline
ID &    RA      &       DEC &     z    &    V  &    R  &    I  & $M_B$   &  A  & C \\
$(1)$ & $(2)$ &$(3)$ & $(4)$ &$(5)$ & $(6)$ &$(7)$ & $(8)$ &$(9)$ & $(10)$ \\
\hline
C1 & 04:54:2.2   & -02:57:10 & 0.5421 & 21.94 & 20.85 & 19.95 & -21.06  & 0.25219    &     0.32179 \\
C2 & 04:54:17.2  & -03:01:56 & 0.5486 & 22.02 & 21.00 & 20.07 & -20.91  & 0.15682    &     0.34844 \\
C3 & 04:54:18.6  & -03:01:03 & 0.5465 & 22.16 & 21.20 & 20.49 & -20.68  & 0.07922    &     0.27806 \\
C4 & 04:54:17.6  & -02:59:23 & 0.5324 & 22.34 & 21.41 & 20.72 & -20.37  & 0.36244    &     0.33291 \\
C5 & 04:54:1.5   & -02:59:24 & 0.5312 & 22.42 & 21.01 & 19.81 & -20.82  & 0.15670    &     0.52285 \\
C6 & 04:54:1.3   & -02:59:22 & 0.5305 & 21.83 & 21.15 & 20.59 & -20.59  & 0.32947    &     0.40650 \\
C7 & 04:54:5.0   & -02:59:40 & 0.5277 & 19.75 & 19.23 & 18.86 & -22.45  & 0.32994    &     0.20636 \\
C8 & 04:54:4.4   & -03:00:14 & 0.5325 & 20.80 & 20.10 & 19.46 & -21.68  & 0.27380    &     0.27789 \\
C9 & 04:54:15.0  & -03:00:22 & 0.5246 & 20.56 & 19.89 & 19.46 & -21.76  & 0.33459    &     0.26970 \\
C10 & 04:54:17.7 & -03:02:29 & 0.5312 & 21.92 & 21.61 & 21.21 & -20.05  & 0.19672    &     0.21096 \\
C11 & 04:54:10.3 & -03:00:46 & -      &   -    &   -    &   -    &   -	&   -	     &  	-  \\
F1 & 04:54:2.7  & -02:58:41 & 0.9009 & 23.27 & 22.75 & 21.46 & -20.97  & 0.18516    &	  0.18578 \\
F2 & 04:54:3.8  & -02:59:19 & 0.5795 & 20.88 & 20.11 & 19.42 & -21.96  & 0.27311    &	  0.22909 \\
F3 & 04:54:14.9 & -02:58:17 & 0.5667 & 21.88 & 21.24 & 20.64 & -20.66  & 0.33834    &	  0.20427 \\
F4 & 04:54:7.8  & -03:00:33 & 0.1867 & 20.54 & 20.37 & 20.00 & -18.78  & 0.17365    &	  0.32308 \\
F5 & 04:54:10.0 & -03:00:28 & 0.1573 & 19.05 & 18.72 & 18.26 & -19.95  & 0.19844    &	  0.30783 \\
F6 & 04:54:18.9 & -03:00:05 & 0.0982 & 19.23 & 18.93 & 18.55 & -18.56  & 0.14823    &	  0.27114 \\
F7 & 04:54:2.3  & -02:58:10 & 0.9125 & 22.22 & 21.91 & 20.73 & -21.64  & 0.42715  &	 0.16334 \\
F8 & 04:54:10.4 & -03:00:47 & 0.4443 & - & - & - &    -     &   -	    &	       -  \\
F9 & 04:54:10.2 & -03:01:57 & 0.3259 & 20.83 & 20.03 & 19.08 & -20.15  & 0.29565    &     0.53731 \\
F10 & 04:54:0.5  & -02:58:15 & 0.4947 & 22.47 & 21.86 & 21.38 & -19.67  & 0.20119    &     0.23420 \\
F11 & 04:54:20.6 & -03:00:16 & -      & 21.25 & 20.67 & 20.04 & -	& 0.18512   &	 0.27344 \\

\hline
\end{tabular}
\end{center}

Column (1) Object ID. \\
Column (2,3) RA and DEC (J2000). \\
Column (4) Object redshift. \\
Column (5,6,7) $V$, $R$, $I$ extinction corrected total magnitudes from
the FORS2/VLT images. \\
Column (8) Rest frame Johnson-$B$ magnitudes k-corrected with {\tt kcorrect} \citep{BR07}. \\
Column (9) Asymmetry index. \\
Column (10) Concentration index. \\

{\it For F11, the redshift is uncertain as explained in
Sect.2.3.  Redshift of galaxy F7 and C11 are taken from \citet{M08}.  Galaxy C11
and F8 are so close together on the FORS images that their magnitudes could not be
measured separately.}  \\

\end{table*}

\begin{table*}[h]
\caption{Morphological parameters for the MS0451 sample}
\label{tab4}
\begin{center}
\begin{tabular}{llccccccc}

\hline\hline
ID & comp & F814W & $R_{e}/R_{d} (kpc)$ & n & q & PA & $\chi^{2}$ & B/D\\
$(1)$ & $(2)$ &$(3)$ & $(4)$ &$(5)$ & $(6)$ &$(7)$ & $(8)$  & $(9)$  \\
\hline      
       C1 & exp disk &20.04$\pm$0.01 	& 3.182   $\pm$ 0.016  & -		  & 0.31$\pm$0.00	 & 6.68$\pm$0.16    & 	1.95    & -  	\\    
      C2 & exp disk &20.35$\pm$0.01 	& 4.057   $\pm$ 0.029  & -		  & 0.68$\pm$0.00	 & 130.45$\pm$0.56  & 	1.62    & 0.20      \\
   	 & S\'{e}rsic bulge &22.08$\pm$0.07 	& 0.840   $\pm$ 0.071  & 2.37$\pm$0.20	  & 0.99$\pm$0.02	 & -	   & 	  -  	& 	   \\    
      C3 & exp disk &20.44$\pm$0.00 	& 3.643   $\pm$ 0.016  & -		  & 0.23$\pm$0.00	 & 107.48$\pm$0.10  & 	1.89    &  - \\
      C4 & exp disk & 20.67$\pm$0.00	& 2.334   $\pm$ 0.013  & -		  & 0.16$\pm$0.00	 & 351.81$\pm$0.08  & 	2   	&  - \\
       C5 & exp disk & 21.33$\pm$0.03	& 2.646   $\pm$ 0.044  & -		  & 0.19$\pm$0.00	 & 133.62$\pm$0.16 & 	1.93   &  2.05     \\
  	  & S\'{e}rsic bulge & 20.55$\pm$0.02	& 1.619   $\pm$ 0.035  & 3.73$\pm$0.09	  & 0.47$\pm$0.01	 & -	   & 	   -	&   \\    
       C6 & exp disk & 21.08$\pm$0.00	& 1.070   $\pm$ 0.006  & -		  & 0.90$\pm$0.01	 & 91.42$\pm$2.85  & 	2.56	&  - 	 \\   
       C7 & exp disk & 18.87$\pm$0.00	& 5.086   $\pm$ 0.019  & -		  & 0.78$\pm$0.00	 & 104.2$\pm$0.51  & 	2.72	&  - 	 \\
       C8 & exp disk & 19.62$\pm$0.00	& 4.983   $\pm$ 0.025  & -		  & 0.64$\pm$0.00	 & 116.19$\pm$0.40  & 	2.24       &  0.05 \\
	  & S\'{e}rsic bulge & 22.86$\pm$0.02	& 0.419   $\pm$ 0.016  & 1.43$\pm$0.19	  & 0.74$\pm$0.03	 & -	   & 	   -  &   	\\
      C9 & exp disk & 19.48$\pm$0.00	& 3.693   $\pm$ 0.016  & -		  & 0.50$\pm$0.00	 & 110.27$\pm$0.23  & 	3.56  &  -   \\
      C10 & exp disk  &21.62$\pm$0.01 	& 1.071  $\pm$ 0.009  & -		  	& 0.80$\pm$0.01	 & 30.22$\pm$2.13  & 	2.65  &  -   \\ 
      F1 & exp disk &21.43$\pm$0.01	& 3.351   $\pm$ 0.039  & -		  & 0.51$\pm$0.00	 & 19.77$\pm$0.58   & 	1.54  &  -   \\
       F2 & exp disk & 19.41$\pm$0.00      & 3.880   $\pm$ 0.016  & -		 & 0.90$\pm$0.00	& 137.13$\pm$1.32 &    2.06   &  -  \\
       F3 & exp disk & 20.66$\pm$0.01      & 4.489   $\pm$ 0.026  & -		 & 0.22$\pm$0.00	& 123.95$\pm$0.12  &    1.61  &  -   \\
       F4 & exp disk & 20.08$\pm$0.00      & 2.062   $\pm$ 0.011  & -		 & 0.33$\pm$0.00	& 96.37$\pm$0.15   &    2.13  &  -   \\
       F5 & exp disk & 18.28$\pm$0.00      & 3.453   $\pm$ 0.007  & -		 & 0.36$\pm$0.00	& 129.11$\pm$0.07  &    2.41  &  0.05   \\
       	  & S\'{e}rsic bulge & 21.64$\pm$0.03	& 0.400   $\pm$ 0.018  & 1.84$\pm$0.11	  & 0.89$\pm$0.02	 & -	   & 	  -  &	\\
      F6 & exp disk &18.47$\pm$0.00 	& 2.432   $\pm$ 0.012  & -		  & 0.65$\pm$0.00	 & 179.96$\pm$0.24  & 	2.19 &  0.04   \\
    	 & S\'{e}rsic bulge &21.88$\pm$0.07 	& 0.500   $\pm$ 0.039  & 1.62$\pm$0.12	  & 1.00$\pm$0.03	 & -	   & 	  -  & 	\\    
      F7 & exp disk &20.96$\pm$0.01	& 3.175   $\pm$ 0.031  & -		  & 0.71$\pm$0.01	 & 47.28$\pm$0.93   & 	1.73 & -		\\ 
      F9 & exp disk & 20.36$\pm$0.02	& 2.451   $\pm$ 0.021  & -		  & 0.29$\pm$0.00	 & 9.36$\pm$0.18	   & 	2.05  	& 1.51   \\
    	 & S\'{e}rsic bulge & 19.91$\pm$0.02	& 1.602   $\pm$ 0.033  & 3.09$\pm$0.05	  & 0.60$\pm$0.00	 & -	   & 	  -  &	\\    
      F10 & exp disk &21.33$\pm$0.01 	& 3.581   $\pm$ 0.039  & -		  & 0.28$\pm$0.00	 & 61.08$\pm$0.26   & 	1.81 & -		\\
      F11 & exp disk &20.14$\pm$0.01 	& 0\farcs7   $\pm$  0\farcs007 & -		  & 0.37$\pm$0.00	 & 134.39$\pm$0.18  & 	1.92  &   0.12   \\
    	 & S\'{e}rsic bulge &22.44$\pm$0.12 	& 0\farcs245   $\pm$  0\farcs025 & 1.33$\pm$0.10	  & 0.85$\pm$0.03	 & -	   & 	  -  &	\\    

\hline					
\end{tabular}				
\end{center}

Column (1): Object ID. \\
Column (2): Component. \\
Column (3): Total magnitude. \\
Column (4): Effective radius of the bulge/scale length of the disk. \\
Column (5): S\'{e}rsic index of the bulge profile. \\
Column (6): Flattening. \\
Column (7): Position angle of the disk measured from North through East. \\
Column (8): $\chi^{2}$ of the fit. \\
Column (9): Bulge to disk ratio. \\

{\it The photometric zero point of the magnitude measurements is $Z_{p}^{F814W}=25.492$
\citep{BCCPASMP05}.  The redshift of galaxy F11 is uncertain as explained in Sect.2.3.  As a result, for
this galaxy, column (4) is given in arcseconds.  For galaxy F7, we used the redshift from \citet{M08}.  The
surface brightness profile of Galaxy C10 is much more extended on one side of the galaxy than on the other
side.  Therefore the scale length and total magnitude of the disk given in the table are not reliable. 
This galaxy was excluded while calculating the correlations between the disk scale length and the
irregularity in the gas kinematics.}
\end{table*}

\end{document}